\documentclass[preprint, astrosymb,%linenumbers,trackchanges,
resetfootnote]{aastex7}
\usepackage{comment}

\usepackage [english]{babel}
\usepackage [autostyle, english = american]{csquotes}
\MakeOuterQuote{"}

\usepackage{changepage}
\usepackage[flushleft]{threeparttable}

\newcommand{\fluxcgs}{erg~s$^{-1}$~cm$^{-2}$}
\newcommand{\lumcgs}{erg~s$^{-1}$}

\def\simgt{\lower.5ex\hbox{$\; \buildrel > \over \sim \;$}}
\def\simlt{\lower.5ex\hbox{$\; \buildrel < \over \sim \;$}}

\newcommand\chandra{{\it Chandra}}
\newcommand\xmm{{\it XMM-Newton}}

\newcommand\swift{{\it Swift\/}}
\newcommand\nustar{{\it NuSTAR}}

\newcommand\ixpe{{\it IXPE}}
\newcommand\xrism{{\it XRISM}}
\newcommand\maxi{{\it MAXI}}
\newcommand\mkat{{\it MeerKAT}}
\newcommand\nicer{{\it NICER}}

\newcommand\src{{MAXI J1744-294}}
\newcommand\axj{{AX J1745.6-2901}}
\newcommand\Swft{{Swift J174540.2-290037}}

\shorttitle{MAXI J1744-294}
\shortauthors{Mandel et al.}
\submitjournal{ApJ}
\graphicspath{{./}{}}

\begin{document}

\title{A multiwavelength study of the Galactic center black hole candidate MAXI J1744-294 }

\correspondingauthor{Shifra Mandel}
\email{ss5018@columbia.edu}

\author[0000-0002-6126-7409]{Shifra Mandel}
\altaffiliation{National Science Foundation Fellow}
\affiliation{Columbia Astrophysics Laboratory, Columbia University, New York, NY 10027, USA}
\email{ss5018@columbia.edu}

\author[0000-0002-9709-5389]{Kaya Mori} 
\affiliation{Columbia Astrophysics Laboratory, Columbia University, New York, NY 10027, USA}
\email{km211@columbia.edu}

\author[0000-0002-2218-2306]{Paul A. Draghis}
\affiliation{MIT Kavli Institute for Astrophysics and Space Research, Massachusetts Institute of Technology, Cambridge, MA 02139, USA}
\email{pdraghis@mit.edu}

\author[0000-0003-1621-9392]{Mark Reynolds} 
\affiliation{Department of Astronomy, Ohio State University, 140 West 18th Ave., Columbus, OH 43210}
\affiliation{Department of Astronomy, University of Michigan, 1085 S. University Ave., Ann Arbor, MI 48109}
\email{reynolds.1362@osu.edu}

\author[0000-0002-2006-1615]{Chichuan Jin} 
\affiliation{National Astronomical Observatories, Chinese Academy of Sciences, Beijing 100101, China}
\affiliation{School of Astronomy and Space Science, University of Chinese Academy of Sciences, Beijing 100049, China}
\affiliation{Institute for Frontier in Astronomy and Astrophysics, Beijing Normal University, Beijing 102206, China}
\email{ccjin@bao.ac.cn}

\author[0009-0003-8610-853X]{Maxime Parra} 
\affiliation{Department of Physics, Ehime University, 2-5, Bunkyocho, Matsuyama, Ehime 790-8577, Japan}
\email{maxime.parrastro@gmail.com}

\author[0000-0002-6789-2723]{Gaurava K. Jaisawal}
\affiliation{DTU Space, Technical University of Denmark, DK-2800 Lyngby, Denmark}
\email{gaurava@space.dtu.dk}

\author[0009-0008-1132-7494]{Benjamin Levin} 
\affiliation{Columbia Astrophysics Laboratory, Columbia University, New York, NY 10027, USA}
\email{bsl2134@columbia.edu}

\author[0009-0009-6455-3804]{Eric Miao} 
\affiliation{Columbia Astrophysics Laboratory, Columbia University, New York, NY 10027, USA}
\email{em3928@columbia.edu}

\author[orcid=0009-0001-4644-194X]{Lorenzo Marra}
\affiliation{INAF Istituto di Astrofisica e Planetologia Spaziali, Via del Fosso del Cavaliere 100, 00133 Roma, Italy}
\email{lorenzo.marra@inaf.it}  

\author[0000-0001-5800-3093]{Anna Ciurlo} 
\affiliation{Department of Physics and Astronomy, UCLA, Los Angeles, CA 90095-1547}
\email{ciurlo@astro.ucla.edu}

\author[0009-0005-0419-4038]{Sean~A.~Granados} 
\affiliation{Department of Physics and Astronomy, UCLA, Los Angeles, CA 90095-1547}
\email{seangranados@astro.ucla.edu}

\author[0009-0005-4358-5146]{Noa Grollimund} 
\affiliation{Université Paris Cité, Université Paris-Saclay, CEA, CNRS, AIM, F-91191 Gif-sur-Yvette, France}
\email{noa.grollimund@cea.fr}

\author[0000-0002-4576-9337]{Matteo Bachetti} 
\affiliation{INAF-Osservatorio Astronomico di Cagliari, via della Scienza 5, 09047 Selargius (CA), Italy}
\email{matteo.bachetti@inaf.it}

\author[0000-0002-6384-3027]{Fiamma Capitanio} 
\affiliation{INAF Istituto di Astrofisica e Planetologia Spaziali, Via del Fosso del Cavaliere 100, 00133 Roma, Italy}
\email{fiamma.capitanio@inaf.it}

\author[0000-0002-0092-3548]{Nathalie Degenaar} 
\affiliation{Anton Pannekoek Institute for Astronomy, University of Amsterdam, %Science Park 904, NL-1098 XH 
Amsterdam, the Netherlands}
\email{degenaar@uva.nl}

\author[0000-0002-3681-145X]{Charles~J.~Hailey}
\affiliation{Columbia Astrophysics Laboratory, Columbia University, New York, NY 10027, USA}
\email{chuckh@astro.columbia.edu}

\author[0000-0002-6089-5390]{JaeSub Hong}
\affiliation{Center for Astrophysics | Harvard \& Smithsonian, Cambridge, MA 02138, USA}
\email{jhong@cfa.harvard.edu}

\author[0000-0002-6154-5843]{Sara Motta} 
\affiliation{INAF-Osservatorio Astronomico di Brera, Via Bianchi 46, I-23807, Merate (LC), Italy}
\email{sara.motta@inaf.it}

\author[0000-0003-0293-3608]{Gabriele Ponti} 
\affiliation{INAF-Osservatorio Astronomico di Brera, Via Bianchi 46, I-23807, Merate (LC), Italy}
\affiliation{Max-Planck-Institut für extraterrestrische Physik, Garching, Germany}
\affiliation{Como Lake Center for Astrophysics (CLAP), DiSAT, Università degli Studi dell’Insubria, %via Valleggio 11, 
22100 Como, Italy}
\email{gabriele.ponti@inaf.it}

\author[0000-0003-0155-2539]{Michael M. Shara} 
\affiliation{Department of Astrophysics, American Museum of Natural History, New York, NY, USA}
\email{mshara@amnh.org}

\author[0000-0001-8195-6546]{Megumi Shidatsu} 
\affiliation{Department of Physics, Ehime University, 2-5, Bunkyocho, Matsuyama, Ehime 790-8577, Japan}
\email{shidatsu.megumi.wr@ehime-u.ac.jp}

\author[0000-0001-5506-9855]{John A. Tomsick}
\affiliation{Space Sciences Laboratory, University of California, Berkeley, CA 94720-7450, USA}
\email{jtomsick@ssl.berkeley.edu}

\author[0000-0002-3289-5203]{Randall Campbell} \affiliation{W.M. Keck Observatory, Kamuela, HI 96743}
\email{randyc@keck.hawaii.edu}

\author[0000-0001-5538-5831]{Stéphane Corbel} 
\affiliation{Université Paris Cité, Université Paris-Saclay, CEA, CNRS, AIM, F-91191 Gif-sur-Yvette, France}
\email{stephane.corbel@cea.fr}

\author[0000-0002-5654-2744]{Rob Fender} 
\affiliation{Astrophysics, Department of Physics, University of Oxford, Keble Road, Oxford OX1 3RH, UK}
\email{rob.fender@physics.ox.ac.uk}

\author[0000-0003-3230-5055]{Andrea Ghez} 
\affiliation{Department of Physics and Astronomy, UCLA, Los Angeles, CA 90095-1547}
\email{ghez@astro.ucla.edu}

\author[0000-0002-1323-5314]{Jonathan Grindlay}
\affiliation{Center for Astrophysics | Harvard \& Smithsonian, Cambridge, MA 02138, USA}
\email{jgrindlay@cfa.harvard.edu}

\author[0000-0001-6803-2138]{Daryl Haggard}
\affiliation{Department of Physics, McGill University, 3600 rue University, Montréal, Quebec City H3A 2T8, Canada}
\email{daryl.haggard@mcgill.ca}

\author[0000-0003-2874-1196]{Matthew W. Hosek Jr.}
\altaffiliation{Brinson Prize Fellow} 
\affiliation{Department of Physics and Astronomy, UCLA, Los Angeles, CA 90095-1547}
\email{mwhosek@astro.ucla.edu}

\author[orcid=0000-0002-7518-337X]{Ziqian Hua}
\affiliation{School of Astronomy and Space Science, Nanjing University, Nanjing 210023, China}
\affiliation{Key Laboratory of Modern Astronomy and Astrophysics (Nanjing University), %Ministry of Education, 
Nanjing 210023, China}
\email{zqhua@smail.nju.edu.cn}

\author[0000-0001-8670-4575]{Ole König}
\affiliation{Center for Astrophysics | Harvard \& Smithsonian, Cambridge, MA 02138, USA}
\email{ole.koenig@cfa.harvard.edu}

\author[0009-0003-0653-2913]{Kai Matsunaga}
\affiliation{Department of Physics, Graduate School of Science, Kyoto University, %Kitashirakawa Oiwake-cho, Sakyo-ku, 
Kyoto 606-8502, Japan}
\email{matsunaga.kai.i47@kyoto-u.jp}

\author[0000-0001-7374-843X]{Romana Miku\v{s}incov\'{a}}
\affiliation{INAF Istituto di Astrofisica e Planetologia Spaziali, Via del Fosso del Cavaliere 100, 00133 Roma, Italy}
\email{romana.mikusincova@inaf.it}

\author[0000-0002-3310-1946]{Melania Nynka} 
\affiliation{MIT Kavli Institute for Astrophysics and Space Research, Massachusetts Institute of Technology, Cambridge, MA 02139, USA}
\email{mnynka@mit.edu}

\author[0000-0002-4463-2902]{Grace Sanger-Johnson} 
\affiliation{Department of Physics and Astronomy, Michigan State University, East Lansing, MI 48824, USA}
\email{sangerjo@msu.edu}

\author[0009-0007-0585-9462]{Giovanni Stel} 
\affiliation{INAF-Osservatorio Astronomico di Brera, Via Bianchi 46, I-23807, Merate (LC), Italy}
\email{giovanni.stel@inaf.it}

\author[0009-0007-0537-9805]{Antonella Tarana}
\affiliation{INAF Istituto di Astrofisica e Planetologia Spaziali, Via del Fosso del Cavaliere 100, 00133 Roma, Italy}
\email{Antonella.tarana@inaf.it}

\author[0000-0002-3516-2152]{Rudy Wijnands} 
\affiliation{Anton Pannekoek Institute for Astronomy, University of Amsterdam, %Science Park 904, NL-1098 XH 
Amsterdam, the Netherlands}
\email{r.a.d.wijnands@uva.nl}

\author[0000-0002-2967-790X]{Shuo Zhang} 
\affiliation{Department of Physics and Astronomy, Michigan State University, East Lansing, MI 48824, USA}
\email{zhan2214@msu.edu}

\begin{abstract}

For the first time in nearly a decade, a bright transient was detected in the central parsec (pc) of the Galaxy.  MAXI J1744-294, or -- as it was known in its previous life -- Swift J174540.2-290037, was discovered in outburst by the \maxi\ telescope in January 2025.  
We present the results of a broadband, multi-wavelength study of \src, including data from the \nustar, \chandra, \xmm, \swift, and \nicer\ X-ray telescopes, as well as complementary radio and near-infrared observations.  We analyze the changing X-ray emission as the outburst evolved from the high/soft to the low/hard state.  
Using relativistic reflection features in the data, we estimate a spin of $a>0.92$ and viewing inclination $\theta=28^{+3}_{-4}$ deg, consistent with the parameters measured for Swift J174540.2-290037.  
Based on the spectral and temporal characteristics of \src, 
we reaffirm its classification 
as a candidate black hole (BH) low-mass X-ray binary (LMXB) -- the third
candidate BH transient discovered within 20 arcsec of the Galactic supermassive black hole Sgr~A*.   
This work provides further evidence for a cusp of BH-LMXBs in the central pc of our Galaxy, as argued for in previous observational studies 
and suggested by analytical and theoretical work.  
Our ongoing multi-wavelength study, involving a complementary range of observatories and spanning different outburst states, can serve as a model for future time domain astrophysics research.

\end{abstract}


\section{Introduction} \label{sec:intro}

The 1999 launch of the \chandra\ and \xmm\ X-ray observatories, with their superior angular resolution and 
sensitivity,  heralded a new era of Galactic center X-ray population studies.  
Indeed, 
a remarkably high number of X-ray transients have 
since been detected in the Galactic center 
by \chandra\ and \xmm, later joined by the \swift\
X-ray telescope \citep{Muno2005, Degenaar2012, Degenaar2015, Mori2019, Mori2021}.  
From a theoretical standpoint, the high abundance of X-ray transients was hardly surprising; predictions for an overdensity of stellar-mass black holes in the vicinity of Sgr A* -- the supermassive black hole (SMBH) at the center of our Galaxy -- predated the discovery of the first sources detected in central pc by almost a decade \citep{Morris1993}. Later analytical work also supported a large concentration of low-mass X-ray binaries (LMXBs) in the inner Galactic center region \citep{Generozov2018, Panamarev2019}.  Studies of faint X-ray sources in the central pc also suggest a large population of quiescent LMXBs \citep{Muno2006, Muno2009, Hailey2018, Mori2021}.

LMXBs consist of a compact object -- typically a black hole (BH) or neutron star (NS) -- that accretes material from a low-mass stellar companion.  Typically, mass transfer is achieved through Roche-lobe overflow.  LMXBs are known to undergo transient outbursts, during which their X-ray luminosities may increase by more than six orders of magnitude \citep{Tomsick2005, Belloni2016, Mori2019}. Transient LMXBs are key laboratories for studying high-energy astrophysical phenomena. Their  
outbursts can produce bright, multiwavelength signals that probe accretion physics, jet formation, and extreme 
relativistic effects.  
LMXBs are often discovered during outbursts; hence transients dominate the Galactic LMXB population.  Monitoring their outbursts provides critical insight into binary evolution, recurrence behavior, and the demographics of BHs and NSs.

Understanding the LMXB population in the Galactic center is especially important, as it offers key insights into binary formation and evolution in the deep gravitational well 
around a SMBH.  
Approximately 15\% of all known LMXBs in our Galaxy have been detected in the Galactic center \citep{Mori2021, Fortin2024}, which hosts by far the highest concentration of X-ray binaries in our Galaxy.  Between 1999 and 2016, a total of 20 new X-ray transients were observed within a radius of $\sim50$ pc from Sgr A* \citep{Muno2005, Degenaar2012, Degenaar2015, Mori2019, Mori2021}.  Many of these sources, including all that were identified as NS-LMXBs, underwent multiple outbursts in the last two-and-a-half decades \citep{Degenaar2015, Mori2021}.  However, for over eight years following the 2016 outbursts of two new BH-LMXB candidates, no bright X-ray transients were observed in the Galactic center.

This quiet spell came to an end on January 2, 2025, when the \maxi\ all-sky X-ray observatory detected a bright X-ray outburst in the Galactic center \citep{Kudo2025}.  A follow-up observation by \swift/XRT 
indicated that the source -- designated \src\ -- was a new X-ray transient, hitherto undetected \citep{Heinke2025}.  Further target of opportunity (ToO) observations by the \nustar, \xrism\, \xmm\, and \chandra\ X-ray observatories quickly followed \citep{Mandel2025b, Mandel2025c, Mandel2025d}, allowing us to analyze the outburst properties and localize the source to $\sim18$\arcsec\ south of Sgr A*.  Additional X-ray observations were conducted with \nicer\ \citep{Jaisawal2025}, \ixpe\ \citep{Marra2025}, and the Einstein Probe Follow-up X-ray Telescope (EP/FXT) \citep{Wang2025}.

In the radio band, \src\ was observed  
by \mkat\ starting on January 5, 2025 \citep{Grollimund2025} and 
monitored through September 2025 (Grollimund et al., in prep).  The Karl G. Jansky Very Large Array (VLA) also detected \src\ in April 2025 observations of the Galactic center \citep{Michail2025}.  Additionally, near-infrared (NIR) observations with Keck were pursued in an attempt to identify the NIR counterpart to \src.

\src's location in the crowded Galactic center presents unique challenges, including high absorption ($N_H \gtrsim 10^{23}$ cm$^{-2}$), bright diffuse background emission \citep{Mori2015}, and contamination from nearby sources, including Sgr A* flares \citep{Baganoff2001, Barriere2014}, the supernova remnant (SNR) Sgr A East, and the pulsar wind nebula (PWN) G359.95-0.04 (Figure \ref{fig:bkg}).  Of particular concern is the nearby transient NS-LMXB \axj, which was active during the outburst of \src.  And like \axj\ \citep{Jin2017, Jin2018}, \src\ is surrounded by a dust scattering halo that extends out to hundreds of arcsec.  \nustar\ data helps mitigate those effects, since high-energy ($>10$ keV) X-rays are not very sensitive to absorption or dust scattering.  

The combination of broadband ($3-79$ keV) \nustar\ data and softer ($<10$ keV) X-ray data from \swift, \xmm, and \chandra\ allows us to overcome the degeneracy between absorption and spectral hardness that plagues X-ray sources in regions with high line-of-sight $N_H$. Together, the complementary data allows for a comprehensive study of \src, comprising broadband spectral properties -- including reflection features (\nustar); source location and dust scattering halo (\chandra); soft X-ray coverage to constrain $N_H$ and the accretion disk temperature (\xmm\ and \swift); fast time resolution (\nustar); and exquisite energy resolution to reveal the neutral and highly ionized Fe emission line features in fine detail (\xrism, publication in prep).

\begin{figure}
    \centering
    \includegraphics[width=0.9\linewidth]{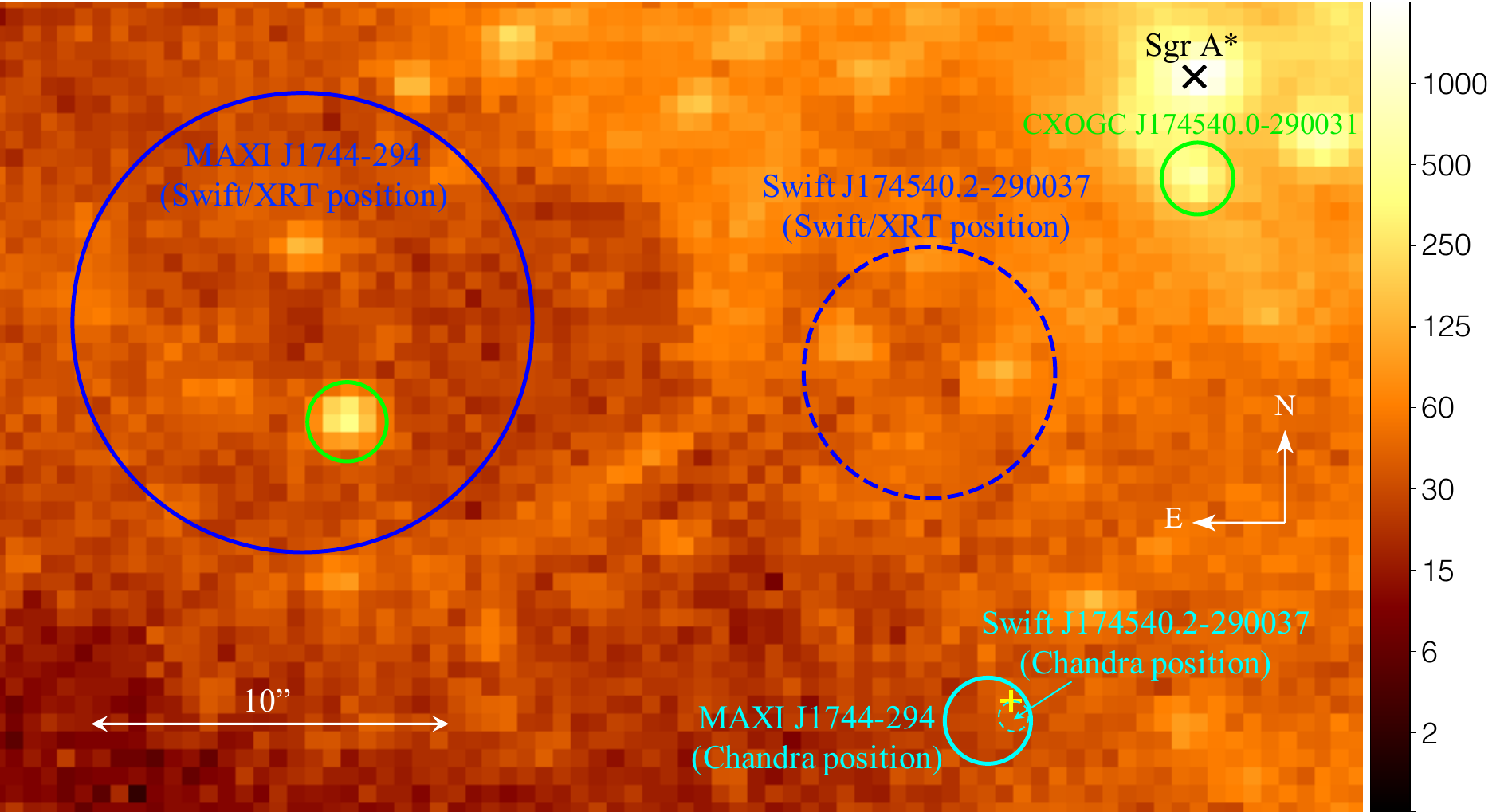}
    \caption{\chandra\ image of the region around \src.  The \swift/XRT positions of \src\ and \Swft\ are marked by blue solid and dashed circles, respectively.  Their respective \chandra\ positions are shown by cyan solid and dashed circles.  The radii of the circles correspond to position uncertainties.  The refined MeerKAT position for \src\ is marked by a yellow cross.  Other sources in the field of view (FoV) are outlined in green. }
    \label{fig:chandra_sw}
\end{figure}

Another consequence of \src's location in the crowded central pc is the difficulty in determining whether the outburst originates from a recurrent transient or a new source.  \src\ was initially classified as the latter based on its \swift/XRT position (R.A.$=$17:45:41.93, decl.$=$-29:00:35), which was far enough removed from other known transients to rule out an association \citep{Heinke2025}.  However, \src's \chandra\ coordinates (R.A.$=$17:45:40.476, decl.$=$-29:00:46.10; \cite{Mandel2025d}) are offset by $<1$ arcsec from the \chandra\ coordinates obtained for the 2016 transient \Swft\ (R.A.$=$17:45:40.42, decl.$=$-29:00:45.93 (J2000); \cite{Mori2019})\footnote{The latter is itself offset from the \swift/XRT coordinates for \Swft\ (R.A.$=$17:45:40.60, decl.$=$-29:00:36.4; \cite{Degenaar2016}) by almost $10$ arcsec.  Note that the \chandra\ and \swift/XRT coordinates for \src\ are offset by $>22$ arcsec.}.  Furthermore, the refined MeerKAT position for \src\ is $<0.5$ arcsec offset from \Swft's \chandra\ position.  Figure \ref{fig:chandra_sw} illustrates the \swift/XRT (blue) and \chandra\ (cyan) positions for both \src\ (solid outline) and \Swft\ (dashed outline).  On balance of the evidence, we conclude that \src\ is most likely a repeat outburst of the transient \Swft.  This makes \src/\Swft\ the first Galactic center BH candidate to be detected in outburst more than once.  We refer to the transient as \src\ throughout the rest of this paper.    

We present a comprehensive broadband X-ray study of the characteristics and evolution of the \src\ outburst, starting with the first \nustar\ observation in early February and continuing through September 2025.  
Our observations span a range of outburst states, reflected in the spectral changes we observed in the data.

In Section \ref{sec:data}, we describe the many X-ray observations of \src, as well as the initial data processing.  Section \ref{sec:analysis} provides a meticulous account of the data analysis.  We analyze the results and compare them to \Swft in Section \ref{sec:results}.  Finally, Section \ref{sec:concl} summarizes our findings.  Additional details and figures describing the data analysis are provided in the Appendix. \\

\section{X-ray observations and data reduction} \label{sec:data}

\setlength{\tabcolsep}{8pt}
\begin{table}
\caption{X-ray observations of \src.  Not listed below are near-daily observations by \swift/XRT and 
approximately a dozen additional observations by \nicer.  See Section \ref{sec:data} for more.  \label{tab:obs}}  
\begin{center}
\begin{tabular}{l l l c c c} 
\hline\hline
Instrument & Start date & Obs. ID & Exposure & $f_X$$^*$ (2-10 keV) & Outburst state \\  [0.11ex]
 & [mm/dd/yyyy]  &   &  [ks] & [\fluxcgs] &   \\ [0.5ex] 
\hline
\nustar & 02/06/2025 & 81001323001 & 22.7 & $1.4\times10^{-9}$ & soft \\
\nustar & 03/03/2025 & 81001323003 & 20.5 & $3.8\times10^{-9}$ & soft  \\
\xmm & 03/03/2025 & 0944100201 & 9.2 & $4.0\times10^{-9}$ & soft  \\
\nustar & 03/09/2025 & 81102302002 & 19.2 & $3.8\times10^{-9}$ & soft  \\
\chandra & 03/09/2025 & 29749  & 9.3 & -- & soft  \\
\nustar & 04/04/2025 & 31002004002 & 17.7 & $4.6\times10^{-9}$ & high/intermediate \\
\nustar & 04/06/2025 & 31002004004 & 18.9 & $4.5\times10^{-9}$ & high/intermediate\\
\nustar & 04/07/2025 & 31002004006 & 24.8 & $4.2\times10^{-9}$ & transition?  \\
\nustar & 04/09/2025 & 31002004008 & 18.9 & $4.5\times10^{-9}$ & high/intermediate\\
\nustar & 06/09/2025 & 51102002001 & 19.3 & $5.8\times10^{-10}$ & low/intermediate  \\
\nustar & 06/14/2025 & 51102002002 & 19.1 & $3.1\times10^{-10}$ & low/intermediate  \\
\nustar & 07/06/2025 & 51102002004 & 19.2 & $5.1\times10^{-11}$ & low/hard   \\
\nustar & 07/10/2025 & 51102002006 & 20.7 & $6.4\times10^{-11}$ & low/hard  \\
\nustar & 08/06/2025 & 51102002008 & 20 & $5.7\times10^{-11}$ & low/hard  \\
\nustar & 08/07/2025 & 31101013002 & 41 & $3.9\times10^{-11}$ & low/hard  \\
\nustar & 08/09/2025 & 31101013004 & 20 & $5.1\times10^{-11}$ & low/hard  \\
\nustar & 08/12/2025 & 31101013006 & 33 & $4.6\times10^{-11}$ & low/hard  \\
\nustar & 09/06/2025 & 81102328002 & 20 & $6.3\times10^{-10}$ &  bright/hard \\
\xmm & 09/06/2025 & 0944100101 & 40.5 & $4.9\times10^{-10}$ & bright/hard  \\
\hline 
\end{tabular} 
\end{center}
$^*$unabsorbed.  
\end{table}

\begin{figure}
    \centering
    \includegraphics[width=0.9\linewidth]{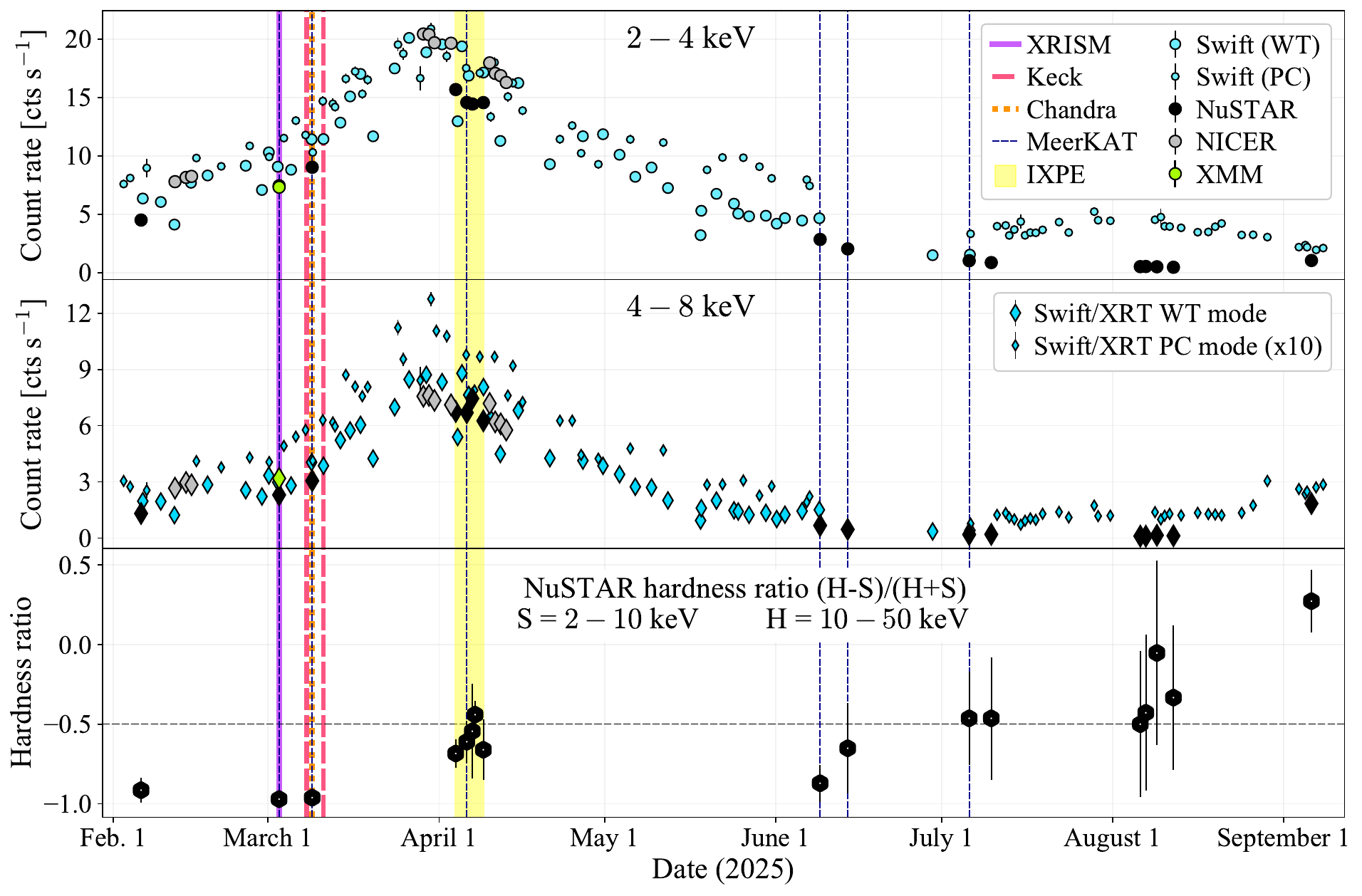}
    \caption{Top and center: \src\ light curves in the 2-4 keV and 4-8 keV energy bands, respectively.  \swift/XRT data are shown 
    in cyan, with larger markers for WT-mode observations and smaller markers for PC-mode data (scaled $\times10$ for visualization). Also shown are \nustar\ (black), \nicer\ (silver; scaled down by a factor of 10), and \xmm/EPIC (green) count rates.  The dates of \xrism, \chandra, \ixpe, MeerKAT, and Keck observations are marked by purple solid, orange dotted, yellow solid, dark blue dashed, and pink dashed lines, respectively.  Though plotted for all data points, errorbars are smaller than most markers.  
    Bottom panel: hardness ratios calculated from \nustar\ fluxes in the $2-10$ keV (soft) and $10-50$ keV (hard) bands.  A dashed horizontal line denotes the 
    hard state threshold.}
    \label{fig:lc_sw}
\end{figure}

Table \ref{tab:obs} lists the \src\ observations we conducted with \nustar, \xmm, and \chandra.  Additional X-ray data was collected with \swift/XRT, \nicer, \ixpe, and \xrism; multiwavelength observations were completed with Keck (NIR) and MeerKAT (radio).  A light curve illustrating the count rates and hardness ratios of \src\ observed between early February and September of 2025 is shown in Figure \ref{fig:lc_sw}.

\subsection{\nustar}\label{subsec:nustar}

A ToO observation targeting \src\ was first carried out on Feb. 6, 2025, followed by two others on March 3 (coincident with the \xmm\ and \xrism\ observations) and March 9 (simultaneous with \chandra). Additional GO datasets were obtained throughout April, June, July, and August  
2025.  Another \nustar\ ToO observation was performed on September 6, simultaneously with a second \xmm\ observation. 
Altogether, we collated 16 \nustar\ observations for this study, with a typical exposure time of $\sim20$ ks.

Each \nustar\ observation was processed using NuSTARDAS v2.1.4a.  To remove artifacts left by the SAA passage in some of the observations, the \texttt{saafilter} parameter in \texttt{nupipeline} was set to \texttt{STRICT}.

\begin{figure}
    \centering
    \includegraphics[width=\linewidth]{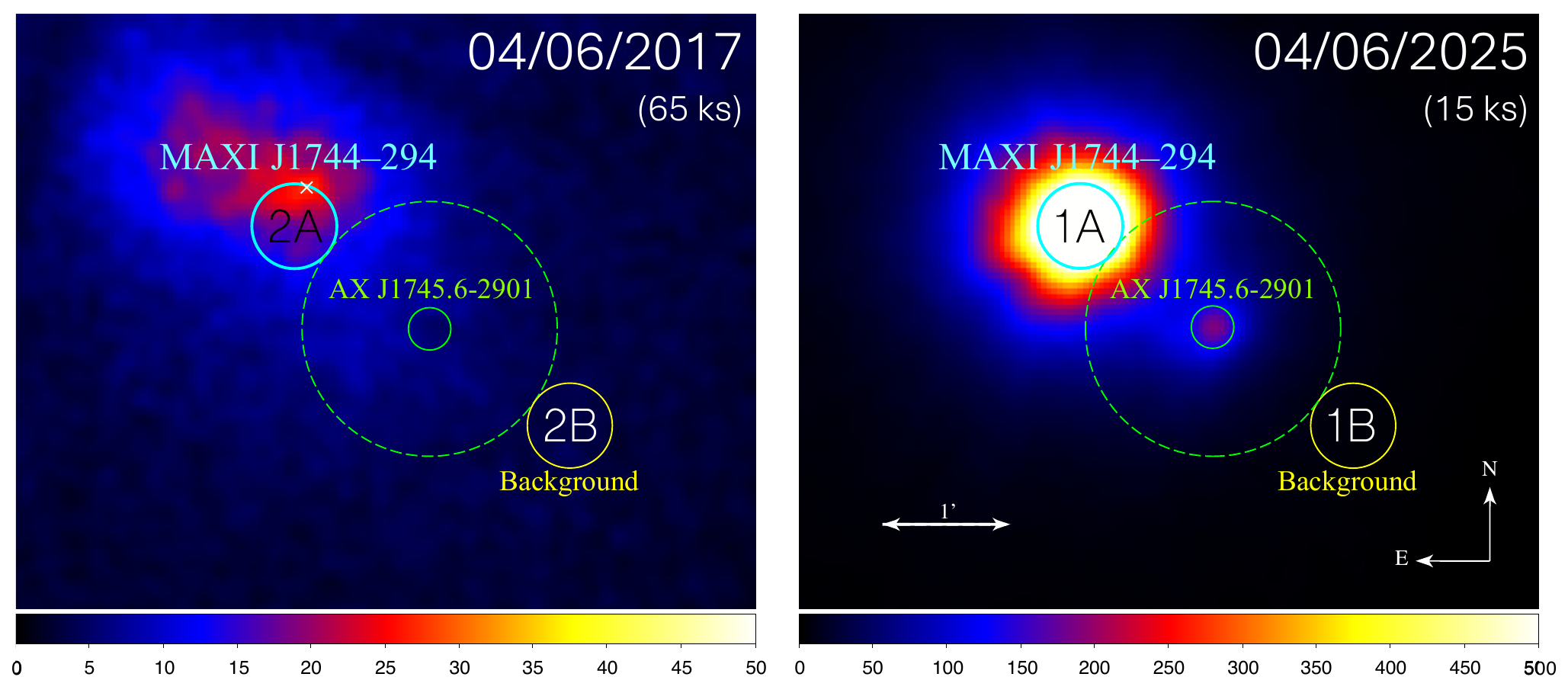}
    \caption{\nustar\ FPMA images of the Galactic center region.  The 2017 observation (left) was used to model the underlying background at the location of \src\ and the \axj\ background region (regions 2A and 2B, respectively).  Region 1B (right) was used to model the contamination from the \axj\ outburst.  See Section \ref{sec:bkg} for more.}
    \label{fig:bkgreg}
\end{figure}

Source regions were selected with a 20\arcsec\ 
radius circular aperture.  To account for contamination from the nearby NS-LMXB \axj, we selected a background region of equal size, located at an offset of $180^\circ$ around the centroid of \axj.  To account for the underlying emission around \src, which is considerably brighter than the quiescent emission at the aforementioned background region location (cf. Figure \ref{fig:bkgreg}), we also extracted a background spectrum -- using the source region defined above -- from an archival observation.  The background spectra were combined using the \texttt{MATHPHA} task in \texttt{FTOOLS}.  Section \ref{sec:bkg} in the Appendix contains a detailed description of the background modeling process.

\nustar\ spectra were grouped using the \texttt{FTGROUPPHA} task in \texttt{FTOOLS}.  We selected the optimal binning method by \cite{Kaastra2016}, with a minimum of 20 cts/bin.  Due to a recent tear in the multi-layer insulation (MLI) on the \nustar/FPMB optic module, 
FPMB spectra currently display an excess flux at low energies 
($<6$ keV)\footnote{\url{https://nustarsoc.caltech.edu/NuSTAR_Public/NuSTAROperationSite/mli.php}}.   
Consequently, flux offsets between the two \nustar\ spectra are apparent in the soft band.  We therefore fit the \nustar\ spectra only above 5 keV, where the discrepancy is negligible.

\subsection{\xmm}\label{subsec:xmm}

The \xmm/EPIC instrument, comprising the PN and two MOS CCD detectors, provides X-ray coverage in the 0.2–12 keV range with spectral resolution $\simlt150$ eV at 6 keV.  A $\sim10$ ks ToO observation of \src\ was obtained with \xmm\ on 03/03/2025, followed by another $\sim40$ ks observation on 09/06/2025.  The observations were performed in small window mode with the thick filter applied to minimize pile-up. Event files were processed with the XMM Science Analysis System (SAS v21.0) to remove background flares.

\begin{figure}
    \centering
    \includegraphics[width=\linewidth]{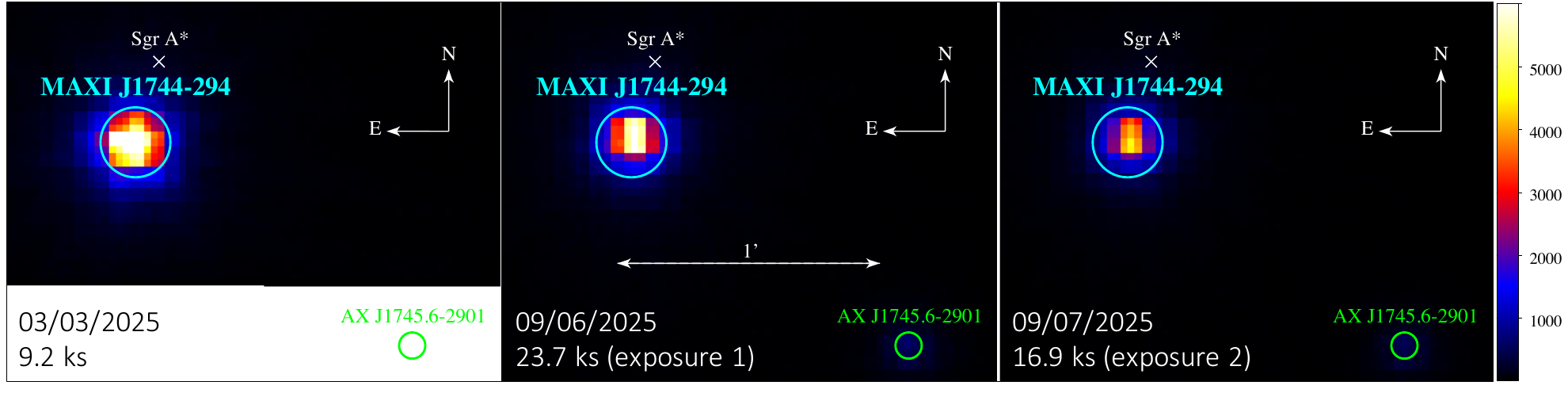}
    \caption{\xmm/EPIC PN detector images from observations conducted on March 3 (left) and Sept. 6 (center and right), 2025.}
    \label{fig:xmm}
\end{figure}

Background spectra were simulated based on spectra extracted from corresponding regions in archival \xmm\ observations of the Galactic center, conducted before the outburst of \src.  Unlike the \nustar\ background modeling, we could not use the archival spectra directly since the older \xmm\ observations were performed in a different observing mode, and the position of \src\ fell on a different detector chip.  To account for these differences, we generated a simulated \xmm/EPIC PN spectrum based on the archival data, but assuming the 2025 \src-observation response files.  A detailed description of the process used to simulate the \xmm\ background can be found in Parra et al. (in prep).

We do not utilize the \xmm/RGS data at this time given the challenges posed by the high absorption at the Galactic center, which particularly affects softer X-ray data ($\simlt2$ keV).

\subsubsection{March observation}

A $\sim10$ ks ToO observation of \src\ was obtained with \xmm\ on 03/03/2025.  No proton flares were apparent in the $10-12$ keV light curve.  All three \xmm/EPIC detectors showed signs of heavy pile-up due to the brightness of \src.  Excising the central 12\arcsec\ from the source region reduced the pile-up level in the PN spectrum to $<2\%$, as indicated by the SAS \texttt{epatplot} task.  We opted not to use the MOS data, as their pileup was more severe and attempts to correct for it resulted in the elimination of most photons $>5$ keV.  

\xmm/EPIC PN source spectra and light curves were subsequently extracted from an annular region centered on \src\ with inner/outer radii of $12/25$\arcsec\ (Figure \ref{fig:xmm}).  The outer radius was selected to maximize the number of source photons included while avoiding the mirror supports' outlines. 
Response matrices and ancillary response files were generated using the SAS tasks \texttt{rmfgen} and \texttt{arfgen}.

The simultaneously active NS-LMXB \axj\ was unfortunately outside the FoV of the \xmm/EPIC detectors, making its contamination challenging to model.  Given the \xmm/EPIC PN PSF (HEW$=16.6$\arcsec) and \axj's position offset of $\sim1.3$\arcmin\ from \src, the contribution from \axj\ should be minimal at low energies, where \src\ is much brighter.  However, $\simgt6$ keV (in the \xmm\ band) the \axj\ emission dominates -- as seen in the \nustar\ data (Figure \ref{fig:obs2_3to10}) -- and contamination may be significant.  We therefore only use the \xmm\ spectrum below 6 keV for this observation (Section \ref{subsec:bb_fits}).

\subsubsection{September observation}

A second \xmm\ observation of \src\ was performed on 09/06/2025.  This observation was interrupted by a ground station outage and split between two exposures of approximately $23.7$ and $16.9$ ks, respectively.  Both exposures exhibited background flaring, which was removed with GTIs generated with the \texttt{tabgtigen} tool.  EPIC spectra were extracted from a  
region of radius $r=15\arcsec$, with the central $6\arcsec$ excised to avoid pile-up.  
As before, we generated RMF and ARF files using the \texttt{rmfgen} and \texttt{arfgen} SAS tasks.  

Unlike the earlier \xmm\ observation, \axj\ was clearly visible in the FoV and much fainter than \src\ (Figure \ref{fig:xmm}); we consider the \axj\ contamination negligible here, and therefore fit these \xmm/EPIC spectra in the wider $2-9$ keV range.

\subsection{\chandra}\label{subsec:chandra}

\begin{figure}
    \centering
    \includegraphics[width=0.3\linewidth]{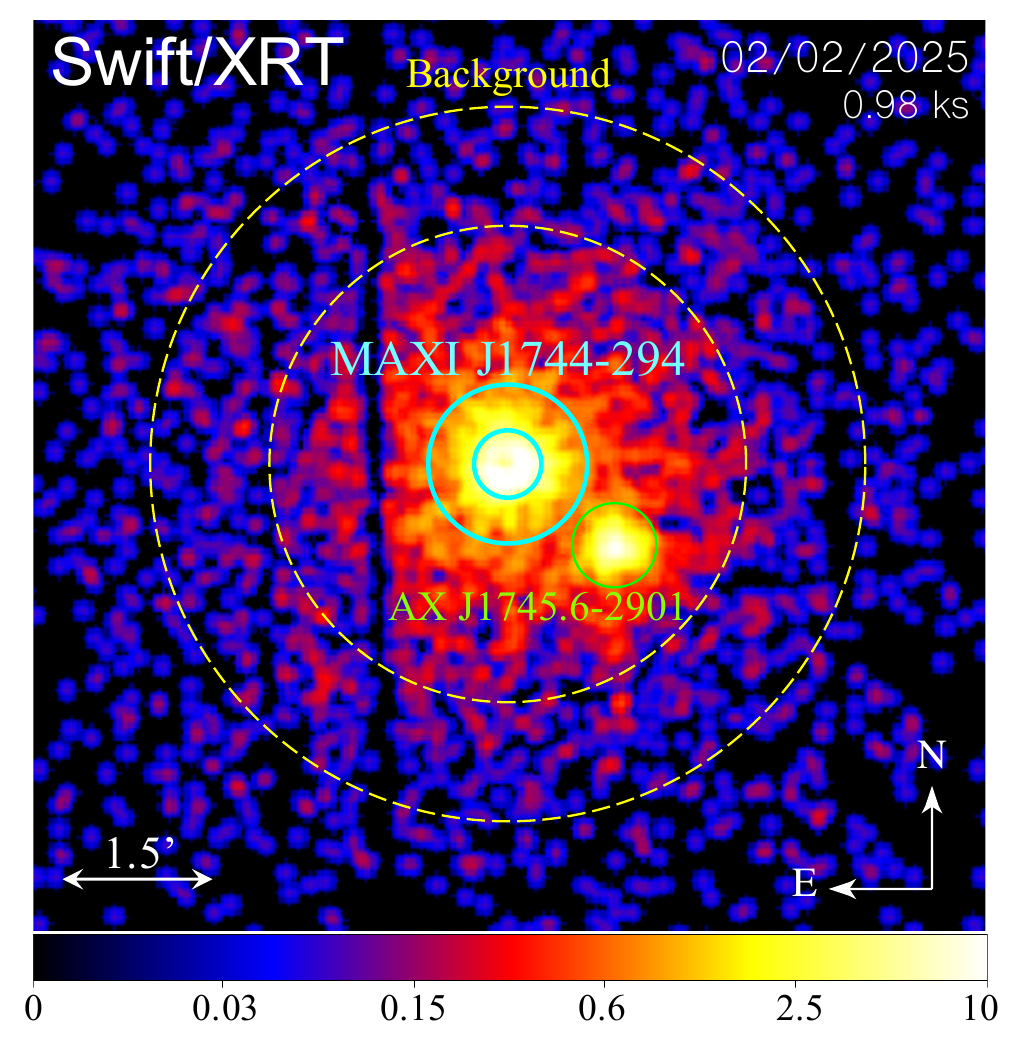}
    \hspace{-0.25cm}
    \includegraphics[width=0.7\linewidth]{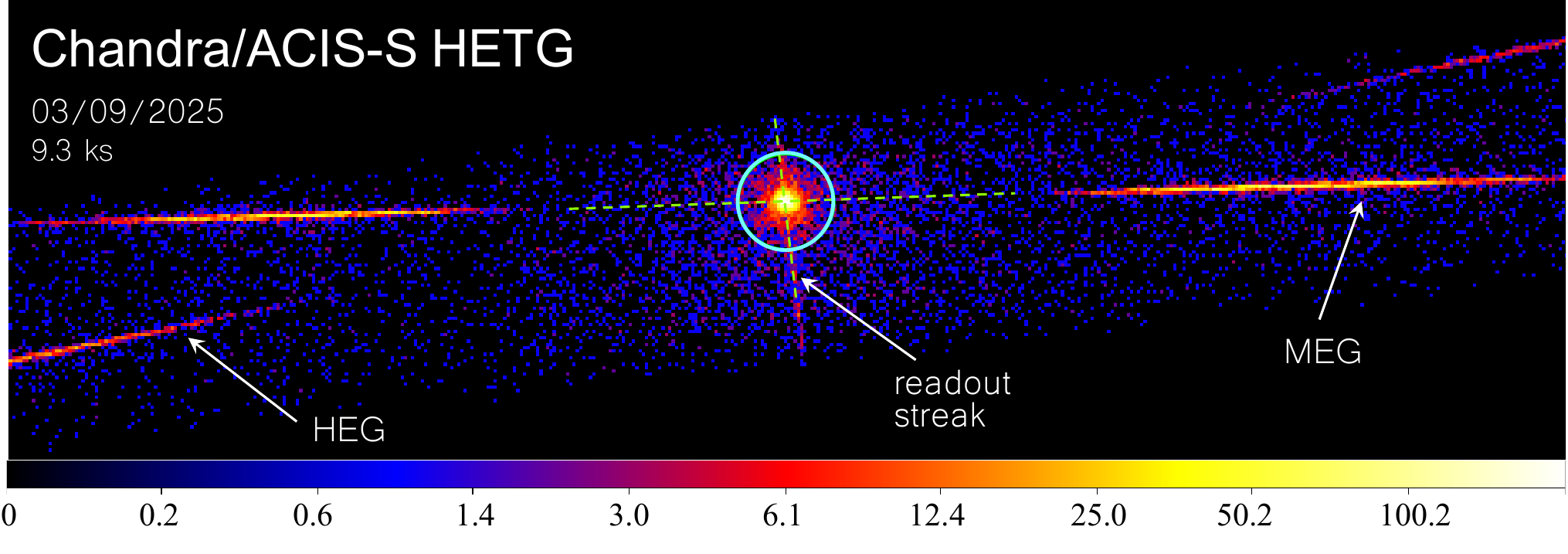}
    \caption{Left: \swift~XRT PC-mode image of the Galactic center from 2025 February 2. \src~is the bright source in the center of the image with \axj~lying $\sim80$\arcsec\ to the SW. 
    Annular extraction regions 
    are shown for \src\ (cyan) and background (yellow) spectra (see Section \ref{subsec:swift} for details).  Right: \chandra/ACIS-S HETG image of \src.  The location of \src\ was determined by finding the intersection between the readout streak and the MEG dispersion streak (dashed lines).  See Section \ref{subsec:chandra} for more.}
    \label{fig:sw_chandra}
\end{figure}

A ToO observation of \src\ was completed with \chandra/ACIS-S on March 9, 2025.  Due to the high source flux, we used the 1/8th subarray mode (S2+S3) in combination with the High-Energy Transmission Grating (HETG), for a 0.4 sec frame time.  
The primary purpose of the \chandra\ observation was to determine an accurate X-ray position for \src.  This was accomplished using the \textit{CIAO} \texttt{tg\_findzo}
algorithm, which locates the intersection between the readout streak and dispersion streak (Figure \ref{fig:sw_chandra}, right).  It yielded a position of RA = 17:45:40.476, Dec = -29:00:46.10.  This position is consistent with that measured by \textit{MeerKAT} for \src\ \citep{Grollimund2025}, and with the \chandra\ position of \Swft\ \citep{Mori2019}.  The details of the \chandra\ observation and X-ray source location were first reported in \cite{Mandel2025d}.

A search of the Chandra Source Catalog (CSC 2.1; \cite{Evans2024}) turned up a faint point source, 2CXO J174540.4-290046, within the \chandra\ uncertainty region.  It is not clear whether 2CXO J174540.4-290046 is the quiescent counterpart of \src, since the X-ray source density within this region is very high (making a random overlap likely), and the source catalog is incomplete within the central arcminutes due to a combination of high background and source confusion.    2CXO J174540.4-290046 has a reported flux of $\sim5\times10^{-14}$ \fluxcgs, corresponding to a luminosity of $3.8\times10^{32}$ \lumcgs\ for an assumed distance of 8 kpc. %\\
However, on closer inspection, 2CXO J174540.4-290046 was only detected with high significance during the outburst decay of \Swft, with other CSC detections at large off-axis angles.  This calls into question both the quiescent flux and its possible association with \src/\Swft.  On the other hand, we note that \cite{Mori2019} searched 45 archived Galactic center \chandra/ACIS-I observations for the quiescent counterpart to \Swft, finding none; they used ACIS Extract to derive an upper limit for the quiescent luminosity of $L_X<1.6\times10^{31}$ \lumcgs.

\subsection{\swift/XRT}\label{subsec:swift}

The Galactic center has been regularly observed with the \textit{Neil Gehrels Swift observatory} \citep{Burrows2005}, hereafter \swift, since 2006 as part of a long term monitoring campaign \citep{Degenaar2015}. For the 2024/2025 observing seasons, the Galactic center was observed on a quasi-daily cadence\footnote{The plan is for daily observations, but this can be interrupted due to, e.g., rapid response ToOs etc.}. In an observation on 2 February 2025, a bright new X-ray source was detected, confirming the earlier detections of \src~ \citep{Kudo2025,Heinke2025}.

The monitoring program observations typically occur in photon-counting (PC) mode; however, as the source rapidly brightened, significant pile-up was evident in XRT imaging {e.g., \citet{Davis2001}}. As such, the observational strategy was modified to include an observation in windowed-timing (WT) mode every third day. These observations would not be subject to pile-up and would thus present unblemished spectra for analysis and also enable a valuable cross-check of the pile-up corrections required for the PC-mode observations. As \src~is piled up in the PC-mode observations, spectra were extracted from annular regions with an inner  
8.5 pixel radius circle centered on the PSF core excluded (Figure \ref{fig:sw_chandra}, left). The subsequent spectra are found to be consistent with neighboring observations in WT-mode, though some excess of hard X-rays ($\gtrsim$ 8 keV) is still present in observations during the brightest portions of the outburst. These are readily identified as a pile-up related excess given that the emission is consistent with unphysical power-law indices ($\Gamma \leq 1$).

An additional complication is presented by the  
NS-LMXB \axj~\citep{Degenaar2009, Degenaar2015, Ponti2018}, which lies only $\sim$ 80\arcsec~from the position of \src\ (Figure \ref{fig:sw_chandra}, left) and was observed to be active throughout the outburst of \src. \axj~is highly variable when active and our PC-mode observations allow us to accurately track the flux from \axj~throughout the outburst from \src. 
As the \axj\ light curve in Figure \ref{fig:axj_lc} highlights, the flux from \axj\ remained $<10\%$ that observed from \src\ (Figure \ref{fig:lc_sw}).  Contamination from \axj\ is therefore considered negligible, particularly $<6$ keV where \axj\ is faintest relative to \src\ (Figure \ref{fig:obs2_3to10}).

The XRT data was downloaded from the HEASARC\footnote{https://heasarc.gsfc.nasa.gov} and reprocessed using the latest version of the \swift/XRT calibration products (v20250609), and the relevant XRT pipeline tools (\texttt{xrtpipeline}). \src~spectra were extracted from a 20 pixel radius circular region, and a 40 by 66 pixel rectangular region centered on the source position for the PC and WT-mode observations respectively. In PC-mode \src~and \axj~are well separated and the wings of the PSF from \axj~only negligibly contaminate the chosen \src~extraction region for the chosen extraction radius. In WT-mode, the separation of these sources is dependent on the space-craft roll angle. \axj~is visible in the WT-mode trace for the majority of observations under consideration herein and the projected position of \axj~was masked on WT-mode observations with a 12.7 pixel wide region (or 30$\arcsec$).

Background spectra consist of two primary components. Firstly, we consider the non-Galactic center diffuse background and utilize an annular region from $60-90$ pixels to extract this. The second  background component originates in the diffuse Galactic center gas at the source position and is estimated via an analysis of archival \swift~observations of the Galactic center obtained in November and December 2024. Spectra are extracted from a 20-pixel radius region at the position of the source. These were fit with a simple absorbed powerlaw model. This background component is well described by a highly absorbed powerlaw model ($N_H = 13\times10^{22}~\rm cm^{-2}, \Gamma = 3$), with a flux $f_X = 8.9\times 10^{-12}$ 
\fluxcgs\ ($2.0-10.0$~keV). At this level, the diffuse background makes a negligible contribution to the observed \src~flux for the majority of the outburst.

Appropriate responses and exposure maps were generated (\texttt{xrtexpomap, xrtmkarf}). All spectra are grouped to 20 ct/bin and were subsequently analyzed in \textsc{xspec} v12.14.0h \citep{Arnaud1996}.

\subsection{\nicer}\label{nicer}

The \nicer\ X-ray Timing Instrument (XTI; \cite{Gendreau2012}) is a non-imaging soft X-ray telescope mounted on the International Space Station. It consists of 56 co-aligned concentrator optics, each coupled to a silicon-drift detector \citep{Prigozhin2012}. It records X-rays in the 0.2–12~keV range with timing and spectral resolutions of approximately 100~ns and 100~eV, respectively. 

\nicer\ observed the transient outburst from \src\ between 2025-02-11 and 2025-04-13 \citep{Jaisawal2025}. The observations, obtained across ObsIDs 7205140104–7205140107 and 8205140116, 8205140117, 8205140121, 8205140128–8205140131, were carried out during nightside orbits and are unaffected by optical light leak. The total exposure for these good-time intervals is 8.1~ks. The data were processed with the standard {\tt nicerl2} pipeline in {\tt HEASoft}~v6.35, applying standard filtering criteria. Light curves were extracted in the 0.5–10, 2–4, and 4–8~keV energy bands using the {\tt nicerl3-lc} task to study the outburst evolution and construct the hardness–intensity diagram. 

Figure \ref{fig:lc_sw} shows the count rates detected by \nicer\ in the $2-4$ and $4-8$ keV bands; the \nicer\ count rates were reduced by a factor of 10 for better visualization with other telescope detections. During the observation period, \nicer\ measured source count rates between 110 and 300 counts~s$^{-1}$ in the 0.5–10~keV band. Being a non-imaging instrument, \nicer\ light curves may be subject to contamination from AX J1745.6-2901 and diffuse Galactic center emission.  However, as Figure \ref{fig:lc_sw} illustrates, the \nicer\ data follows a similar evolutionary trend as \nustar\ and \swift\ in the same energy bands, indicating that this contamination may not be significant.  %\\

\subsection{Additional multiwavelength observations}\label{subsec:other}

\subsubsection{\xrism}\label{subsec:xrism}

We obtained a $\sim71$ ks Director’s Discretionary Time (DDT) observation of \src\ with \xrism\ on March 3, 2025 (ObsID 901002010), concurrent with our \xmm\ ToO observation. 
\xrism/Resolve provides high-resolution ($5–7$ eV) spectroscopy that allows us to study the Fe line complex in exquisite detail.  
A comprehensive study of the Resolve and Xtend data -- including in-depth analysis of the background contributions -- will be presented in Parra et al. 2025 (in prep).

\subsubsection{\ixpe}

The Imaging X-ray Polarimetry Explorer (\ixpe) is the first X-ray observatory dedicated to measuring X-ray polarization. It operates in the 2–8 keV energy range and consists of three identical co-aligned telescope modules, designed to detect X-ray polarization with high sensitivity \citep{Weisskopf2022}. \ixpe\ observed \src\ between April 5–8, 2025, yielding a total of $\sim150$ ks in livetime \citep{Marra2025}.  A detailed study of the data was conducted by \cite{Marra2025}, who found no significant polarization.  \cite{Marra2025} constrained the inclination to $\simlt38^{\circ}-72^{\circ}$ (with the upper limit dependent on BH spin and disk atmosphere albedo), consistent with our results from broadband spectral fitting (Section \ref{subsec:refl}).

\subsubsection{MeerKAT}\label{subsec:mkat}

We performed multi-epoch radio observations of MAXI J1744--294 with MeerKAT using L-band (856-1712 MHz) and S4-band (2625-3500 MHz)
receivers as part of the X-KAT program. We used PKS J1939--6342 as a primary calibrator to set the absolute flux and bandpass, 
and J1830--3602 for complex gain calibration. Each L-band run consisted of a single 15-min scan of MAXI J1744--294, interleaved by 2-min scans 
of the phase calibrator, as well as a two 5-min scans of the primary calibrator. The deeper S-band observation included six 30-min scans
of the target, for a total exposure time of 3 hrs. The correlator was configured to deliver 1024 channels across the total bandwidth,
with an 8-second integration time per visibility point. We reduced and imaged the data using the \textsc{oxkat\footnote{\url{https://github.com/IanHeywood/oxkat}}}
data reduction scripts \citep{oxkat}. The visibilities were initially flagged and calibrated using the Common Astronomy Software Application package
\citep[\textsc{casa};][]{casa}. Additional flagging on the target data was conducted using the \textsc{tricolour} package \citep{tricolour}.
For imaging the field of MAXI J1744--294, we used \textsc{wsclean} \citep{wsclean} with a Briggs weighting scheme (robust parameter of $-0.3$). 
We generated full-band, multifrequency synthesis (MFS) images by deconvolving in 8 sub-bands. After a step of direction-independent
self-calibration with the \textsc{cubical} package \citep{cubical}, the target was imaged a second time using masked deconvolution.
The final images were obtained by applying a hard uv-cut (20 $\text{k}\lambda$) in order to filter out bright diffuse emission from the Sgr A complex.
This reduction process yielded radio maps with an angular resolution of $\sim 6''$ (resp. $\sim 3''$). To extract the positions and flux densities,
we fitted the source in the image plane with an elliptical Gaussian model. The best radio position was obtained by averaging
the fitted source positions over all observations: RA = 17:45:40.427, Dec = -29:00:45.55 (J2000, $1\sigma$ error of $\sim 0.1''$).  
A detailed study of the MeerKAT data will be presented in Grollimund et al. (in prep).

\subsubsection{Keck}

We conducted follow-up observations (PI: R. Campbell and J. Lyke) in the near-infrared (NIR) with the AO-assisted NIRC2 imager on W. Keck Observatory on March 11th 2025.
We obtained broad-band observations in the Kp filter (central wavelength 2.124~$\mu$m and width 0.351~$\mu$m) for a total of 70~seconds (7 frames with integration time 1~sec and 10 coadds, the average FWHM was 100~mas) and narrow-band observations with the Br-$\gamma$ and K-cont filters (central wavelengths and 2.169 and  2.271~$\mu$m respectively and width $\sim$0.03~$\mu$m) for a total of 100~seconds (10 frames with integration time 10~sec and 1 coadd, the average FWHM were 200 and 255~mas respectively).

The data was reduced using the KAI pipeline \citep{Lu2021} where raw images were flat-fielded, sky subtracted, masked for bad pixels and cosmic rays, and finally corrected for distortion using the distortion solution reported in \cite{Service2016}. 
For the narrow-band Br-$\gamma$ and K-cont images, we rescaled the broad-band Kp skies so that the ratio between science and sky frame is similar to the Kp one to avoid oversubtraction. 
Only Kp, Br-$\gamma$, and K-cont individual frames with a FWHM smaller than 1.25 times the minimum FWHM (calculated on a bright isolated source near the center of the images) were used in the final combined image. 
Further details of this process are given in \cite{Gautam2019} (but using a different set of stars). 
The lowest FWHM K-cont image was subtracted from each Br-$\gamma$ frame before combining them.

\begin{figure}
    \centering    \includegraphics[width=0.6\linewidth]{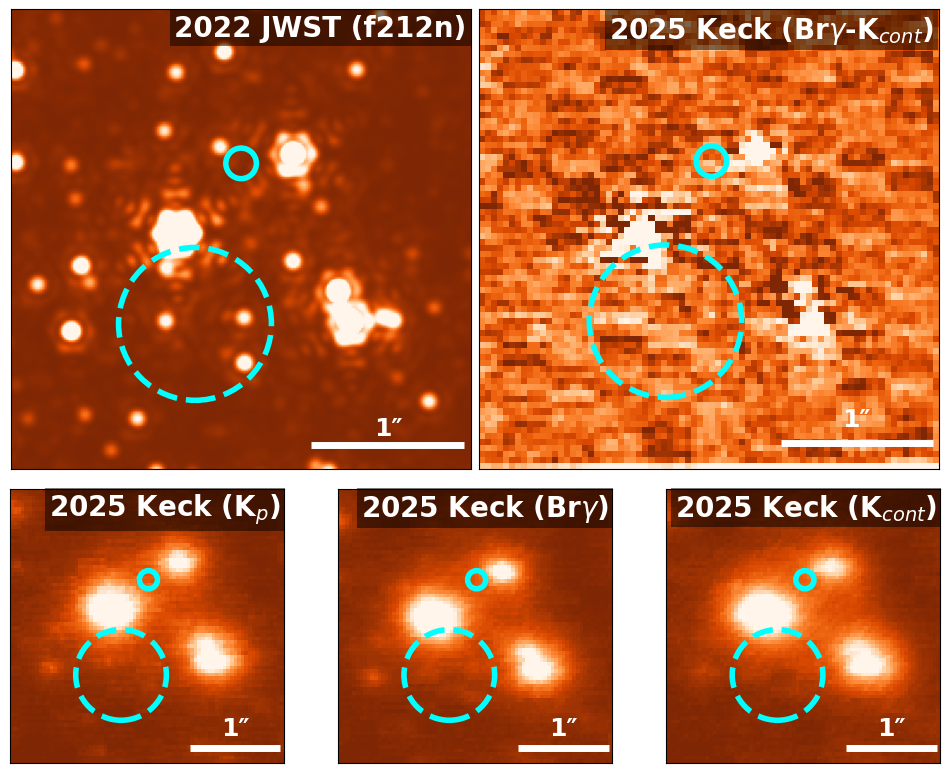}
    \caption{NIR follow-up observations. Top left: JWST NIRCam image from 2022. Top right: Keck continuum-subtracted Br-$\gamma$ subtracted. Bottom: Keck Kp (left), Br-$\gamma$ (center) and K-cont (right).  
    The large dashed cyan circle represents the original MeerKAT position of \src\ based on the earliest observations \citep{Grollimund2025}, while the small solid circle marks the refined MeerKAT position after months of monitoring.  
    There is no evidence of any IR excess in either location with respect to the JWST reference image.}
    \label{fig:Keck}
\end{figure}

Figure~\ref{fig:Keck} (top right) shows the final continuum-subtracted Br-$\gamma$ image.  We consider two regions: 1) the position  
published in \cite{Grollimund2025} based on the earliest MeerKAT observations of \src\ (large dashed circle), and 2) a refined position determined after months of MeerKAT monitoring (small solid circle), as described in Section \ref{subsec:mkat}.  Neither location shows evidence of excess emission.  
We compared the Keck observations with an earlier NIR image (Figure~\ref{fig:Keck}, top left) of the same region obtained in 2022 with NIRCam on JWST (PI: J. Lu) in the F212N filter (central wavelength 2.121~$\mu$m and width $\sim$0.03~$\mu$m). 
This comparison shows no newly arisen emission visible anywhere in the region.

We note that the Keck NIRC2 data are of poor quality with respect to the average performances in this region, with a FWHM significantly above diffraction limit (48~mas). 
This is due to the fact that engineering upgrades were being performed to the AO system at the time of this target of opportunity observation.  
Additionally, the total integration time was limited since the target field was rising in the morning twilight at this time of the year. 
In spite of these limitations, the Keck observations provide an upper limit on the NIR flux that is useful for constraining the SED.

We use the Kp observations to provide an upper limit at the most recent MeerKAT location.
We used an isolated bright star (at 17:45:40.098~$\alpha$, -29:00:27.403~$\delta$) as a reference PSF model, extracted using a customized version of the PSF fitting code \textit{StarFinder} \citep{Diolaiti1999}. 
IRS~16C ($m_{Kp}=$9.88; \cite{Blum1996}) is used as a photometric calibrator. 
We performed systematic star-planting simulations at fainter and fainter fluxes until \textit{StarFinder} is no longer able to detect the simulated source.
Considering a region within 0.1'' of the MeerKAT location, this leads to an upper limit of 17.73 magnitudes at Kp.
The same process in a less crowded nearby region $\sim$3'' away from the MeerKAT position leads to an upper limit of 16.14 magnitudes. \\

\section{Spectral and timing analysis} \label{sec:analysis}

\subsection{Broadband spectral modeling}\label{subsec:bb_fits}

LMXBs generally exhibit emission with one or more of the following components:

\begin{itemize}\setlength{\itemsep}{-0.25pt}
    \item Thermal emission from the accretion disk.  This 
    takes the form of a multi-temperature blackbody spectrum, as the disk temperature increases closer to the inner edge.  Disk blackbody emission is especially intense during the high/soft state.
    \item Non-thermal emission from Comptonization, generated when seed photons (e.g. from thermal disk emission) are up-scattered by a corona of hot electrons.  It is dominant in the low/hard state.
    \item Reflection off the accretion disk.  This manifests as Fe K$\alpha$ emission, as well as a Compton reflection hump (though the latter is not always pronounced enough to be observed).  
\end{itemize}

In addition to the above, NS-LMXBs 
often show an additional thermal component from either the boundary layer (the region where the gas from the accretion disk reaches a weakly magnetized NS's surface and slows to co-rotate with the NS, releasing some of its energy in the form of X-rays \citep{Popham2001, Suleimanov2006}) or from "hot spots" on the surface of the NS \citep{Ponti2018}.  While we observe the signatures of thermal disk emission, comptonization, and reflection in the spectra of \src\ (Section \ref{subsec:model1}), we see no evidence of an additional thermal X-ray component.  

\nustar's excellent broadband coverage is 
especially well-suited for studying higher-energy reflection and Comptonization processes.  
On the other hand, soft X-ray instruments like \swift/XRT are particularly effective at determining the hydrogen column density $N_H$.  
We therefore jointly fit spectra extracted from \swift/XRT or (where available) \xmm/EPIC observations -- taken concurrently with the \nustar\ data -- to complement the broadband \nustar\ spectra.

Spectral fitting was performed using the \texttt{XSPEC} package v12.14.1 \citep{Arnaud1996}.  All spectral models included a multiplicative constant, fixed at $c=1$ for FPMA and allowed to vary for FPMB and \swift/XRT (or \xmm), for cross-normalization between the telescopes.  Interstellar absorption $N_H$ was modeled with the multiplicative \texttt{tbabs} component, and abundances set to \cite{Wilms2000}.  In addition, we incorporated a model to account for the dust scattering halo present around \src, as described in 
Appendix \ref{sec:dust}.

\subsubsection{Phenomenological models}\label{subsec:model1}

All \nustar\ spectra showed a strong thermal component in the soft ($<10$ keV) band -- fit with a multi-temperature disk blackbody model -- as well as a non-thermal power-law tail that dominated the emission above $\sim10$ keV.  The relative strengths of these two components varied throughout the outburst.  In addition to the continuum, Fe emission lines were present in 
most \nustar\ spectra, also displaying varying intensity and width.

\begin{figure}
    \centering
    \includegraphics[width=\linewidth]{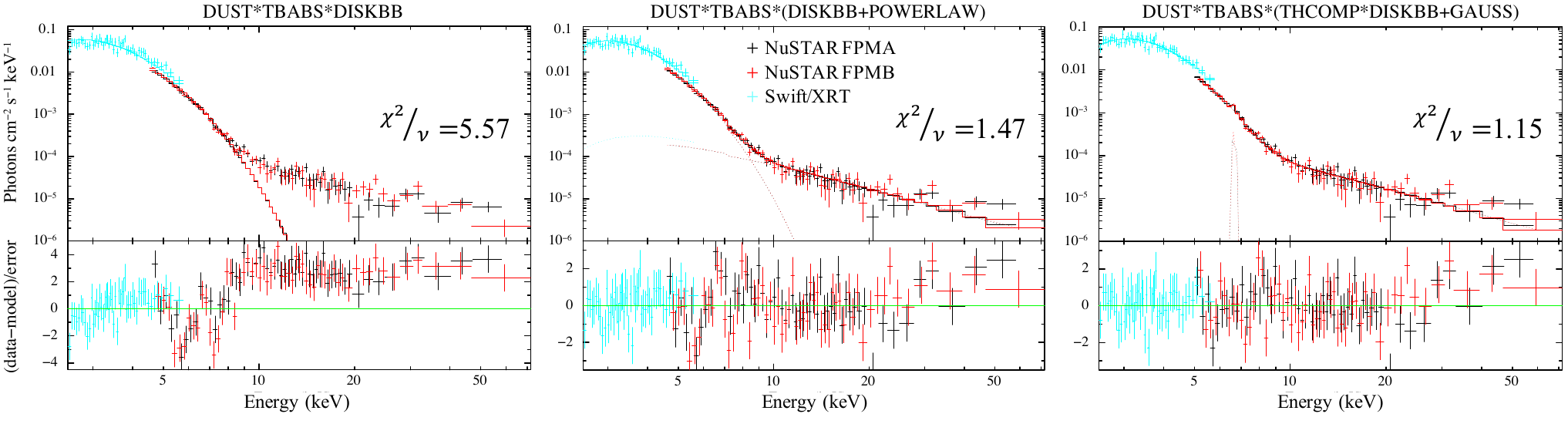}
    \caption{\swift/XRT (light blue) and \nustar\ FPMA/FPMB (black and red, respectively) spectra of \src\ from observations dated Feb. 6, 2025.  The data were fit to a range of spectral models: disk blackbody (left), disk blackbody + power-law (center), 
    and Comptonized disk blackbody + Gauss (right). 
    Note the residuals at energies $>20$ keV, which indicate the presence of a Compton reflection hump.}
    \label{fig:spec2x2}
\end{figure}

To obtain estimates of the flux and hardness of individual observations in a consistent manner conducive to juxtaposition, we fit all \nustar\ data with a phenomenological model that accounts for the main emission components.  The softer thermal emission was fit with \texttt{diskbb}, which models the thermal emission from a multi-temperature, optically thick, geometrically thin (Shakura–Sunyaev) accretion disk \citep{Shakura1973}.  
We initially modeled the high-energy ($\simgt10$ keV) emission using a simple power-law.  This was subsequently replaced with the \texttt{thcomp} model, which was applied as a convolution to the \texttt{diskbb} component to account for Comptonization of the disk photons.  An additional Gaussian component was then added to model the Fe K$\alpha$ emission between $6-7$ keV 
where residuals indicated the presence of such a line.  
The fitting process is illustrated in Figure \ref{fig:spec2x2}, where we highlight the first \nustar\ observation, taken on Feb. 6, 2025.  
Tables \ref{tab:spectra} and \ref{tab:spectra2} list the best-fit spectral parameters, X-ray fluxes, and reduced $\chi^2$ values for the combined \texttt{constant*dust*tbabs*(thcomp*diskbb[+Gauss])} model.  Additional details on these "baseline" spectral fits, along with plots showing 14 of the observed spectra (Figures \ref{fig:spec_obs2_go2} and \ref{fig:spec_obs8to15}), are provided in Appendix \ref{sec:spec2}.

\setlength{\tabcolsep}{6pt}
\renewcommand{\arraystretch}{1.3}
\begin{sidewaystable}
\begin{center}
    \caption{Results from joint \swift/XRT, \nustar\ and (for 3/3) \xmm/EPIC spectral fitting 
    for the bright part of the \src\ outburst.  Uncertainties are quoted for \nustar/FPMA spectra 
    at the 90\%
    C.L.  Corresponding spectral plots can be found in Figures \ref{fig:spec2x2}, \ref{fig:go3_spec}, and \ref{fig:spec_obs2_go2}.
        }\label{tab:spectra}
\begin{tabular}{l|c|cc|c|c|c|cc|c}
\hline\hline
Date & Feb. 6 & \multicolumn{2}{c|}{March 3} & March 9 & April 4 & April 6 & \multicolumn{2}{c|}{April 7} & April 9 \\
\hline
Parameter & & \swift$^a$ & \textit{XMM}$^a$ & & & & Part I$^b$ & Part II$^b$ & \\
\hline
$N_H$ [$10^{23}$ cm$^{-2}$)] & 1.41$^{+0.08}_{-0.13}$ & 1.75$^{+0.14}_{-0.13}$ & 1.72$^{+0.05}_{-0.02}$ & 1.77$^{+0.13}_{-0.07}$ & 1.46$^{+1.54}_{-^c}$ & 1.46$^{+0.09}_{-^c}$ & 1.46$^{+1.54}_{-^c}$ & 1.46$^{+0.08}_{-^c}$ & 1.48$^{+1.52}_{-^c}$	\\
$kT_{\rm in}$ [keV]  & 0.67$^{+0.02}_{-0.01}$ & 0.64$^{+0.01}_{-0.01}$ & 0.61$^{+0.01}_{-0.01}$ & 0.65$^{+0.01}_{-0.01}$ & 0.80$^{+0.01}_{-0.01}$ & 0.80$^{+0.01}_{-0.01}$ & 0.83$^{+0.01}_{-0.03}$ & 0.88$^{+0.01}_{-0.01}$ &  0.79$^{+0.01}_{-0.01}$ \\
$\Gamma$	  & 2.00$^{+0.90}_{-^c}$ & 2.23$^{+0.23}_{-^c}$ & 2.74$^{+^c}_{-0.32}$ & 1.81$^{+0.14}_{-^c}$ & 1.81$^{+0.03}_{-0.02}$ & 1.89$^{+0.03}_{-0.02}$ & 1.88$^{+0.06}_{-0.04}$ & 1.97$^{+0.01}_{-0.01}$  & 1.87$^{+0.08}_{-0.03}$	\\
$kT_e$ [keV] $^d$   & 500 & 268 & 499 & 498 & 150 & 188 & 499 & 500 & 498 \\
cov$_{\rm frac}$ [$10^{-3}$] & 3.0$^{+0.4}_{-0.2}$ & 1.5$^{+0.3}_{-0.6}$	& 3.3$^{+0.7}_{-1.8}$ & 0.6$^{+0.7}_{-0.2}$ & 9.8$^{+1.6}_{-0.5}$ & 19.9$^{+1.7}_{-3.0}$ & 28.1$^{+2.7}_{-8.7}$ & 56.2$^{+1.0}_{-2.2}$ & 16.3$^{+1.0}_{-0.4}$\\
E$_{\rm line}$ [keV]   & 6.66$^{+0.08}_{-0.09}$ & 6.68$^{+0.25}_{-0.22}$	& 6.69$^{+0.06}_{-0.05}$ & 6.73$^{+0.09}_{-0.09}$ & 6.60$^{+0.05}_{-0.07}$ & 6.60$^{+0.07}_{-0.07}$ & 6.65$^{+0.17}_{-0.17}$ & 6.59$^{+0.05}_{-0.05}$ & 6.61$^{+0.05}_{-0.04}$ \\
$\sigma_{\rm line}$  [keV] & 0.05$^{+0.75}_{-^c}$  & 0.01$^{+0.12}_{-^c}$ & 0.01$^{+0.23}_{-^c}$ & 0.10$^{+0.25}_{-^c}$ & 0.15$^{+0.06}_{-^c}$ & 0.20$^{+0.07}_{-0.07}$ & 0.10$^{+0.13}_{-^c}$ & 0.20$^{+0.06}_{-0.06}$ & 0.19$^{+0.03}_{-^c}$ \\
EqW [eV]   & 74$^{+52}_{-37}$ & 62$^{+86}_{-42}$ & 102$^{+75}_{-48}$  & 50$^{+26}_{-27}$ & 48$^{+25}_{-24}$ & 55$^{+25}_{-31}$ & 22$^{+49}_{-22}$ & 68$^{+24}_{-25}$ & 63$^{+39}_{-28}$\\
$\chi^2/\nu$ (dof)   & 1.15 (150) & 1.14 (151) & 1.24 (121) & 0.96 (166) & 1.19 (232) & 1.38 (235) & 1.04 (177) & 1.32 (269) & 1.19 (222) \\
$F_{\rm 2-10\, keV} \ ^e$  & 3.68$^{+0.42}_{-0.40}$ & 6.37$^{+0.33}_{-0.46}$ & 7.24$^{+0.15}_{-0.17}$ & 8.19$^{+0.35}_{-0.49}$ & 14.4$^{+0.22}_{-0.16}$ & 14.9$^{+1.16}_{-1.47}$ & 13.1$^{+0.76}_{-2.54}$ & 15.4$^{+0.30}_{-1.07}$ & 13.6$^{+0.76}_{-1.81}$ \\
$F_{\rm 10-50\, keV} \ ^f$   & 1.75$^{+1.00}_{-0.71}$ & 1.27$^{+3.15}_{-1.25}$ & 1.05$^{+0.66}_{-0.59}$ & 1.87$^{+2.73}_{-1.85}$ & 28.6$^{+9.40}_{-28.3}$ & 36.7$^{+18.6}_{-14.9}$ & 38.5$^{+32.3}_{-38.2}$  & 59.8$^{+15.3}_{-12.8}$ & 28.9$^{+18.9}_{-28.6}$ \\
$L_{X,\rm \, 2-50\, keV}$$\ ^{g}$ 
   & 3.03 &	5.10 & 5.76 & 6.57 & 13.5 & 14.6 & 13.3 & 16.8 & 12.9 \\[2pt]
\hline 
\end{tabular}
\end{center}
\vspace{-0.2cm}
$^a$Spectrum fit in the $2.5-6$ keV range. \\
$^b$Split as described in Section \ref{subsec:go3}. \\
$^c$Poorly constrained at the $90\%$ C.L. \\
$^d$$kT_e$ was poorly constrained across the board, therefore we do not list the uncertainties. \\
$^e$[$10^{-10}$ \fluxcgs], 
absorbed.  cf. Table \ref{tab:obs} for unabsorbed $2-10$ keV flux. \\
$^f$[$10^{-11}$ \fluxcgs], absorbed.\\
$^g$[$10^{36}$ \lumcgs], assuming a distance of 8 kpc.\\
\end{sidewaystable}

\setlength{\tabcolsep}{6pt}
\renewcommand{\arraystretch}{1.3}
\begin{sidewaystable}
\begin{center}
    \caption{Results from \nustar\ spectral fitting for the second part of the outburst, following its peak.  Uncertainties are quoted for \nustar/FPMA at the 90\% 
    C.L.  Corresponding spectral plots can be found in Figure \ref{fig:spec_obs8to15}.    }\label{tab:spectra2}
\begin{tabular}{l|c|c|c|c|c|c|c|c|c}
\hline\hline
Date & June 9 & June 14 & July 6 & July 10 & August 6 & August 7 & August 9 & August 12 & Sept. 6 \\
\hline
Parameter & & &  & & & &  &  & \\
\hline
$N_H$ [$10^{23}$ cm$^{-2}$)] & $1.40^{+0.59}_{-1.40}$ & $1.64\pm^{a}$ & $1.01\pm^{a}$ & $1.00\pm^{a}$ & $2.50\pm^{a}$ & $1.40\pm^{a}$ & $1.40\pm^{a}$ & $1.40\pm^{a}$ & $2.07^{+0.82}_{-0.50}$	\\
$kT_{\rm in}$ [keV]  & 0.70$^{+0.02}_{-0.04}$ & 0.82$^{+0.03}_{-0.04}$ & $0.94^{+0.04}_{-0.07}$ & $0.91^{+0.08}_{-0.07}$ & 0.89$^{+0.09}_{-0.06}$ & $0.91^{+0.07}_{-0.07}$ & 1.07$^{+0.11}_{-0.10}$ & $1.03^{+0.07}_{-0.08}$ &  $1.24^{+0.04}_{-0.10}$ \\
$\Gamma$	  & 1.76$^{+0.17}_{-0.21}$ & 1.30$^{+0.21}_{-0.19}$ & $1.10^{+0.24}_{-1.10}$ & 1.36$^{+0.18}_{-0.21}$ & 1.18$^{+0.39}_{-1.18}$ & $1.00^{+0.16}_{-1.00}$ & 1.41$^{+0.16}_{-0.17}$ & 1.30$^{+0.19}_{-0.20}$  & 1.81$^{+0.02}_{-0.02}$	\\
$kT_e$ [keV] $^b$   & 297 & 51 & 176 & 18 & 500 & 499 & 143 & 500 & 500 \\
cov$_{\rm frac}$ [$10^{-2}$] & $0.2^{+0.1}_{-0.1}$ & 0.3$^{+0.13}_{-0.06}$	& $0.9^{+4.4}_{-0.5}$ & $0.6^{+0.6}_{-0.3}$ & $0.6^{+1.1}_{-0.2}$ & $2.1^{+1.2}_{-0.7}$ & $4.1^{+2.6}_{-1.4}$ & $1.7^{+1.7}_{-0.5}$ & $43.6^{+4.48}_{-14.3}$\\
E$_{\rm line}$ [keV]   & 6.62$^{+0.14}_{-^a}$ & --	& -- & -- & -- & 6.44$^{+0.17}_{-^a}$ & $6.42^{+0.09}_{-6.42}$ & $6.48^{+0.12}_{-6.48}$ & 6.40$^a$ \\
$\sigma_{\rm line}$  [keV] & $0.17^{+0.14}_{-0.17}$  & -- & -- & -- & -- & $0.20^a$ & $0.05^{+0.15}_{-^a}$ & $0.06^{+0.16}_{-^a}$ & 0.40$^{+0.09}_{-0.14}$ \\
EqW [eV]   & $133^{+276}_{-96}$ & -- & -- & -- & -- & $337^{a}_{-293}$ & $275^{+407}_{-201}$ & $177^{+430}_{-137}$ & $140^{+196}_{-134}$ \\
$\chi^2/\nu$ (dof)   & 1.11 (76) & 1.23 (82) & 1.06 (83) & 1.04 (84) & 0.83 (78) & 1.02 (100) & 1.11 (96) & 1.02 (102) & 0.94 (257) \\
$F_{\rm 2-10\, keV} \ ^c$  & $1.61^{+0.03}_{-0.90}$ & $0.92^{+0.03}_{-0.45}$ & $0.38^{+0.06}_{-0.23}$ & $0.38^{+0.05}_{-0.25}$ & $0.13^{+0.02}_{-0.10}$ & $0.16^{+0.03}_{-0.09}$ & $0.23^{+0.16}_{-0.15}$ & $0.18^{+0.02}_{-0.12}$ & $3.20^{+0.21}_{-0.81}$ \\
$F_{\rm 10-50\, keV} \ ^d$   & $1.07^{+5.22}_{-1.06}$ & $1.93^{+2.41}_{-1.91}$ & $1.15^{+3.39}_{-1.14}$ & $1.18^{+0.04}_{-1.16}$ & $0.49^{+1.97}_{-0.49}$ & $0.60^{+3.49}_{-0.60}$ & $1.77^{+10.3}_{-1.75}$ & $0.98^{+1.83}_{-0.97}$ & $56.4^{+8.64}_{-18.7}$ \\
$L_{X,\rm \, 2-50\, keV}$$\ ^{e}$ & 1.35 & 0.87 & 0.39 & 0.39 & 0.14 &
       0.17 & 0.32 & 0.22 & 6.94  \\[2pt]
\hline 
\end{tabular}
\end{center}
\vspace{-0.2cm}
$^a$Poorly constrained at the $90\%$ C.L. \\
$^b$$kT_e$ was poorly constrained across the board, therefore we do not list the uncertainties. \\
$^c$[$10^{-10}$ \fluxcgs], absorbed.  cf. Table \ref{tab:obs} for unabsorbed $2-10$ keV flux. \\
$^d$[$10^{-11}$ \fluxcgs], absorbed.\\
$^e$[$10^{36}$ \lumcgs], assuming a distance of 8 kpc.
\end{sidewaystable}

\subsubsection{Reflection and spin}\label{subsec:refl}

To account for the relativistic reflection features present in the \nustar\ spectra of \src, we used the \texttt{relxill v2.3} family of models (\cite{2014ApJ...782...76G, 2014MNRAS.444L.100D}). We selected the \texttt{relxillCp} variant of the model, as unlike 
the basic version, it allows for variable density of the accretion disk atmosphere  
($n=10^{15}-10^{20}\;\rm cm^{-3}$) 
and assumes an incident flux 
from an \texttt{nthcomp} (thermally comptonized) spectrum. Furthermore, unlike the \texttt{relxilllp} variant, \texttt{relxillCp} does not assume a lamp-post coronal geometry, and simply parametrizes the flux incident on the accretion disk as a broken power law of index $q_1$ up to a breaking radius $R_{\rm br}$, and of index $q_2$ at higher radii. In the spectral fitting, we set the reflection fraction in the model to only take positive values, in order to ensure that the model simultaneously accounts for both the coronal emission and the reflected spectrum. In \texttt{XSPEC} parlance, the complete model is \texttt{constant * dust * TBabs * (diskbb + relxillCp)}. We fit all spectra with the same model.

To account for calibration uncertainties between the two \nustar\ FPM detectors at low energies, we only fit the spectra above 4 keV, and truncated them at the high-energy end where they became dominated by the background. We fixed the equivalent hydrogen column density along the line of sight in the \texttt{TBabs} component to 
$1.7\times10^{23}\;\rm cm^{-2}$, 
the value determined previously through joint \swift/XRT, \xmm, and \nustar\ analysis (Table \ref{tab:spectra}). 

Below we describe the spectral fits performed for the soft/intermediate state observations conducted between February--April 2025, as well as the harder (yet bright) observation dated September 2025.  We do not attempt to model the reflection in the low-hard state observations collected between June and August of 2025 due to poor statistics.  Even the "baseline" spectral model parameters showed large uncertainties for those datasets (Table \ref{tab:spectra2}), and we could not hope to extract meaningful constraints given the large number of parameters involved in reflection modeling.  While combining some of those datasets would have 
reduced statistical uncertainties, the X-ray emission might very well have changed between observations, rendering unreliable any results obtained from jointly fit non-contemporaneous spectra.  We therefore limit the reflection modeling to observations that can be individually fit (using both the \nustar\ FPMA and FPMB spectra).

\paragraph{Reflection in the soft state (February--April 2025)}

\begin{figure}
    \centering
    \includegraphics[width=\linewidth]{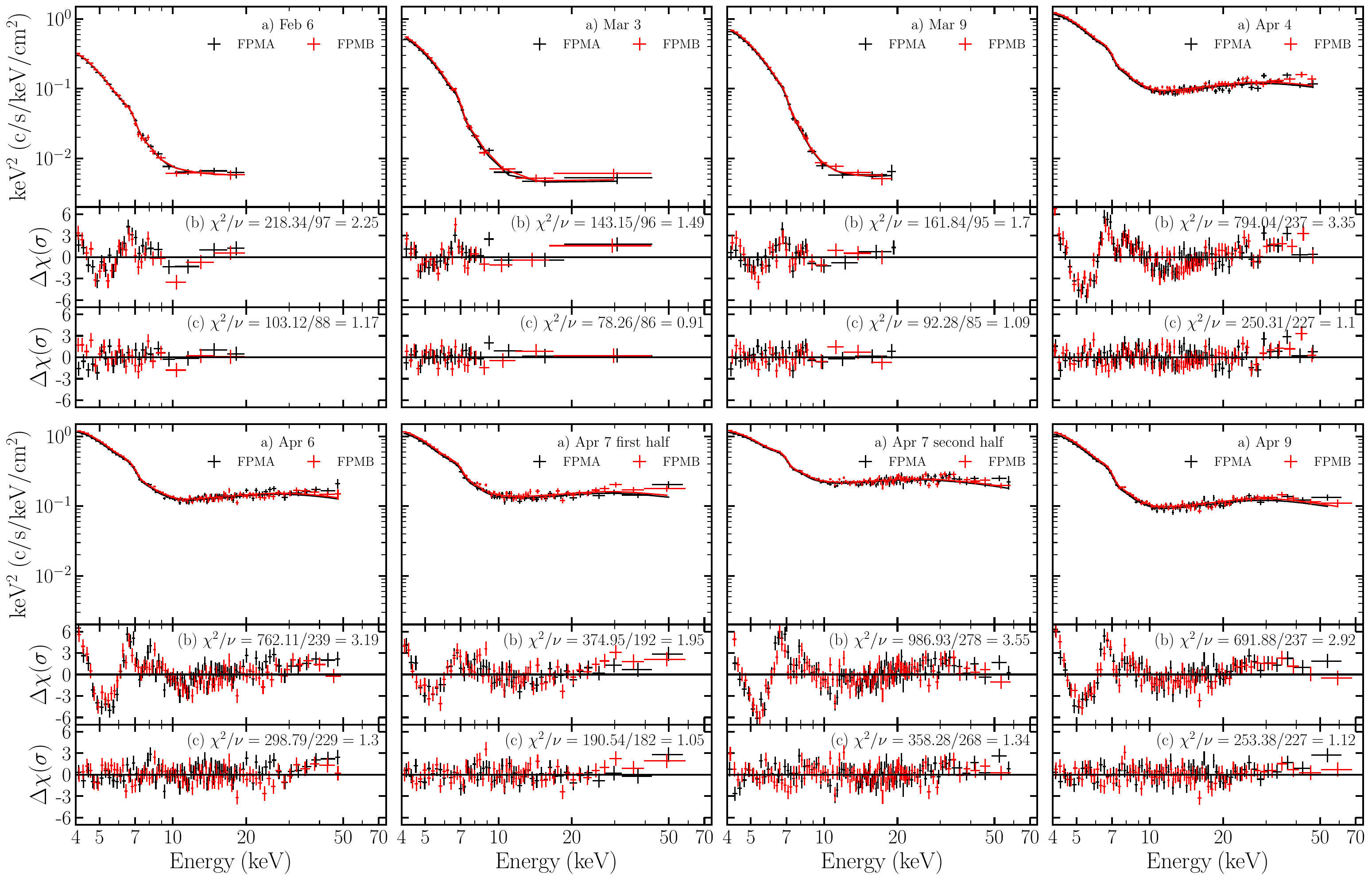}
    \caption{The \nustar\ spectra obtained from the eight observations treated in this work. The top panels show the spectra, as plotted through the \texttt{eedata} function in \texttt{Xspec}. Panels b) and c) show the residuals in terms of $\sigma$ produced when fitting the spectra with a \texttt{diskbb+powerlaw} model (panels b) and with a \texttt{diskbb+relxillCp} model (panels c), together with the associated fit statistics.}
    \label{fig:relxillCp_spectra}
\end{figure}

Figure \ref{fig:relxillCp_spectra} shows the 
first 
eight pairs of \nustar\ FPMA\&FPMB spectra analyzed, together with the residuals produced when fitting the spectra with a \texttt{powerlaw} model (panels b) and with the \texttt{relxillCp} model (panels c). The best-fit statistic resulting from the spectral fits is included in the panels in Figure \ref{fig:relxillCp_spectra}. Throughout the analysis, we find that even though the model produces good spectral fits and effectively accounts for the relativistic reflection features, the limited sensitivity of the observations prohibits precise simultaneous measurements of the spin of the compact object, of the inner radius, and of the inclination of the inner accretion disk through the \texttt{relxillCp} component. We attempted spectral fits with the spin of the compact object fixed at multiple values: $a=0$, $a=0.5$, and $a=0.99$, but found that the spectral fits are insensitive to the spin parameter. Therefore, we fixed the spin parameter to $a=0$ in our analysis and allowed the other parameters of the \texttt{relxillCp} model to vary freely, including the inner disk radius, the viewing inclination, the Fe abundance, the disk density, the ionization, and the coronal emissivity profile. 

\begin{figure}
    \centering
    \includegraphics[width=0.45\linewidth]{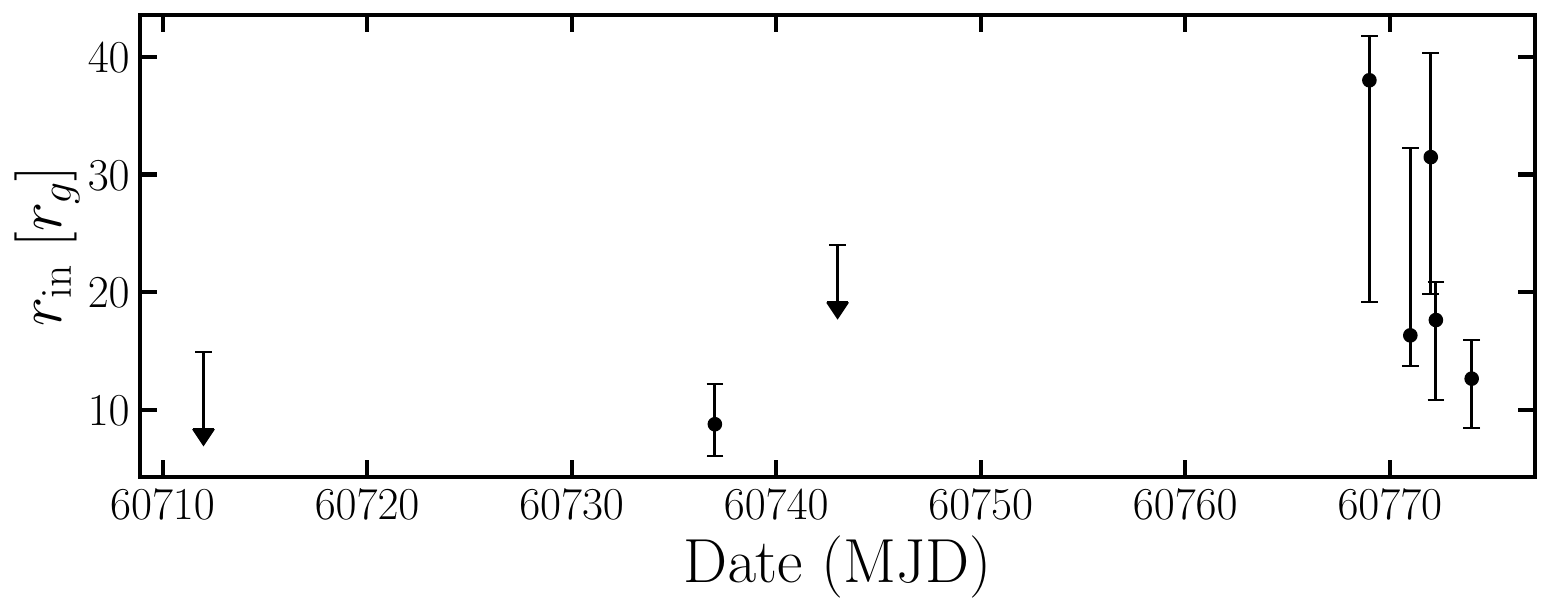}
    \caption{Inner disk radius measured using \texttt{relxillCp} in units of gravitational radii $r_g$ vs. time of the \nustar\ observation. The arrows indicate upper limits.}
    \label{fig:inner_disk_radius}
\end{figure}

Even with the spin fixed, the lack of robust coverage of the Compton hump prohibits breaking degeneracies between parameters that simultaneously influence the blue wing of the relativistically broadened Fe line (see, e.g., \cite{Draghis2025}), rendering inclination measurements uncertain. The red wing of the Fe line, not being influenced by the Fe K edge at 7.1 keV and not being as sensitive to parameter degeneracies, offers the opportunity to study the degree of disk truncation in the system. Figure \ref{fig:inner_disk_radius} shows the radius of the inner disk measured in units of gravitational radii $r_g$ as a function of the observation time, along with the $1\sigma$ uncertainties obtained through the \texttt{error} function in \texttt{XSPEC}. The arrows in Figure \ref{fig:inner_disk_radius} indicate upper limits on the measurements. A qualitative interpretation of the measured inner disk radii suggests that in the softer spectral state (first few observations), the disk extended closer to the compact object, and as the source evolved to a soft-intermediate state with higher coronal contribution, the degree of disk truncation increased.

We attempted a crude estimate of the mass of the compact object in the system by simultaneously leveraging information from parameters measured through the \texttt{relxillCp} and \texttt{diskbb} components in the model. The normalization of the \texttt{diskbb} component is defined as $(R_{in}/D_{10})^2\;\cos(\theta)$, where $R_{in}$ is an apparent disk inner radius in km, $D_{10}$ is the distance to the source in units of 10 kpc, and $\theta$ is the viewing inclination of the inner disk. \cite{Kubota1998} give the relation between the apparent disk inner radius and the true inner radius as $r_{in}=\xi \; f^2 \; R_{in}$, where $\xi=\sqrt{3/7} \times (6/7)^3$ is a correction factor, $r_{in}$ is the physical inner disk radius, and $f$ is the spectral hardening factor (\cite{Shimura1995}), taking values between 1.5-1.9 for accretion disks around stellar-mass BHs. This gives $r_{in}=\xi\;f^2\;D_{10} \sqrt{\frac{\rm norm_{diskbb}}{\cos(\theta)}}$, measured in km. By taking $r_g=1.5\;\rm km\times\frac{M}{M_\odot}$, we can express the mass of the compact object as 
\begin{equation}
    \frac{M}{M_\odot}=\frac{\xi\;f^2\;D_{10} \sqrt{\frac{\rm norm_{diskbb}}{\cos(\theta)}}}{1.5\times r_{in}}
\end{equation}
with $r_{in}$ expressed in units of $r_g$, as measured by \texttt{relxillCp}. Assuming a distance of 8 kpc ($D_{10}=0.8$), and making educated guesses for the inclination $\theta$ (based on \ixpe\ results) and the spectral hardening factor $f$, we can use this method to estimate the mass of the compact object by taking the values of $r_{in}$ and $\rm norm_{diskbb}$ as measured in the spectral fits to the 8 \nustar\ spectra. 

\begin{figure}
    \centering
    \includegraphics[width=0.75\linewidth]{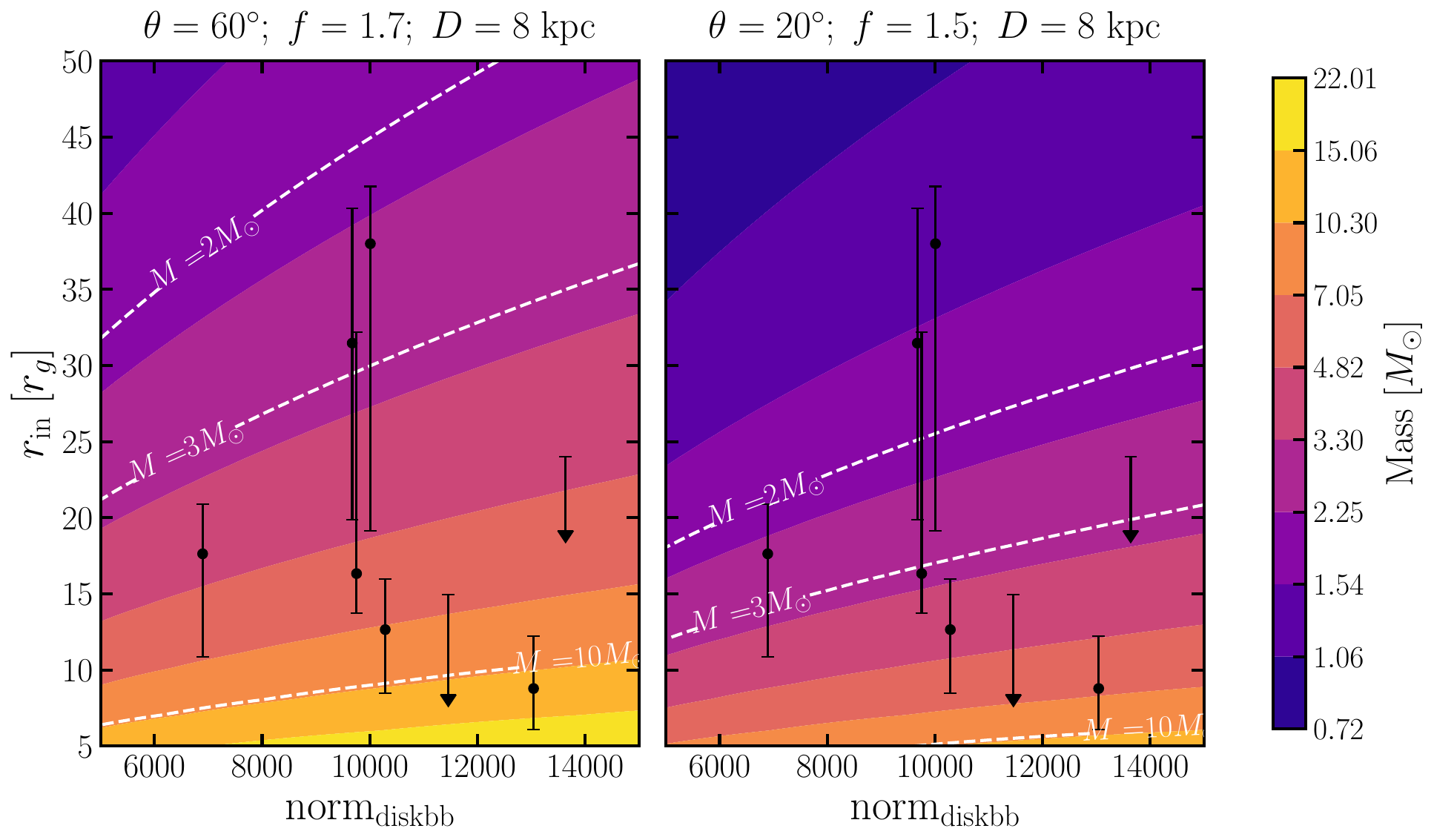}
    \caption{Inner disk radii and \texttt{diskbb} normalizations measured from spectral fits of the eight \nustar\ observations, plotted on contours of equal mass. Both panels assume a distance of 8 kpc to the system. The left panel assumes an inclination of $\theta=60^\circ$ and a hardening factor $f=1.7$. The right panel assumes values that would minimize the mass of the compact object, $\theta=20^\circ$ and $f=1.5$, while remaining within the bounds of a reasonable estimate. The measurements indicate a preference for masses larger than $2-3\;M_\odot$, suggesting that the compact object might be a BH.}
    \label{fig:mass_estimate}
\end{figure}

Figure \ref{fig:mass_estimate} shows the combinations of measured inner disk radius and \texttt{diskbb} normalization obtained from the analysis of the eight \nustar\ spectra, overlaid on contours of equal mass. The left panel calculates the contours assuming an inclination of $\theta=60^\circ$ and a spectral hardening factor of $f=1.7$, while the right panel uses values that would minimize the mass: low inclination $\theta=20^\circ$ and low hardening factor $f=1.5$. The white dashed lines indicate the contours of mass equal to 2, 3, and 10 $M_\odot$, respectively. The measurements suggest an overall preference for masses of the compact object greater than $2\;M_\odot$, even in the least favorable combination of assumed parameters (right panel in Figure \ref{fig:mass_estimate}). This analysis provides evidence that the compact object in \src\ is likely a stellar-mass BH rather than a NS. It is, however, important to note that the uncertainties regarding the distance to the system, the viewing inclination, and the assumption of a constant spectral hardening factor throughout the duration of the outburst -- together with correlations between the temperature and normalization of the \texttt{diskbb} component that are difficult to assess given the low-energy threshold of 4 keV used in our analysis --  
prevent us from definitively measuring the mass of the compact object.

\paragraph{Reflection in the hard state (September 2025)}

\begin{figure}
    \centering
    \includegraphics[width=0.49\linewidth]{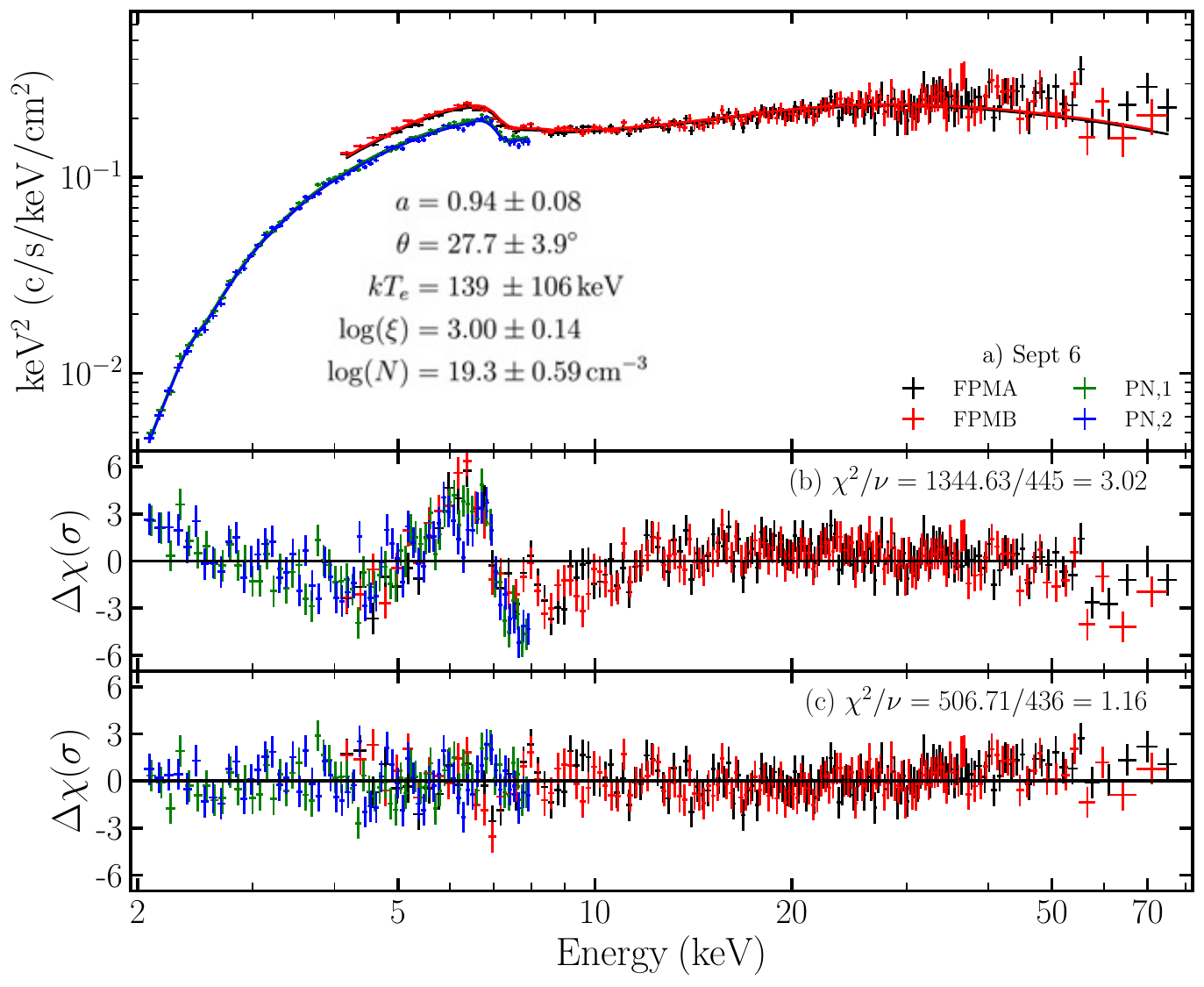}
    \includegraphics[width=0.5\linewidth]{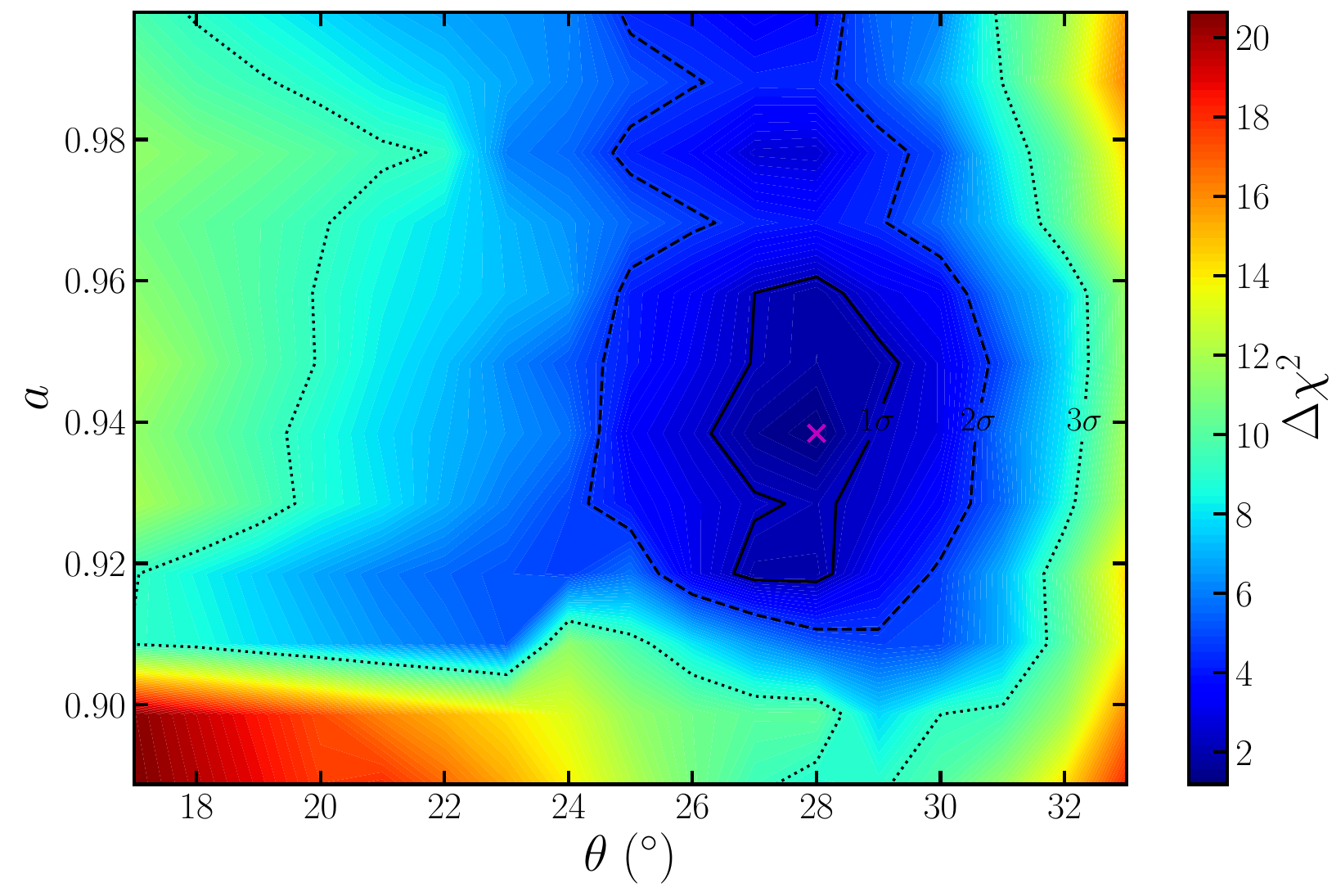}
    \caption{Left: \nustar/FPMA and FPMB (black and red, respectively) spectra jointly fit with contemporaneous \xmm/EPIC-PN (green and blue) spectra collected September 6-7, 2025.  Best-fit spectral parameters with $1\sigma$ uncertainties for the \texttt{const * dust * tbabs * (diskbb + relxillCp)} model are listed in panel a (top).  Panel b (middle) shows residuals from fitting to an absorbed \texttt{diskbb + powerlaw} model, highlighting reflection features.  Panel c (bottom) illustrates the lack of residuals after fitting the spectra to the \texttt{const * dust * tbabs * (diskbb + relxillCp)} model, indicating a good fit.  Right: contour plot showing the spin $a$ vs inclination $\theta$ parameter space.  $1\sigma$, $2\sigma$, and $3\sigma$ contours are outlined in black solid, dashed, and dotted lines, respectively.}
    \label{fig:obs4_joint_results}
\end{figure}

We performed a joint fit of the September 6 \nustar\ FPMA/FPMB spectra and of the \xmm/EPIC-PN spectra. We used the same models as for the analysis of the individual \nustar\ observations presented before. We fit the \xmm\ spectra in the 2--8~keV band, and the \nustar\ spectra in the 4--79~keV band. The spectra are shown in panel a) in Figure \ref{fig:obs4_joint_results} (left). Fitting the spectra with a phenomenological model (\texttt{const * dust * tbabs * (diskbb + powerlaw)}) produces the residuals shown in panel b) in Figure \ref{fig:obs4_joint_results}, indicating features consistent with relativistic reflection
, such as a broadened asymmetric Fe line with gravitationally redshifted wing.
We continued our joint analysis of the \nustar\ and \xmm\ spectra using the \texttt{const * dust * tbabs * (diskbb + relxillCp)} model. We linked all parameters between the four spectra with the exception of a normalization constant that was fixed at 1 for FPMA, and allowed to vary for the other three spectra to account for calibration uncertainties and source variability. Furthermore, we find that the data require different power-law indices for the \nustar\ and \xmm\ spectra, respectively. Therefore, we linked the power-law index $\Gamma$ between the two \nustar\ detectors and between the two \xmm\ spectra, but allowed the two sets to vary independently. In our best performing model, we find that the spectra from the two \nustar\ detectors agree within 1.5\%, but the constant offset for the \xmm\ data takes values greater than 1, in contrast with the expectations based on visually inspecting the spectra in Figure \ref{fig:obs4_joint_results}. However, this effect is compensated by the different power law indices between \nustar\ and \xmm, with \nustar\ favoring $\Gamma\sim2$ while \xmm\ measures $\Gamma\sim1.6$. We note that if fit with the same value of $\Gamma$ 
between all four spectra, the constant offset between \nustar\ and \xmm\ takes a value of $\sim0.7$, in agreement with expectations, but the fit is worse in terms of statistics by $\Delta\chi^2>250$, with significant differences between the detectors in the overlapping 4-8 keV band. Nevertheless, we continued our analysis with this best-fit model.

In the analysis of previous individual \nustar\ observations, we found that the fits statistically favored solutions that allowed the inner disk radius to take values larger than the size of the ISCO, regardless of the assumed spin of the compact object. Therefore, in that analysis, we fixed the spin to zero. However, for these 
hard-state 
observations, solutions with the spin fixed at zero and variable inner disk radius always favored low disk radii, i.e. $r_{\rm in} < r_{\rm ISCO}$ ($6\, r_g$ for a Schwarzschild BH). 
We therefore fixed the inner disk at the ISCO radius and allowed the spin of the compact object to vary freely. This model produced a significant improvement in terms of statistics, with $\Delta\chi^2\sim50$ for no change in the number of free parameters. Given that the spin of the compact object and the inner disk radius cannot be measured simultaneously owing to degeneracies between the parameters, and given the statistical improvement offered by allowing the spin to vary and fixing the inner radius at the ISCO, we continue with this approach.

The residuals produced by this best-fit model are shown in panel c) in Figure \ref{fig:obs4_joint_results} (left). The model measures a spin of $a\sim0.95$ and a viewing inclination of $\theta\sim28^\circ$. Such a high spin measurement suggests that the compact object in 
\src\ must be a BH (see e.g. \cite{Draghis2023}). The measured ionization parameter is $\log\xi\sim3$, the disk density $\log N\sim19.3$, the coronal temperature $kT_e\sim140\;\rm keV$, and a reflection fraction of $\rm refl\_frac\sim0.95$. The Fe abundance is pegged at the highest value allowed by the model ($A_{\rm Fe}=10$) and poorly constrained, likely owing to 
degeneracies between model parameters. Lastly, the coronal emissivity profile is measured to have $q_1\sim3.7$ up to a breaking radius of $R_{\rm br}\sim200\;r_g$, and nearly flat at larger radii.
Given the calibration uncertainties in the data, we refrain from performing a detailed exploration of the shape of the parameter space in order to quantify the uncertainties of the parameters. However, we performed an exploration of the 2-dimensional shape of the spin-inclination space using the \texttt{steppar} command in \texttt{XSPEC}. The results of this analysis are shown in Figure \ref{fig:obs4_joint_results} (right), and indicate that the data, given the model, favor a spin of $a\sim0.94^{+0.06^*}_{-0.03}$ and an inclination of $\theta\sim28^{+3}_{-4}$ deg at the $2\sigma$ level.  This inclination $\theta$ is consistent with the upper limits derived by \cite{Marra2025}.

\subsection{Light curves}

Timing analysis is a powerful tool to help us understand how X-ray transients like \src\ evolve, including the transitions between outburst states.  In some cases, it can allow us to detect identifying features, e.g., type I X-ray bursts, which are unique to NS binaries.  The presence of dips and/or eclipses in the temporal data provides energy-independent constraints on the inclination of the source, and in the case of the latter, the orbital period of the binary system.

\subsubsection{Long-term \swift\ light curves}

Figure \ref{fig:lc_sw} shows the long-term flux evolution of \src, starting from the earliest observations at the beginning of February 2025.  We emphasize the \swift/XRT data from window-timing (WT) mode observations, since the more frequent photon-counting (PC) mode observations are heavily affected by pile-up, particularly in the high/soft state.  The \nustar, \xmm, \chandra, \xrism, \ixpe, and Keck observations are also marked.

From Figure \ref{fig:lc_sw}, it appears that the \src\ outburst reached its peak around the beginning of April 2025.  However, when we consider the high-energy ($>10$ keV) \nustar\ data, \src\ attains similar flux levels in September 2025, months past the time when the softer X-ray emission had declined (Figure \ref{fig:nuim_col}).

\subsubsection{\nustar\ light curves}

We evaluated the light curves for each \nustar\ observation to assess short-term variability, including eclipses or type I X-ray bursts.  We began by applying barycentric correction to the event data to transform photon arrival times into the Solar System barycenter, which compensates for Earth’s motion and enables high-precision analysis, using the \texttt{barycorr} tool in HEASOFT (v6.34). We then extracted photon events from source and background regions corresponding to those used for spectral analysis (regions "1A" and "1B" in Figure \ref{fig:bkgreg}).  

Subsequent timing analysis was performed with the Stingray v2.2.4 software package \citep{Huppenkothen2019, Bachetti2024}. We produced light curves and power spectra 
in four distinct energy ranges: 3–50 keV (broadband), 3–6.3 keV (soft band), 6.3–7.2 keV (Fe line), and 7.2–50 keV (hard band). Events from the FPMA and FPMB detectors were processed independently. 

None of the  
\nustar\ light curves display signs of type I X-ray bursts.  We do not observe any significant dips or eclipses -- or indeed any periodic variability -- in the light curves.  Similarly, the corresponding power spectra did not produce any periodic or quasi-periodic signals (see Section \ref{subsec:pds} for more).  However, we did observe changes in the hard and soft flux in one of the observations, which we discuss below in Section \ref{subsec:go3}.

\subsubsection{The April 7 flux evolution}\label{subsec:go3}

\begin{figure}
    \centering
    \includegraphics[width=0.9\linewidth]{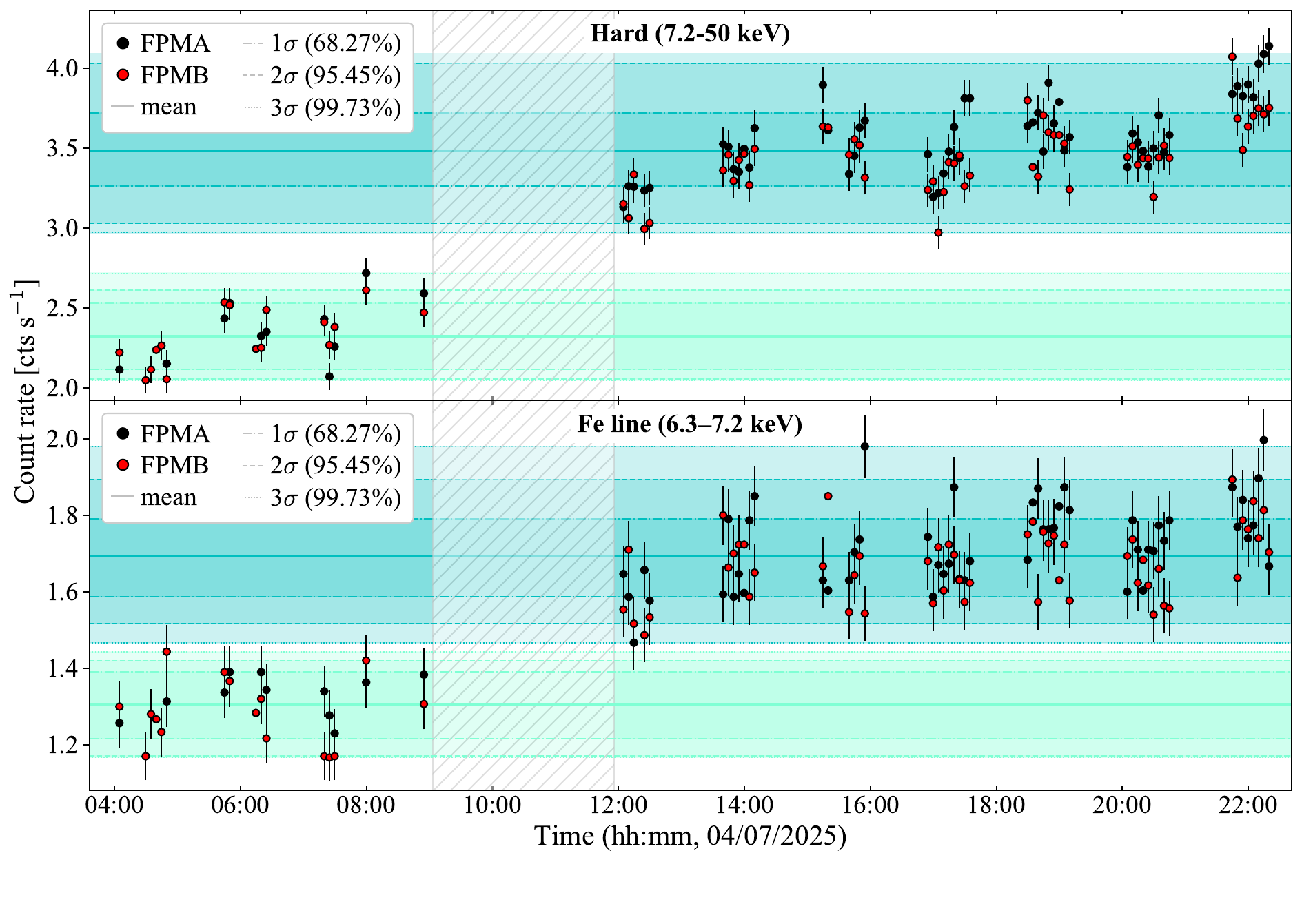}
    \caption{April 7, 2025 \nustar\ FPMA (black) and FPMB (red) light curve in the hard ($7.2-50$ keV; top) and Fe line ($6.3-7.2$ keV; bottom) bands, binned in intervals of 300s.  The shaded silver area marks the \nustar\ observation gap.  The hard X-ray emission appeared to increase significantly partway through the observation, while the soft X-rays experienced a brief drop in flux just prior to the event (cf. Figure \ref{fig:go3_lc_s}). }
    \label{fig:go3_lc_h}
\end{figure}

\nustar\ observation 31002004006 captured a curious change in \src's X-ray emission.  The observation commenced at 2025-04-07T03:59:28.934 and ended at 2025-04-07T22:25:02.209.  
Due to ground station issues, there is a gap in the \nustar\ data approximately 5 hours into the observation, lasting $\sim3$ hours.  
At some point during this gap, between $\sim5-8$ hours after the observation started, there was a marked increase in the hard ($>7$ keV) flux; as Figure \ref{fig:go3_lc_h} shows, there was relatively little corresponding increase in the soft ($<6$ keV) emission.  Instead, the soft data displayed a temporary drop in flux several hours \emph{prior} to the observed increase in hard flux (Figure \ref{fig:go3_lc_s}, top).  It is unclear whether this behavior is reflective of the "flip-flop" phenomenon -- short-term variations in flux, hardness, and PDS that can span timescales from tens of seconds to several ks -- that has been observed around state transitions in some BH transients \citep{Bogensberger2020}, as statistical uncertainties dominate the fluctuations in count rate on smaller timescales ($\simlt100$ sec).  Complicating the situation further, a recent tear in the multi-layer insulation (MLI) on the \nustar/FPMB optic module has resulted in flux offsets between the two \nustar\ detectors in the soft band ($<6$ keV), rendering correlation between their respective count rates impracticable.  

Remarkably, data collected simultaneously by the Karl G. Jansky Very Large Array (VLA) was reported to exhibit a significant increase in flux ($\sim82\%$ and 399\% in the 42.9 GHz and 32.7 GHz bands, respectively) 
at the same time as the \nustar\ hard X-ray flux increased \citep{Michail2025}.  The radio spectrum appears to have softened at the same time as the X-ray spectrum hardened, with the increase in the 32.7 GHz band (by a factor of $\sim5$) larger than the corresponding increase at 42.9 GHz (less than double the earlier measured flux density in the same band).  

To further investigate this apparent transition, we split the data and generated separate spectra from before and after the interval where the change was observed.

\begin{figure}
    \centering
    \includegraphics[width=0.8\linewidth]{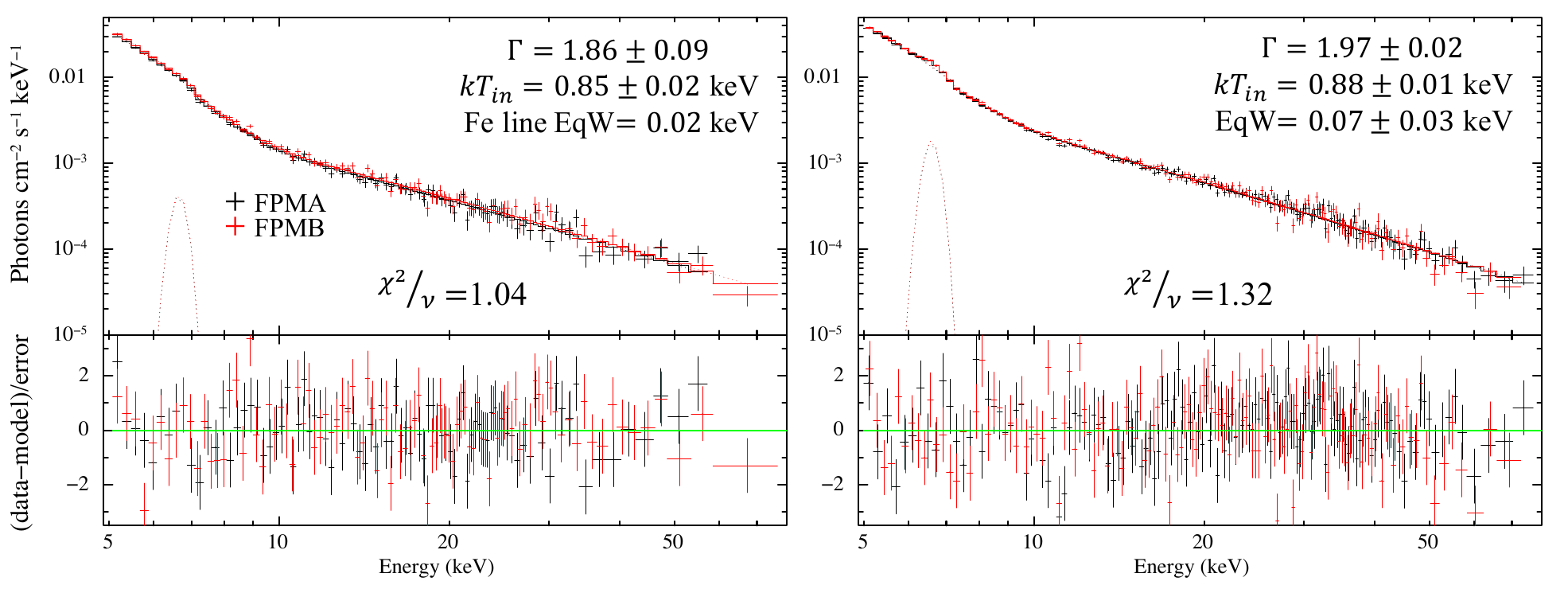}
    \caption{\nustar\ spectra captured in the April 7 observation before (left) and after (right) the observing gap. Both sets of spectra were fit with the \texttt{DUST*TBABS*(THCOMP*DISKBB+GAUSS)} model.  Best-fit photon index $\Gamma$, inner disk temperature $kT_{in}$, and Fe line equivalent width are listed in the upper right corner for each spectrum. The Fe line intensity increased significantly in the latter part of the observation, while changes in $\Gamma$ and $kT_{in}$ were more modest.}
    \label{fig:go3_spec}
\end{figure}

Several features stand out in a side-by-side comparison of the two spectra (Figure \ref{fig:go3_spec}).  Notable among them is the increase in Fe line intensity and spectral hardening during the latter part of the observation; while the best-fit $\Gamma$ was softer for the second set of spectra, there is an obvious increase in flux $>30$ keV, highlighted by the residuals of the poorer fit.  The high-energy residuals in the second spectrum are suggestive of a Compton reflection hump.

To further validate these results and confirm their significance, as well as investigate the flux during the \nustar\ observation gap, we analyzed the data from the \ixpe\ observation during the overlapping period.  We find that in the soft ($3-6.3$ keV) band, the \ixpe\ count rate followed a similar trend as the \nustar\ light curve (Figure \ref{fig:go3_lc_s}). See Section \ref{sec:ixpe_lc} in the Appendix for more.

\subsection{Power density spectra}\label{subsec:pds}

\nustar\ is affected by non-constant and unusually long dead times ($\sim2.5$ ms), which can introduce frequency-dependent distortions in power spectra at the white-noise level that may appear as quasi-periodic features \citep{Bachetti2015b}. To mitigate this effect, \cite{Bachetti2015b} proposed the use of the cospectrum -- the real part of the cross-spectrum -- as a white-noise–subtracted proxy for the “true” power density spectrum (PDS). However, this approach provides only a partial correction; \cite{Bachetti2018} demonstrated that a more comprehensive treatment for dead time requires the use of Fourier Amplitude Difference (FAD)–corrected power spectra and cross-spectra. We used the Stingray FAD function to generate Leahy-normalized, dead-time–corrected spectra from the independent FPMA and FPMB light curves.

Light curves were binned at 1/512 s for each energy band, yielding a Nyquist frequency of 256 Hz. For each light curve, we computed PDS and cross-spectra using three segment lengths: short ($\sim100$ s), medium ($\sim400–600$ s), and long ($\sim1500–3000$ s). These correspond to averages over $\sim5–10$, $\sim20–50$, and $\sim100–250$ segments, respectively, providing a range of frequency resolutions and white-noise variances.

\begin{figure}
    \centering
    \includegraphics[width=0.9\linewidth]{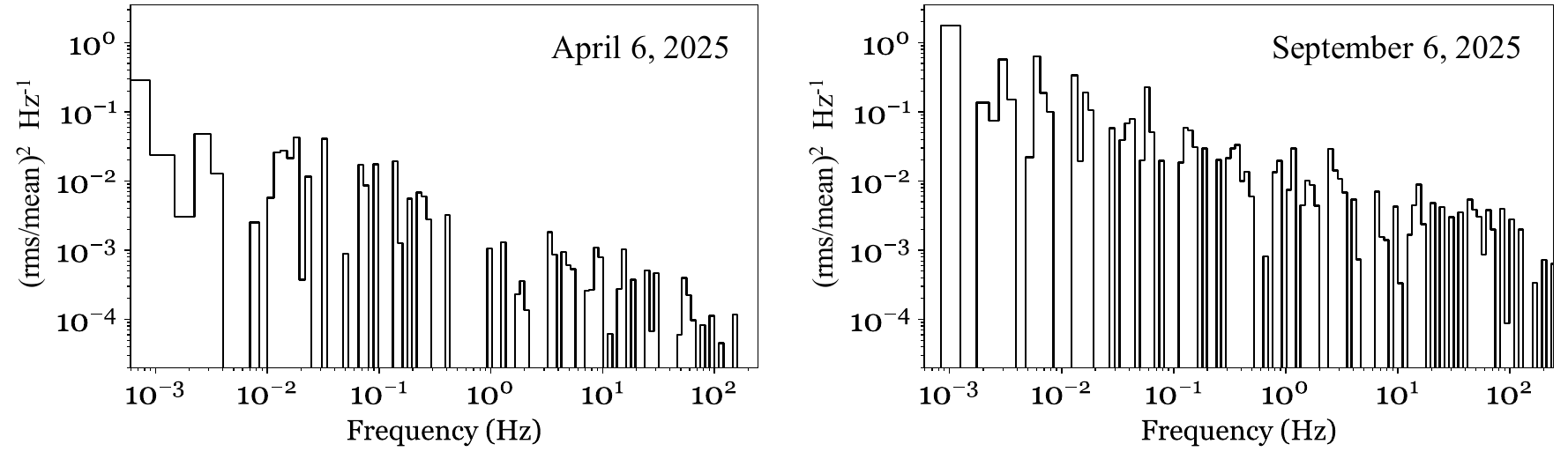}
    \caption{\nustar\ $3–79$ keV PDS (cospectra) of \src\ from the 04/06/2025 (left) and 09/06/2025 (right) observations.  Variability was much higher in the bright hard September data.}
    \label{fig:pds}
\end{figure}

The PDS and co-spectra for most observations appear featureless, with no significant signs of red noise or quasi-periodic oscillations (QPOs).  Subtracting the Poisson noise and rebinning the spectra likewise did not produce any distinguishable features (Figure \ref{fig:pds}, left). 
This lack of observed variability, particularly during the low/hard state when low-frequency QPOs are often seen, is 
likely due to insufficient counts for statistically robust variability detection.

The possible exception is the data collected during September 2025, when \src's spectrum was bright and very hard. The cospectrum shows signs of increasing variability at lower frequencies (Figure \ref{fig:pds}, right), as is expected in the hard state \citep{Belloni2016}.   
\\

\section{Results and Discussion} \label{sec:results}

\subsection{The peculiar outburst evolution of \src}\label{subsec:outburst}

\begin{figure}
    \centering
    \includegraphics[width=\linewidth]{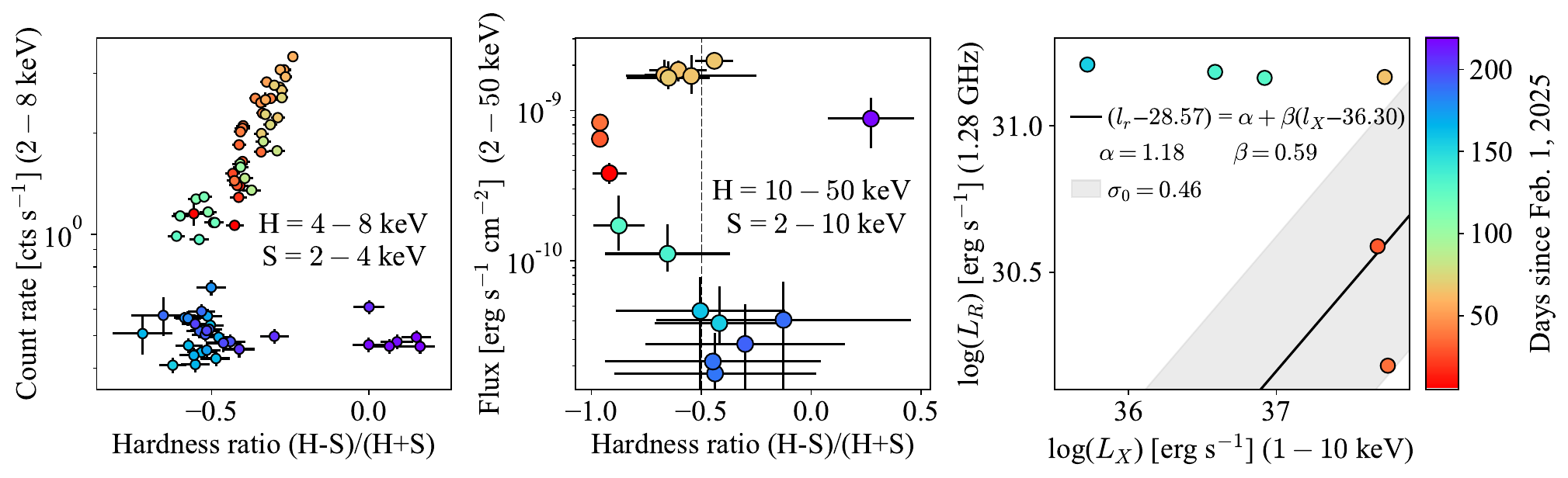}
    \caption{Hardness-intensity diagrams (HIDs) showing the evolution of the \src\ outburst. 
    Left: HID generated from \swift/XRT data collected in PC mode between February--September 2025.  
    Count rates were used to calculate the hardness ratios, with the soft band defined as $2-4$ keV and the hard band as $4-8$ keV.  Center: \nustar\ HID based on broadband flux observed during the same period.  Here, the soft band is defined as $2-10$ keV and the hard band as $10-50$ keV. 
    Right: Projected $1-10$ keV X-ray vs estimated radio luminosity at 1.28 GHz.  
    The black line and shaded region show the best-fit relation and intrinsic scatter for BH XRBs given by \cite{Gallo2018}.  See Section \ref{subsec:outburst} for more. 
    }
    \label{fig:hid}
\end{figure}

Figure \ref{fig:hid} illustrates how \src\ evolved in brightness and hardness as the outburst progressed. We used \swift/XRT observations conducted in window timing (WT) mode to mitigate pileup.  The WT observations were performed every $\sim3$ days starting in February 2025, with a gap in June.  

When \src\ was first detected, it was already in the soft state.  It is unknown when the outburst started or what the spectral hardness was at the time, since it appears to have commenced while the Galactic center was outside telescope observability windows due to solar constraints.  In the three months following the initial detection, the flux climbed steadily (cf. also Table \ref{tab:spectra}).  As the flux peaked in early April 2025, the spectrum grew harder, though the thermal disk emission remained strong.  The outburst began to decline, and the spectrum grew softer as the flux dropped, finally turning towards the low/hard state in July 2025.    

\src\ appeared to remain in the low/hard state through 
August 2025.  An additional \nustar\ ToO observation 
conducted on September 6, 2025 
showed that \src\ had brightened considerably, particularly in the hard band ($>10$ keV; 
Figure \ref{fig:nuim_col}, bottom panel).

In the canonical X-ray transient evolution, outbursts go through a series of transitions that are embodied by a $q$-shaped hardness-intensity diagram (HID) \citep{Belloni2016, Kalemci2022}.  Starting off in the low/hard state (the lower right part of the $q$), with the X-ray spectrum dominated by non-thermal emission, the luminosity increases and the softer disk component becomes more dominant.  The outburst is thought to peak in the high/soft state (upper left part of the $q$), followed by a return to the low/hard state as it fades back into quiescence, completing its circuit on the $q$ track \citep{Belloni2016, Kalemci2022}.  In cases where an outburst fails to cycle through the high/soft state, it is considered a "failed" outburst \citep{Sturner2005, Capitanio2009, Motta2010, Belloni2016}.  Bright radio emission is often observed in the hard state.

\src\ followed an evolutionary path that is unique in several ways.  Even in the high/soft state, it exhibited strong non-thermal emission from Comptonization.  This was particularly remarkable during the first part of April 2025, when \nustar\ observations showed copious emission above $\sim20$ keV (Figure \ref{fig:nuim_col}) even as the flux reached its peak (Figure \ref{fig:lc_sw}).  Over the following months, \src\ followed a more typical pattern as the outburst faded and the emission eventually hardened.  However, by September 2025, it had grown not only bright again but also harder than ever observed before (Figure \ref{fig:hid}, center).  Typically, the bright/hard state is only observed when an outburst transitions from the (initially fainter) hard state into the high/soft state \citep{Kalemci2022}.  While we cannot rule out that \src\ will eventually cycle through a second high/soft state, it has so far shown no inclination to do so based on more recent broadband observations (to be presented in a forthcoming publication).

Outliers in hardness-intensity evolution have previously been observed, including transients that occupy only small regions of the canonical HID, low/soft outburst states, and the aforementioned "failed" outbursts \citep{Kalemci2022}.  However, to the best of our knowledge, there are no known instances of an X-ray transient cycling through the high/soft state, declining to the low/hard state, and \emph{then} entering a bright/hard state -- and remaining bright and hard for a period of at least a month, as \src\ appears to have done.  Further X-ray monitoring in 2026 may reveal whether the HID evolution of the source is indeed distinct from the other BH transients.

\src\ is also an outlier in terms of its radio brightness.  In the hard state, radio emission in X-ray binaries (XRBs) has been observed to increase non-linearly with X-ray luminosity \citep{Fender2001, Belloni2016, Gallo2018}.  Radio flux typically decreases 
with the transition to the soft state.  Figure \ref{fig:hid} (right) shows the estimated radio and absorption-corrected X-ray luminosities from simultaneous MeerKAT ($L-$band) and \nustar\ observations, assuming the Galactic center distance ($\sim8.1$ kpc).  
While the softer March 2025 observations indeed exhibited lower radio fluxes, we find similar radio fluxes across a range of X-ray luminosities and spectral hardness for observations during April--July 2025.  \cite{Gallo2018} parametrize the radio-to-X-ray luminosity correlation as $l_r-28.57=\alpha+\beta(l_X-36.30)$, where $l_r$ and $l_X$ are the logarithmic radio and X-ray luminosities.  For BH XRBs, they find a best-fit slope $\beta=0.59$, intercept $\alpha=1.18$, and intrinsic scatter $\sigma_0=0.46$.  The best-fit relation and scatter are illustrated with a black line and shaded silver region, respectively, in Figure \ref{fig:hid} (right).  While the soft data falls within the expected radio-to-X-ray parameter space, the harder data appears far more radio-loud than is typical at the corresponding X-ray luminosities.

\subsection{Fe line complex and reflection changes}\label{subsec:fe}

\begin{figure}
    \centering
    \includegraphics[width=\linewidth]{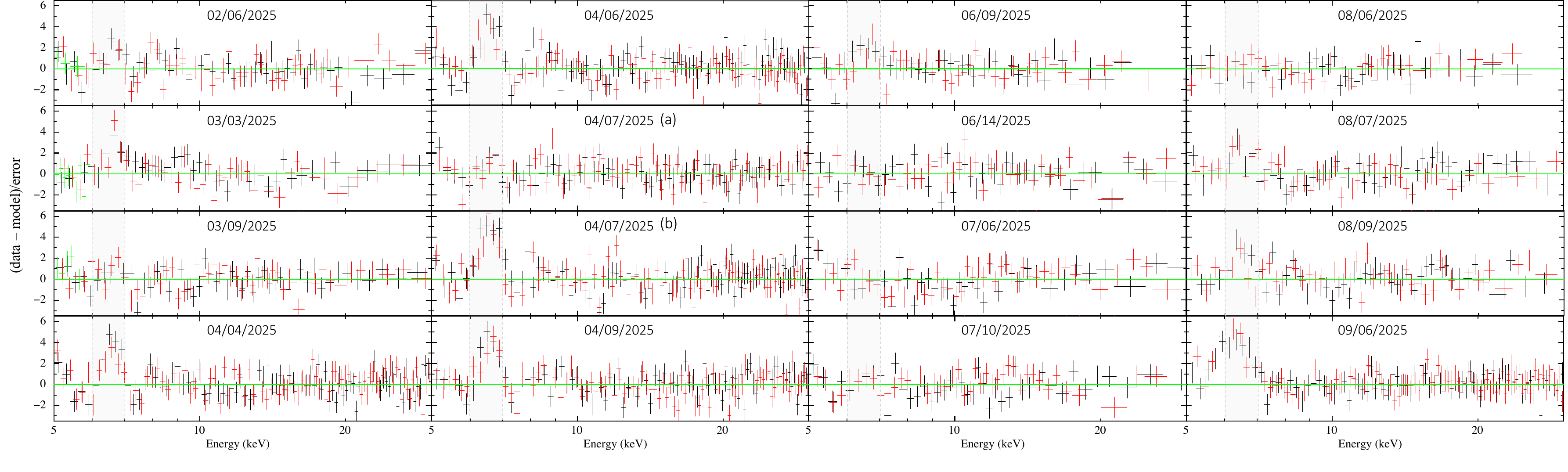}
    \caption{\nustar\ FPMA (black) and FPMB (red) spectral residuals for \src\ after fitting with the \texttt{DUST*TBABS*(THCOMP*DISKBB)} continuum model.  The $6-7$ keV band is highlighted in silver to illustrate the changes observed in the Fe emission lines.  Line widths, centroids, and intensities varied considerably throughout 2025.  See Section \ref{subsec:fe} for more.}
    \label{fig:feres}
\end{figure}

Figure \ref{fig:feres} shows the residuals 
after fitting the spectra 
from the 
\nustar\ observations listed in Table \ref{tab:obs} with a continuum model -- \texttt{DUST*TBABS*(THCOMP*DISKBB)} -- that accounts for thermal disk emission and Comptonization, highlighting the 
reflection features.  
\src\ exhibited varying levels of Fe emission (and absorption) features throughout the outburst, often appearing to change in shape, width $\sigma$ and/or intensity over a period of days 
or even hours (as was the case for the April 7 dataset).  
These variations are indicative of changes in the disk geometry (particularly the inner disk radius $r_{in}$) and plasma ionization, as described in Section \ref{subsec:refl}.  
Four of the spectra (dated between June 14--August 6) show no evidence of Fe emission lines in the $6-7$ keV band, though the line reappears on August 7.  Others show distinctly asymmetric Fe lines, a signature of gravitational redshift.  This effect is particularly stark in the September data (Figure \ref{fig:feres}, bottom right).

Also illustrated by Figure \ref{fig:feres} is the appearance of neutral Fe K absorption edges in some of the spectra at $\sim7.1$ keV.  Much like the other Fe line characteristics, the presence and intensity of the absorption edge appears to vary; e.g., there is little sign of it in the March 3 data, but six days later it appears to be strong.  The absorption edge appears less distinct (or not at all) in the hard-state spectra (August--September 2025), hinting at a potentially 
variable local 
absorber, e.g. a disk wind.

Finally, Figure \ref{fig:feres} highlights the variation in high-energy ($>10$ keV) X-ray emission, where the Comptonized continuum and reflection emission are dominant.  A marked increase in hard X-ray photons is evident in the April 2025 spectra, particularly above $\sim20$ keV.  A similar effect can be seen in the September data.

\subsection{Comparison to \Swft}

\Swft\ outbursted on May 28, 2016 \citep{Degenaar2016, Mori2019}.  It was observed by \nustar\ on June 9, 2016, for 49 ks.  A detailed description of the data processing and analysis can be found in \cite{Mori2019}, hereafter M19.  The 2016 exposure was approximately twice as long as the 2025 observation, and unlike the 2025 \nustar\ spectrum, it was not fit jointly with \xmm; as the 2016 observation preceded the MLI tear, M19 was able to fit the full $3-79$ keV \nustar\ band then.

Remarkably, the best-fit spin and inclination derived by M19 from fitting the June 2016 \nustar\ data are in full agreement with the results of our September 2025 observation: $a>0.94$, $\theta=28\pm2$.  The residuals of the absorbed powerlaw fit to the 2016 \nustar\ spectra (Figure 4 in M19) show a similar relativistically broadened, asymmetric Fe emission line as we observed in the 2025 \nustar\ spectra (Figure \ref{fig:obs4_joint_results}).  However, the continuum appeared different in the 2016 data, both for the thermal disk temperature ($kT=0.46^{+0.15}_{-0.04}$) and the non-thermal powerlaw photon index ($\Gamma=1.59\pm0.01$).  This is not surprising, as \Swft\ was likely in a harder state than \src\ during our September observation.  
Note that the 2016 data were fit using a different relativistic reflection model (\texttt{constant*dust*tbabs*(diskbb+nthcomp+relconv*reflionx)}).  Nevertheless, the parameters that should remain constant over decades (and across reliable spectral models) -- the spin $a$ and inclination $\theta$ of the source -- indeed are consistent.

\begin{figure}
    \centering
    \includegraphics[width=0.95\linewidth]{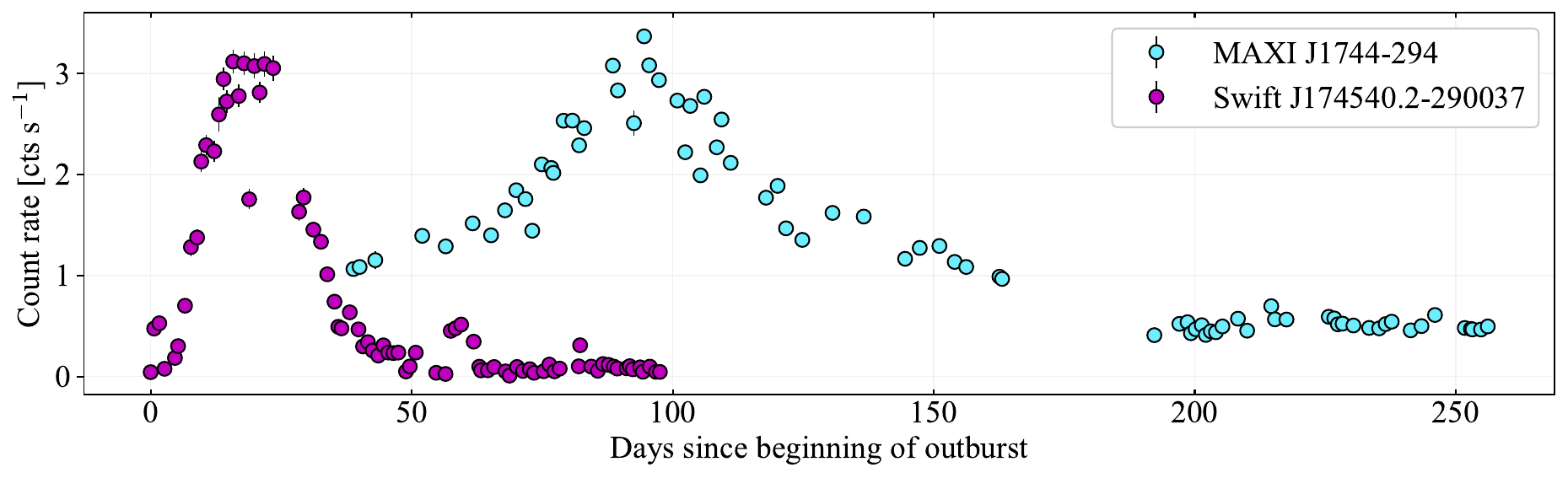}
    \caption{The 2016 \swift/XRT light curve of \Swft\ (magenta), overlaid with the 2025 \swift/XRT light curve of \src\ (cyan).  While the 2016 \Swft\ outburst declined quickly within weeks, \src\ remained bright for at least a year. }
    \label{fig:2016lc}
\end{figure}

The 2016 outburst differed from its 2025 counterpart in several key respects, chief among them the outburst evolution.  Unlike the 2025 outburst of \src, the 2016 transient was consistently monitored starting from its inception.  Also unlike the 2025 outburst, it declined precipitously within weeks.  Figure \ref{fig:2016lc} shows the light curve of the 2016 outburst overlaid with the 2025 outburst.  (Note that the 2025 light curve starts more than a month after \src\ was first detected by \maxi, as \swift/XRT was not able to observe \src\ until it exited the Sun constraint window.)  The 2025 event clearly has a very different evolutionary path, with a much slower rise and decline.  Although the 2016 peak luminosity of \Swft\ ($L_X\sim10^{37}$ \lumcgs) was similar to that observed for \src, the latter may have been brighter before \swift/XRT monitoring commenced.

\subsection{Galactic center LMXBs and \src\ classification}\label{subsec:gc_trans}

Identifying BH systems based purely on their X-ray characteristics is challenging, because NS and BH binaries exhibit many similar spectral characteristics.  Nevertheless, there are some differences in their distributions that can help us distinguish them.

\smallskip
\emph{Spin $a$}:  \hspace{0.1cm}
Rapidly spinning BHs can leave imprints on their X-ray spectra, thanks to relativistic effects that are increasingly apparent at smaller radii.  Since BH spin $a$ is inversely correlated to $r_{\rm ISCO}$, and most of the observed X-ray emission originates outside $r_{\rm ISCO}$ \citep{Reynolds2008}, we can infer the spin $a$ of a BH by measuring the 
relativistic distortions of the reflected components in the X-ray spectra \citep{Reynolds2021}.   
Broadband BH-LMXB studies have often yielded high BH spin $a$ measurements \citep{Mori2019, Draghis2023, Draghis2025}.  
Conversely, \cite{Lo2011} showed that NS spin cannot exceed $a=0.7$ (and none have been observed with $a\simgt0.2$).  
As described in Section \ref{subsec:refl}, the September broadband spectra require a high spin ($a>0.92$) to fit the reflection model (Figure \ref{fig:obs4_joint_results}), indicating a rapidly spinning BH.  While we are unable to fully constrain the BH spin from the soft-state observations, likely because the accretion disk did not extend to $r_{\rm ISCO}$ during those intervals (Figure \ref{fig:inner_disk_radius}), the mass limits we derive from those earlier spectral fits are largely consistent with a BH.

\smallskip
\emph{X-ray vs radio flux}:  \hspace{0.1cm}
While X-ray and radio luminosities are believed to be correlated for both BH and NS transients in the hard state, BH XRBs are far more radio-loud relative to their NS counterparts within the same X-ray luminosity ranges \citep{Fender2001, Belloni2016, Gallo2018}.  An estimate of the radio luminosity using 1.28 GHz MeerKAT data shows that the \src\ radio brightness is more than two orders of magnitude higher than the mean expected value for a NS XRB, given the corresponding $1-10$ keV absorption-corrected X-ray luminosity (from simultaneous \nustar\ observations).  The high ratio of radio to X-ray luminosity $L_r/L_X$ strongly favors a BH designation (Figure \ref{fig:hid}, right).

\smallskip
\emph{Peak $L_X$}:  \hspace{0.1cm}
NS transients do not often exceed $L_X\sim10^{37}$\lumcgs\ (2-10 keV); 
when they do, it is frequently in systems with evolved donors \citep{Heinke2024}, which are characterized by high $\dot{M}$ and usually, frequent outburst recurrence \citep{Lin2019}.  Virtually all NS transients with peak luminosities $L_X>10^{37}$\lumcgs\ have exhibited type I X-ray bursts \citep{Heinke2024}.  Combined with a lack of observed type I X-ray bursts, \src's projected X-ray luminosity -- which peaked at a few $10^{37}$\lumcgs\ -- is therefore more consistent with a BH designation.

\smallskip
\emph{Outburst decay timescales}:  \hspace{0.1cm}
\cite{Yan2015} note that average $e$-folding rise and decay times ($\tau_{\rm rise}$ and $\tau_{\rm decay}$) for BH transients are longer than those for NS outbursts, although significant overlap exists between their respective distributions.  The rise time for \src\ is unknown since  
in the period immediately preceding its detection by \maxi, the Galactic center was not observable due to Sun-angle constraints.  
However, we can estimate the decay timescale $\tau_{\rm decay, 10-90\%}$, under the assumption that the outburst peak occurred during the period when \src\ was monitored.  As Figure \ref{fig:lc_sw} indicates, \src's $2-8$ keV flux peaked around MJD$=60760$ at $\sim2\times10^{-9}$ \fluxcgs, and did not declined by 90\% until $\sim80$ days had passed.  Then, following the definition of \cite{Yan2015} (equation 1), we calculate $\tau_{\rm decay, 10-90\%}\simgt36$ days, which is far more typical of BH transients ($<\!\tau_{\rm decay}\!>\ =25.7^{+31.4}_{-14.1}$ days) compared to their NS counterparts ($<\!\tau_{\rm decay}\!>\ =10.8^{+13.3}_{-5.9}$ days).

\smallskip
\emph{Lack of NS signatures}:  \hspace{0.1cm}
As discussed in Section \ref{subsec:bb_fits}, NS transients often show a thermal emission component emanating from surface hotspots or the boundary layer.  This emission is generally hotter than the disk blackbody emission and emanates from a smaller surface region \citep{Ponti2018}.  We observe no signs of such an emission component.  Likewise, we observe no temporal signatures common to NS transients: no type I X-ray bursts, and no pulsations are evident in the data -- despite extensive monitoring ($>600$ ks with \nustar, \swift, and \xmm\ combined).

\smallskip
\emph{Outburst recurrence}:  \hspace{0.1cm}
NS-LMXB outbursts in our Galaxy typically reach peak luminosities of $L_X\sim10^{36}-10^{37}$\lumcgs, and  
often recur after a few years.  \citep{Yan2015, Degenaar2012, Degenaar2015, Heinke2024}.  In contrast, BH transients have higher average luminosities ($L_X\sim10^{37}-10^{38}$\lumcgs) and often go decades without recurring\footnote{Exceptions abound, often -- but not always -- in systems with evolved donors, which usually undergo frequent outbursts due to the high mass accretion rate $\dot{M}$ \citep{Lin2019}.} \citep{Yan2015, Mori2021}.  This distinction is especially stark in the Galactic center, where known 
transient NS-LMXBs (identified through pulsations or type I X-ray bursts) all have recurrence times of $<5$ years, while 
\src\ is the first BH-LMXB candidate to be 
detected in outburst more than once \citep{Mori2021}.\footnote{Although the sample  
of confirmed Galactic center NS-LMXBs is small, it is still by far the highest-quality, most complete sample of transient LMXBs available \citep{Mori2022}; hence we consider it appropriate to employ it -- as part of a broader metric -- for classifying other Galactic center transients.}  In the Galactic plane, all-sky X-ray monitors are more sensitive to BH transients than to their typically fainter, shorter NS counterparts. Yet recurrent NS transients are still detected far more frequently than BH systems, indicating that NS-LMXBs generally have shorter recurrence times.  This pattern is further supported by the 
transients observed in the Galactic center, where recurrence times for outbursts both bright and faint are well constrained by decades of monitoring with X-ray telescopes including \chandra, \xmm, and \swift.  Assuming that \src\ is, in fact, the same source as \Swft, the recurrence time is almost nine years -- significantly longer than any NS identified in the Galactic center. Moreover, \Swft/\src\ appears to have been quiescent for at least 16 years \emph{prior} to the 2016 outburst, as it was not detected despite frequent monitoring starting in 1999; hence the previous outburst interval appears to be $>16$ years.

\smallskip
\emph{Location}:  \hspace{0.1cm}
Assuming that \src\ is located at the Galactic center (a reasonable assumption given the high $N_H$, dust-scattering halo, and position offset of $<20$ arcsec from Sgr A*), its distance to Sgr A* is $\simlt0.7$ pc.  Based on three-body simulations of stellar binaries orbiting Sgr A*, \cite{Bortolas2017} postulated that supernova (SN) eruptions would result in the ejection of NSs from within the influence radius of the SMBH, while their BH counterparts would remain clustered in the central pc.  Similarly, \cite{Panamarev2019} found that N-body simulations of stellar dynamics in the nuclear star cluster (NSC) resulted in the removal of most NSs through their natal kicks, while $>10^4$ BHs remained; NS binaries are particularly susceptible to disruption even if they do not leave the NSC, since they become wide and therefore don't survive the dense stellar environment.  \cite{Jurado2024} also found that for 
a realistic initial stellar density distribution $\alpha=1.5$, most post-SN BHs remain bound to the SMBH, while a majority of their NS counterparts escape.  This hypothesis is empirically supported by the observed distributions of identified NS-LMXBs and candidate BH-LMXBs in the Galactic center region.  As highlighted by \cite{Mori2021}, BH-LMXB candidates are concentrated around the central pc, while NS-LMXBs are distributed more uniformly in radius out to a distance of $\sim50$ pc, with none identified within the central pc.  The location of \src\ is therefore more consistent with the BH-LMXB population in the Galactic center.

\smallskip
While some of the characteristics outlined above are themselves not definitive in distinguishing BH from NS LMXBs, taken together, they are strongly indicative of a BH-LMXB designation for \src. 
This would 
reaffirm \src/\Swft\ as the third 
candidate BH transient to outburst within a projected distance of $<1$ pc from Sgr A*, the other 
two being CXOGC J174540.0–290031 \citep{Muno2005}  
and SWIFT J174540.7-290015 
\citep{Mori2019}.  Combined with the 12 quiescent BH-LMXB (qBH-LMXB) candidates identified within the central 30 arcsec of our Galaxy \citep{Hailey2018, Mori2021}, these sources are thought to be representative of an even larger population of hundreds of BH-LMXBs that inhabit the central pc \citep{Morris1993, Hailey2018, Generozov2018, Panamarev2019, Mori2021}.  \\

\section{Summary} \label{sec:concl}

We present a comprehensive multi-wavelength study of \src, a new transient detected in the Galactic center.  Our study follows the evolution of the outburst from the soft to intermediate and then hard state, and reveals \src\ to be a likely BH-LMXB.  This conclusion is 
based on the aggregate properties of the source, including the 
\begin{itemize} \setlength{\itemsep}{-1pt}
    \item Spectral properties (e.g. spin $a>0.92$; $2\sigma$ c.l.),
    \item Radio-loud emission in the hard state,
    \item Highest measured luminosity ($L_X> 3.5\times10^{37}$\lumcgs),
    \item Long outburst decay time, 
    \item Lack of pulsations/type I X-ray bursts, 
    \item Long outburst recurrence time, 
    and 
    \item Location within 1 pc of Sgr A*,
\end{itemize}
all of which are consistent with the BH-LMXB scenario.

\begin{acknowledgments}
We thank 
the \nustar\ SOC, \chandra\ and \xmm\ support
teams for their assistance in the execution and analysis of the observations.  Support for SM, KM and the Columbia University team was provided by \nustar\ AO-10 (80NSSC25K0653), \chandra\ AO-26 (SAO GO5-26016X) and \xmm\ AO-23  (80NSSC25K0651) programs. 
SM acknowledges support by the National Science Foundation Graduate Research Fellowship under Grant No. DGE 2036197 and the Columbia University Provost Fellows Program.  
CJ acknowledges the National Natural Science Foundation of China through grant 12473016, and the support by the Strategic Priority Research Program of the Chinese Academy of Sciences (Grant No. XDB0550200). MP acknowledges support from the JSPS Postdoctoral Fellowship for Research in Japan, grant number P24712, as well as the JSPS Grants-in-Aid for Scientific Research-KAKENHI, grant number J24KF0244.  GP acknowledges financial support from the European Research Council (ERC) under the European Union’s Horizon 2020 research and innovation program HotMilk (grant agreement No. 865637), support from Bando per il Finanziamento della Ricerca Fondamentale 2022 dell’Istituto Nazionale di Astrofisica (INAF): GO Large program and from the Framework per l’Attrazione e il Rafforzamento delle Eccellenze (FARE) per la ricerca in Italia (R20L5S39T9).  LM acknowledges the support by the project PRIN 2022 – 2022LWPEXW – “An X-ray view of compact objects in polarized light”, European Union funding – Next Generation EU, Mission 4 Component 1, CUP C53D23001180006. AT acknowledges financial support by the Istituto Nazionale di Astrofisica (INAF) grant 1.05.24.02.04: ``A multi frequency spectro-polarimetric campaign to explore spin and geometry in Low Mass X-ray Binaries''. The work of L.M. is supported by the PRIN 2022 - 2022LWPEXW - “An X-ray view of compact objects in polarized light”, CUP C53D23001180006. 
\end{acknowledgments}

\begin{contribution}
SM was responsible for overall project management, writing the manuscript, X-ray data analysis, and supervising research assistants.  
KM came up with the initial research concept, served as the PI of the \nustar, \xmm\ and \chandra\ ToO programs, obtained the funding and edited the manuscript. 
MR processed the \swift\ data, extracted spectral and timing products, and drafted the relevant text.  
PD performed reflection modeling and drafted the relevant sections in the manuscript.  
BL processed and analyzed the \chandra\ data, generated the \nustar\ background files, and helped complete the manuscript. 
EM conducted timing analysis and drafted the relevant text.  
CJ created the dust scattering models and drafted the relevant text.  
AC supervised Keck data acquisition and analysis, and drafted the relevant section in the manuscript, with assistance from SG.  
GKJ provided and analyzed \nicer\ data and contributed to the manuscript. 
NG provided radio observation details and drafted the relevant text.  
LM generated IXPE light curves and reviewed the draft.  
RC, MH, and AG were involved in obtaining and analyzing NIR data.  
SZ, GS, and OK provided proprietary \nustar\ data.  
MB advised on timing analysis. 
MN assisted with \chandra/ACIS-S HETG observing setup and analysis. 
CH, JT, SM, ND, MP, FC, GP, MS, JH, MS, DH, JG, and RF participated in helpful discussions and reviewed the manuscript.

\end{contribution}

\facilities{NuSTAR, XMM-Newton(EPIC), Swift(XRT), CXO(ASIS-S), XRISM(Resolve and Xtend), NICER, MeerKAT, Keck, JWST}

\software{DS9, 
    HEASoft v6.34 \citep{heasoft2014},
    XSPEC v12.14.1 \citep{Arnaud1996},
    NuSTARDAS v2.1.4a (\url{https://heasarc.gsfc.nasa.gov/docs/nustar/analysis/nustar_swguide.pdf}), %\\
          CIAO \citep{Fruscione2006},
          HENDRICS \citep{Bachetti2018} (\url{https://ascl.net/1805.019}), Stingray \citep{stingray} (\url{https://doi.org/10.5281/zenodo.13974481}),
          astropy \citep{2013A&A...558A..33A,2018AJ....156..123A%,2022ApJ...935..167A
          }, Matplotlib \citep{Hunter2007}  
          } \\

\appendix

\section{\nustar\ images}\label{sec:nu_im}

\begin{figure}
    \centering
    \includegraphics[width=\linewidth]{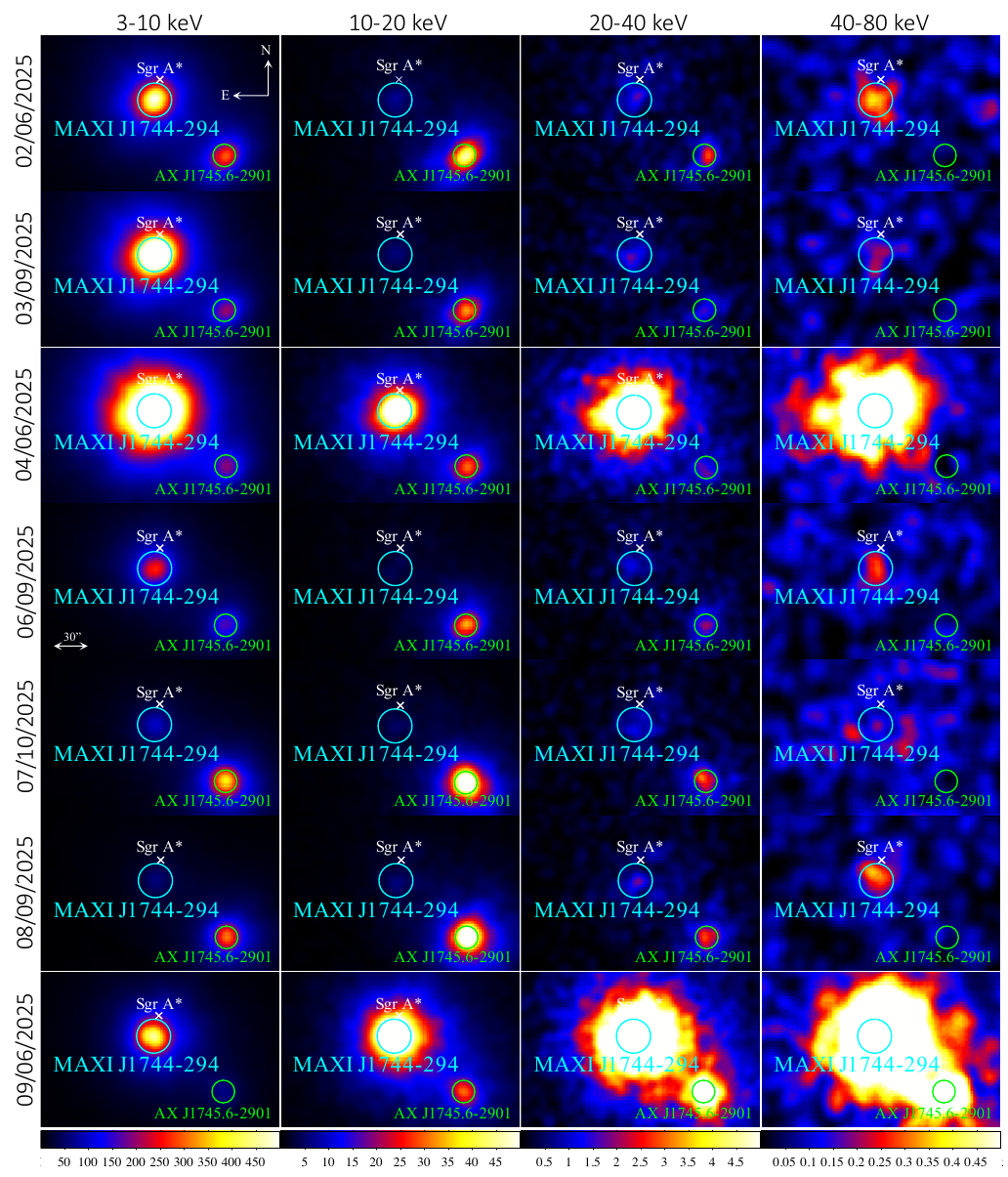}
    \caption{\nustar/FPMA images from a subset of the \src\ observations (cyan circle).  In order from left to right, the filters applied are 3-10 keV, 10-20 keV, 20-40 keV, and 40-60 keV.  The nearby NS-LMXB \axj\ is visible in the lower right (green circle). 
    \src's flux peaked in April; following a decline, it brightened again in September even as it remained hard.}
    \label{fig:nuim_col}
\end{figure}

\begin{figure}[h!]
    \centering
    \includegraphics[width=\linewidth]{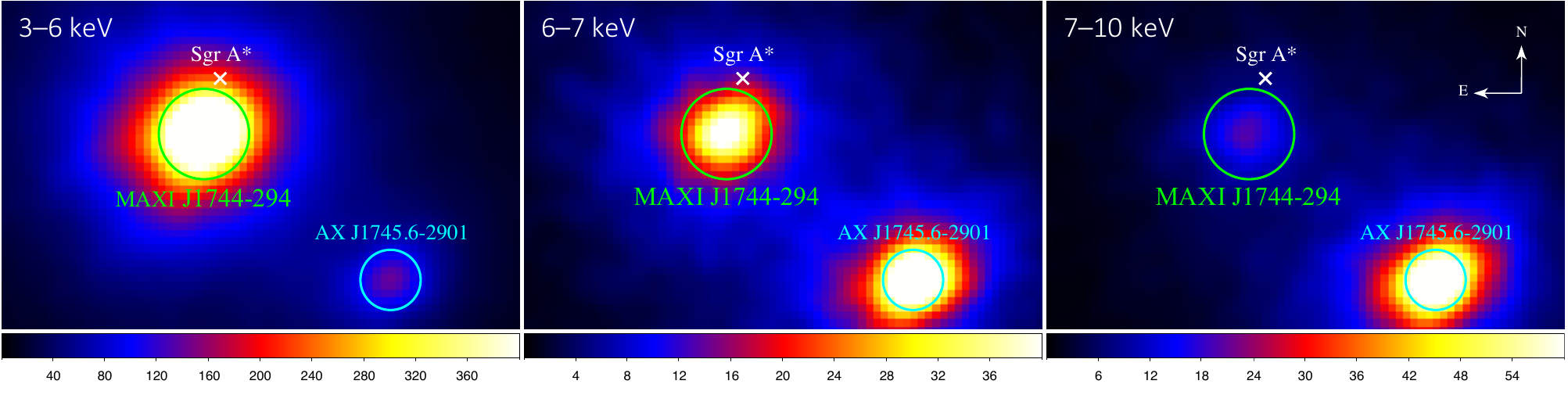}
    \caption{\nustar/FPMA image from Obs. ID 81001323003 in the $3-6$ keV (left), $6-7$ keV (center), and $7-10$ keV (right) bands.  While \src\ (green) clearly dominates the emission in the $3-6$ keV band, \axj\ (cyan, lower right) is $\sim4$ times brighter in the $7-10$ keV band, with a slightly higher flux between $6-7$ keV.  Due to the challenges in modeling the \axj\ contamination in the \xmm/EPIC and \swift/XRT data, we cut off these spectra above 6 keV where the \axj\ emission dominates.  }
    \label{fig:obs2_3to10}
\end{figure}

Figure \ref{fig:nuim_col} shows 
FPMA images for a subset of the \nustar\ observations in several energy bands, illustrating the changes in hardness as the \src\ outburst evolved.  In the first observations (February-March 2025), \axj\ appeared brighter than \src\ in the 10--40 keV energy range.  There was little X-ray emission above 60 keV.  By contrast, in the April observations, \src\ dominated the X-ray flux across the entire 3-79 keV \nustar\ energy band; there was significant X-ray emission even above 60 keV.  
\src\ became much fainter in June and by August, had definitely entered the low/hard state.  However, by the following month, the outburst brightened again, though it remained very hard.    \\

\section{\nustar\ background 
modeling and 
subtraction}\label{sec:bkg}

\begin{figure}
    \centering
    \includegraphics[width=\linewidth]{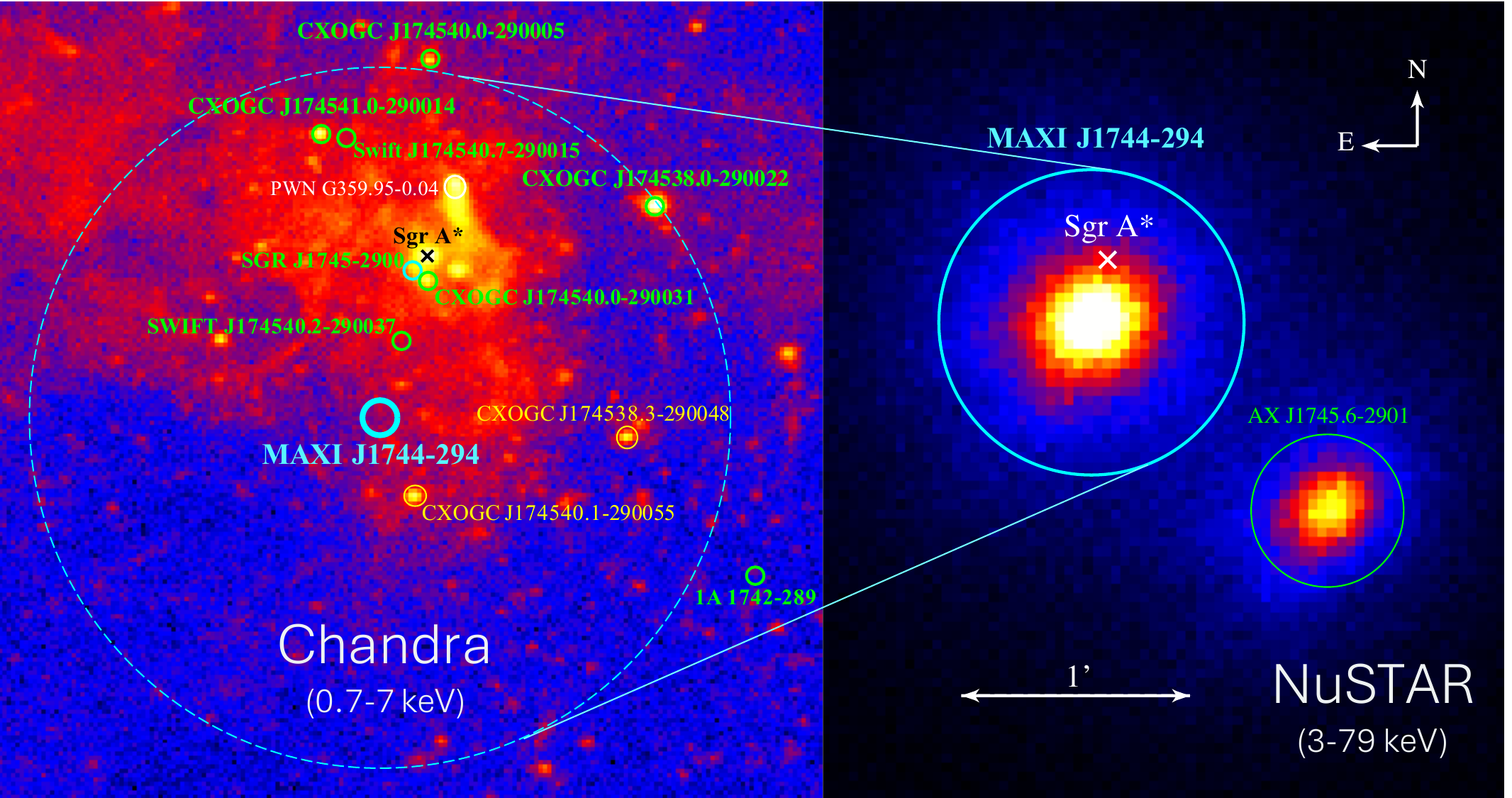}
    \caption{Left: \chandra\ image of $r=40$\arcsec\ region (dashed circle) around \src.  Known X-ray transients (including the \swift/XRT position for \Swft) are shown in green; the location of Sgr A* is marked by a black "X"; yellow circles mark the locations of unclassified X-ray point sources, while the PWN G359.95-0.04 is marked in white.  The diffuse emission in the upper left belongs to the SNR Sgr A East.  Right: \nustar\ image of \src\ (cyan circle) and \axj\ (green).  \nustar's angular resolution (18\arcsec\ FWHM) renders the \src\ spectrum susceptible to contamination from the various sources mapped out %in the top image
    on the left.  \axj\ (lower right) is also a source of contamination.  See Section \ref{sec:bkg} for more.}
    \label{fig:bkg}
\end{figure}

As Figure \ref{fig:bkg} highlights, the main sources of background contamination in the \src\ spectra are: 
\begin{enumerate}\setlength{\itemsep}{-2pt}
    \item Local (persistent) X-ray emission at/around the location of \src, and
    \item X-ray emission from the nearby NS-LMXB \axj, which was active during the \src\ outburst.
\end{enumerate}

Below we discuss the process of modeling the contamination from these two sources.

\smallskip
\subsection{In-situ (underlying) background}

\src\ is located a mere $\sim18$\arcsec\ from Sgr A*.  Consequently, its emission overlays a region that may include contributions from 
\begin{itemize}\setlength{\itemsep}{-2pt}
    \item occasional Sgr A* flares;
    \item diffuse emission from the SNR Sgr A East;
    \item diffuse emission from the PWN G359.95-0.04;
    \item the central hard X-ray emission (CHXE) \citep{Mori2015}; and
    \item quiescent X-ray point sources, including several transients that were observed in outburst over the past two-plus decades (see Figure \ref{fig:bkg}).
\end{itemize}

To account for this contamination, we utilize an archived \nustar\ observation from an epoch when none of the aforementioned transients were detected in the active state.  While we recognize that some of the sources listed above may have variable emission, it is impossible to determine those variations \emph{during the time when \src\ was in outburst}, especially since the flux from \src\ overwhelmingly dominates the region ($\simgt96$\% for most observations).  However, on the flip side, it is reasonable to assume that those variations are negligible in terms of source contamination, since the total contribution from the \emph{in situ} background is limited to $\simlt2$\% (the rest originating from \axj).

\subsection{\axj\ emission}

An equally significant source of contamination is the emission from the nearby NS-LMXB \axj\, which was in outburst throughout the duration of the \src\ observing campaign.  To account for this contamination, we selected a region that was equidistant from the centroid of \axj\ compared to \src\ and extracted a spectrum.  To remove the underlying background contamination at the location of this region (which is distinct from the underlying background at the location of \src), we also extracted a spectrum -- from an equivalent region -- from the above-mentioned archival observation, which we then subtracted from the \axj\ background spectrum, in theory leaving behind only the "pure" \axj\ contamination, without the "local" background contribution.  This remaining \axj\ background spectrum was then merged with the \emph{in situ} \src-underlying background from the archival observation described above, yielding a background spectrum that included both major sources of contamination: 1) the \axj\ emission, and 2) the underlying (mostly diffuse) emission \emph{at the \src\ location}.

Figure \ref{fig:bkgreg} illustrates the three regions that were incorporated into the background spectrum.  We define the \axj\ background, underlying \src\ background, and underlying \axj\ background regions 1B, 2A, and 2B, respectively, where "1" refers to our \src\ outburst observations and "2" refers to archival data; "A" is for \src\ region, "B" is for \axj\ contamination region.  We used the \texttt{MATHPHA} task in \texttt{FTOOLS} to subtract region 2B from region 2A, and combine the resulting spectrum with that of region 1B.  The final result was a background spectrum where the underlying (quiescent) \axj\ background was subtracted from the "active" \axj\ background spectrum, and then combined with the underlying (quiescent) \src\ background spectrum: 

\begin{equation}
    \rm bkg_{Tot} \quad = \underbrace{\rm bkg_{1B}}_{\footnotesize\begin{tabular}{c}\rm underlying \\ background\end{tabular}} + \quad \underbrace{\rm bkg_{2A}-bkg_{2B}}_{\footnotesize\begin{tabular}{c}\rm \axj\ \\contamination\end{tabular}}
\end{equation}

Where the "underlying background" refers to the (persistent/quiescent) background \emph{at the location of \src}, and "\axj\ contamination" refers to the estimated contamination from \axj\ emission, also at the \src\ extraction region. \\

\begin{figure}
    \centering
    \includegraphics[width=0.9\linewidth]{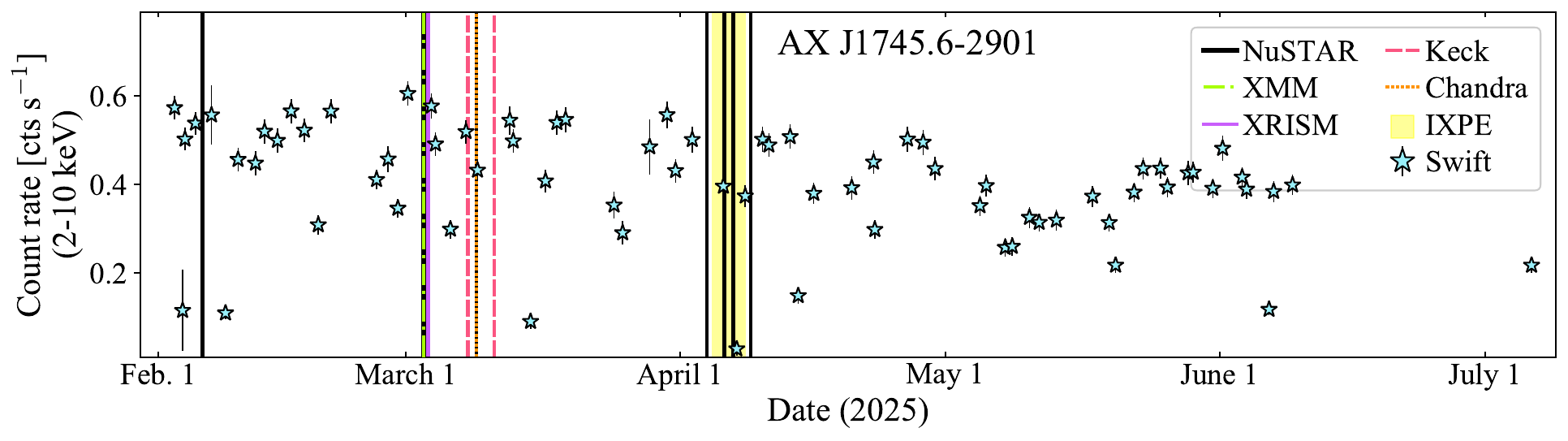}
    \caption{Light curve for \axj\ in the $2-10$ keV band, from \swift/XRT observations.  The dates of other telescope observations are marked by horizontal lines.  The \axj\ flux remained consistently fainter than \src\ in the soft band ($\simlt6$ keV), even when it was brighter in the $10-40$ keV range (Figure \ref{fig:nuim_col}).}
    \label{fig:axj_lc}
\end{figure}

\section{Dust scattering model}\label{sec:dust}

Interstellar dust grains can scatter X-rays, affecting observations of X-ray sources by producing dust scattering halos and rings \citep{Overbeck1965, trumper1973}, modifying the observed spectral shape and timing properties \citep{Corrales2016, Smith2016, Jin2017, Jin2018}. The scattering effects become more pronounced for sources with higher foreground dust column densities and higher X-ray fluxes. Moreover, the scattering optical depth $\tau$ approximately follows $\tau \propto E^{-2}$, making the dust scattering effect most significant in the soft X-ray band \citep{Mathis1991}.

Located toward the Galactic Center, \src\ is a bright source with a high column density of $N_{\rm H} \sim 2 \times 10^{23}$ cm$^{-2}$, and therefore exhibits significant foreground dust scattering. This explains why the {\it Chandra} observation show \src\ with a more extended profile than the typical ACIS PSF. The nearby source \axj, with a similar $N_{\rm H}$, has also been shown to possess a strong dust scattering halo that significantly affects its spectral and timing properties \citep{Jin2017, Jin2018}. Given their small angular separation ($\simlt80$\arcsec) and simultaneous outbursts, the halos of these two sources overlap significantly, introducing additional challenges for disentangling the emission from \src\ and its own halo.

Based on a detailed analysis of \axj's halo profile, \cite{Jin2017}  developed the {\sc axjdust} model to correct spectra from \xmm/EPIC, {\it NuSTAR}, and {\it Swift}/XRT. This correction accounts for the fact that the halo alters the radial profile, making it significantly different from the nominal PSF profile, which in turn leads to biases in the default instrument PSF corrections. A modified version, {\sc fgcdust}, considers only the effect of the Galactic center foreground dust within the Galactic disk.

\begin{figure}
    \centering
    \includegraphics[width=0.6\linewidth]{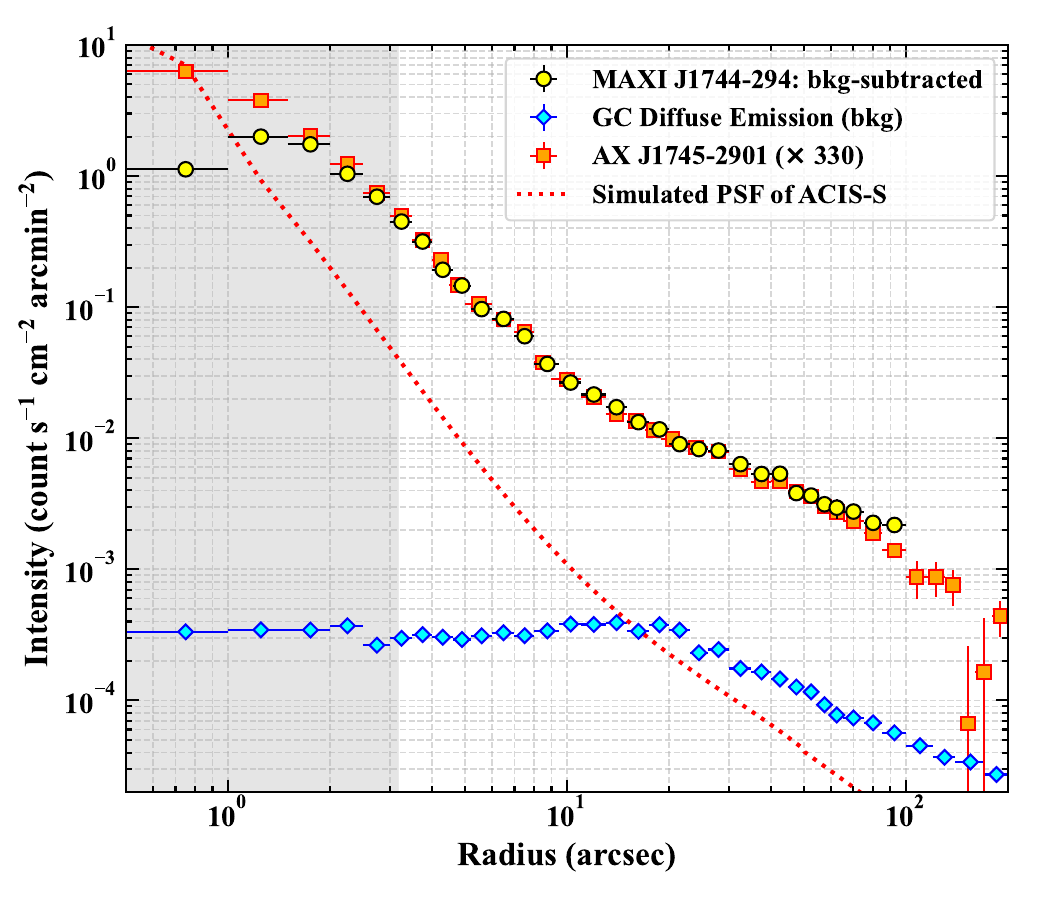}
    \caption{The 2-4 keV radial profile of \src\ (yellow circles), which deviates significantly from the simulated PSF of ACIS-S (red dotted line). The background was estimated from the Galactic center diffuse emission underlying its halo in pre-outburst {\it Chandra} observations (cyan diamonds). The 2–4 keV radial profile of \axj, rescaled by a factor of 330, is also shown for comparison (orange squares); note the deviation of \src\ from the \axj\ profile near $\sim80$\arcsec\ -- corresponding to the position offset between \src\ and \axj\ -- due to the emission from \axj. The shaded region within 3.2\arcsec\ marks the area where the pile-up fraction exceeds 1\%.}
    \label{fig:dust}
\end{figure}

Accurate modeling of \src's halo requires careful removal of contamination from \axj.  This is extremely challenging due to the latter's extended dust scattering halo, which overlaps substantially with that of \src.  
However, if the profile of \src's halo is similar to that of \axj, {\sc axjdust} can provide a reasonable approximation for spectral correction. Using \chandra/ACIS-S observations, we extracted the radial profile of \src\ on the side opposite \axj\ to reduce the halo blending (Figure~\ref{fig:dust}). To account for diffuse emission underneath the halo, we combined {\it Chandra} ACIS-S observations covering the position of \src\ since 2019 with $\ge~10$ ks exposure, when both sources were in the quiescent state, to measure the diffuse emission and subtract it (Jin et al., in prep.). The resulting halo profile of \src\ is indeed similar to that of \axj, consistent with their comparable $N_{\rm H}$. This also indicates that the two sources may have similar foreground dust distributions, 
validating the use of {\sc axjdust} for approximate spectral correction of \src. \\

\section{Additional spectral plots}\label{sec:spec2}

\begin{figure}
    \centering
    \includegraphics[width=0.8\linewidth]{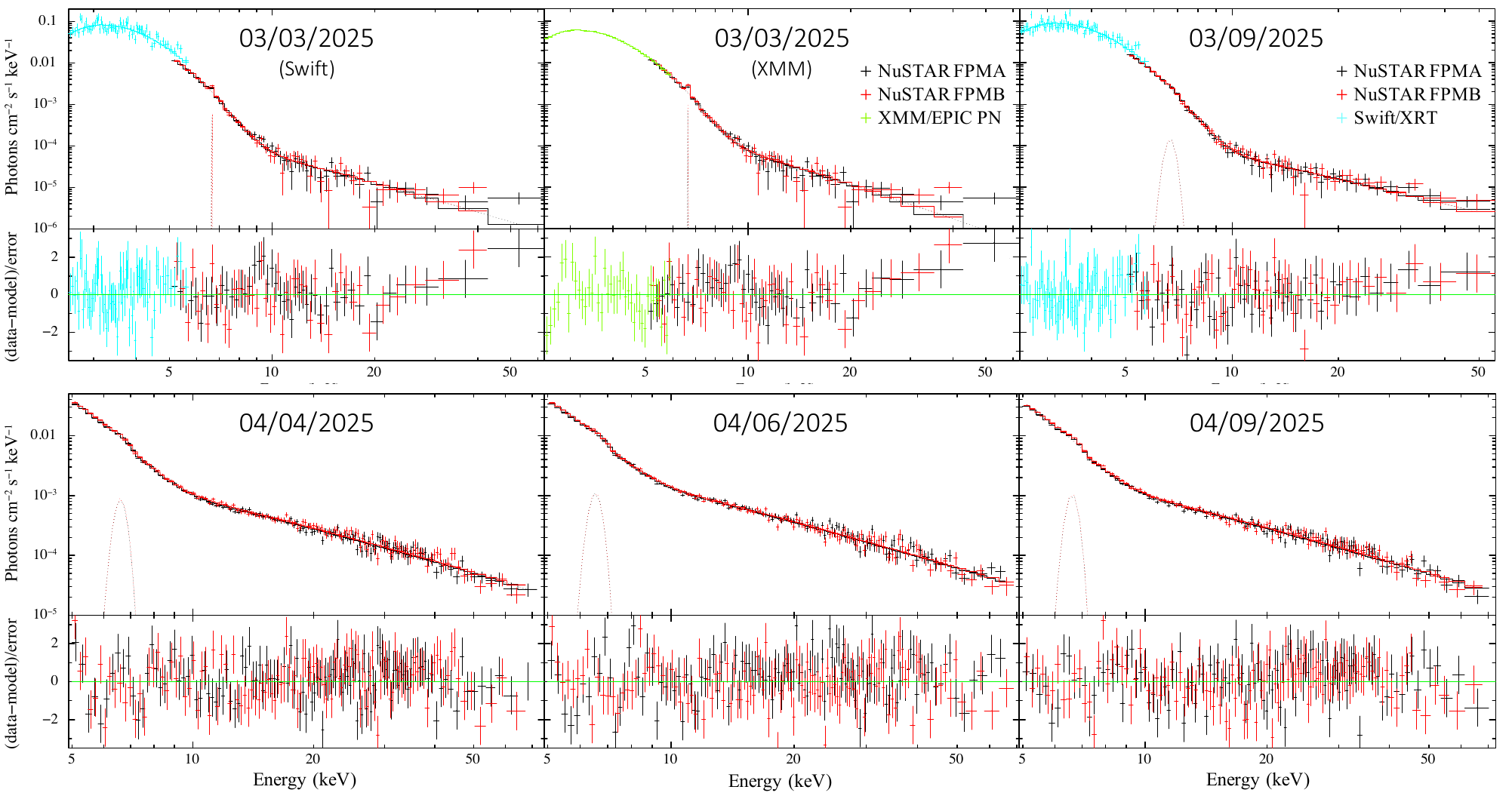}
    \caption{\nustar\ FPMA (black) and FPMB (red) spectra from observations performed between March 3 and April 9, 2025.  The March data were jointly fit with contemporaneous \swift/XRT (cyan) and \xmm/EPIC PN (green) spectra.  All spectra were fit to the model \texttt{DUST*TBABS*(THCOMP*DISKBB+GAUSS)}.  Best-fit spectral parameters are listed in Table \ref{tab:spectra}.}
    \label{fig:spec_obs2_go2}
\end{figure}

Figure \ref{fig:spec_obs2_go2} shows \nustar\ and \swift/XRT spectra collected between March and April 2025.  \swift/XRT (cyan) and \xmm/EPIC PN spectra in the $2.5-6.0$ keV band (green) were fit jointly with the $5.0-79.0$ keV \nustar\ FPMA (black) and FPMB (red) spectra for the March observations (top).  \nustar\ spectra were grouped using the optimal binning method by \cite{Kaastra2016}, with a minimum of 20 cts/bin.  All spectra in Figure \ref{fig:spec_obs2_go2} were fit with the \texttt{DUST*TBABS*(THCOMP*DISKBB+GAUSS)} model.  Changes in the Fe emission line profile and the hard flux are apparent.  See Section \ref{subsec:model1} for more. %\\

\begin{figure}
    \centering
    \includegraphics[width=\linewidth]{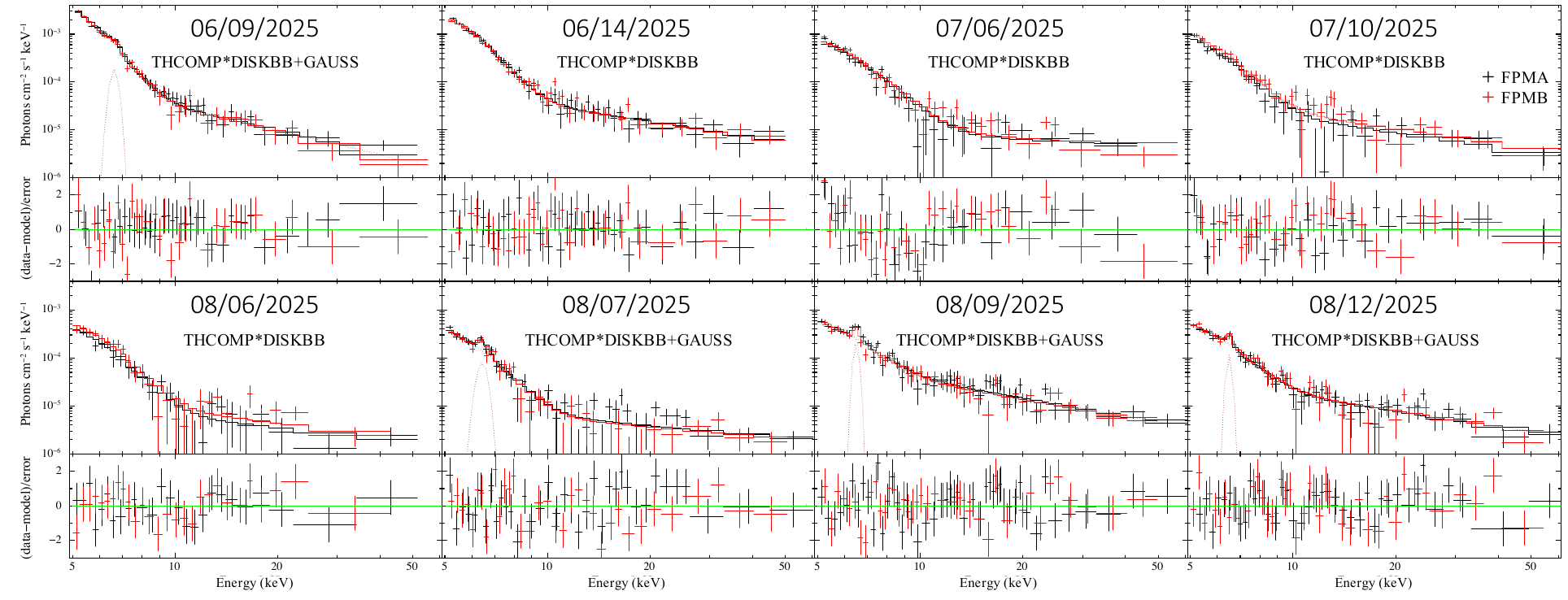}
    \caption{\nustar\ FPMA (black) and FPMB (red) spectra from observations performed between June and August 2025.  All spectra were fit to a disk blackbody + comptonization model, along with absorption and dust scattering (\texttt{DUST*TBABS*(THCOMP*DISKBB)}); half the spectra showed evidence indicating Fe emission lines (Figure xx), for which a Gaussian component was added.  Spectra were rebinned with a minimum $2\sigma$/bin.  Best-fit spectral parameters are listed in Table \ref{tab:spectra2}.}
    \label{fig:spec_obs8to15}
\end{figure}

Eight additional \nustar\ observations of the Galactic center were performed between June 9 and August 12, 2025, all of which detected \src.  \src\ was relatively faint throughout those observations, and starting on June 14, \nustar/FPMB was affected by stray light background (SLB).  The combination of low flux and increased contamination (from \axj\ and the diffuse background, as well as SLB) complicated spectral modeling.  Unlike the earlier observations, some of the spectra taken in the low/hard state did not show evidence of Fe emission lines (Figure \ref{fig:spec_obs8to15}).  However, all spectra showed evidence of both a thermal (disk) component and a non-thermal (comptonization) power-law component.  We therefore fit all spectra to a base \texttt{DUST*TBABS*(THCOMP*DISKBB)} model, and added a Gaussian component where the residuals indicated the presence of an Fe line.  The unfolded spectra, along with residuals from the best-fit model, are shown in Figure \ref{fig:spec_obs8to15}; best-fit spectral parameters can be found in Table \ref{tab:spectra2}. \\

\section{\ixpe\ flux evolution on 04/07}\label{sec:ixpe_lc}

As described in Section \ref{subsec:go3}, the \src\ outburst appeared to enter a transition phase on April 7, 2025.  In the \nustar\ data, this transition manifested in a temporary reduction in the soft band ($3.0-6.3$ keV) flux, followed within hours by an increase in the hard ($7.2-50.0$ keV) flux (Figure \ref{fig:go3_lc_h}).  An unfortunately timed gap in the \nustar\ data -- a result of ground station issues -- prevents us from pinpointing the precise time of this change.  We therefore turned to the \ixpe\ data that was collected concurrently, to see if the same trends appear.

\begin{figure}
    \centering
    \includegraphics[width=0.95\linewidth]{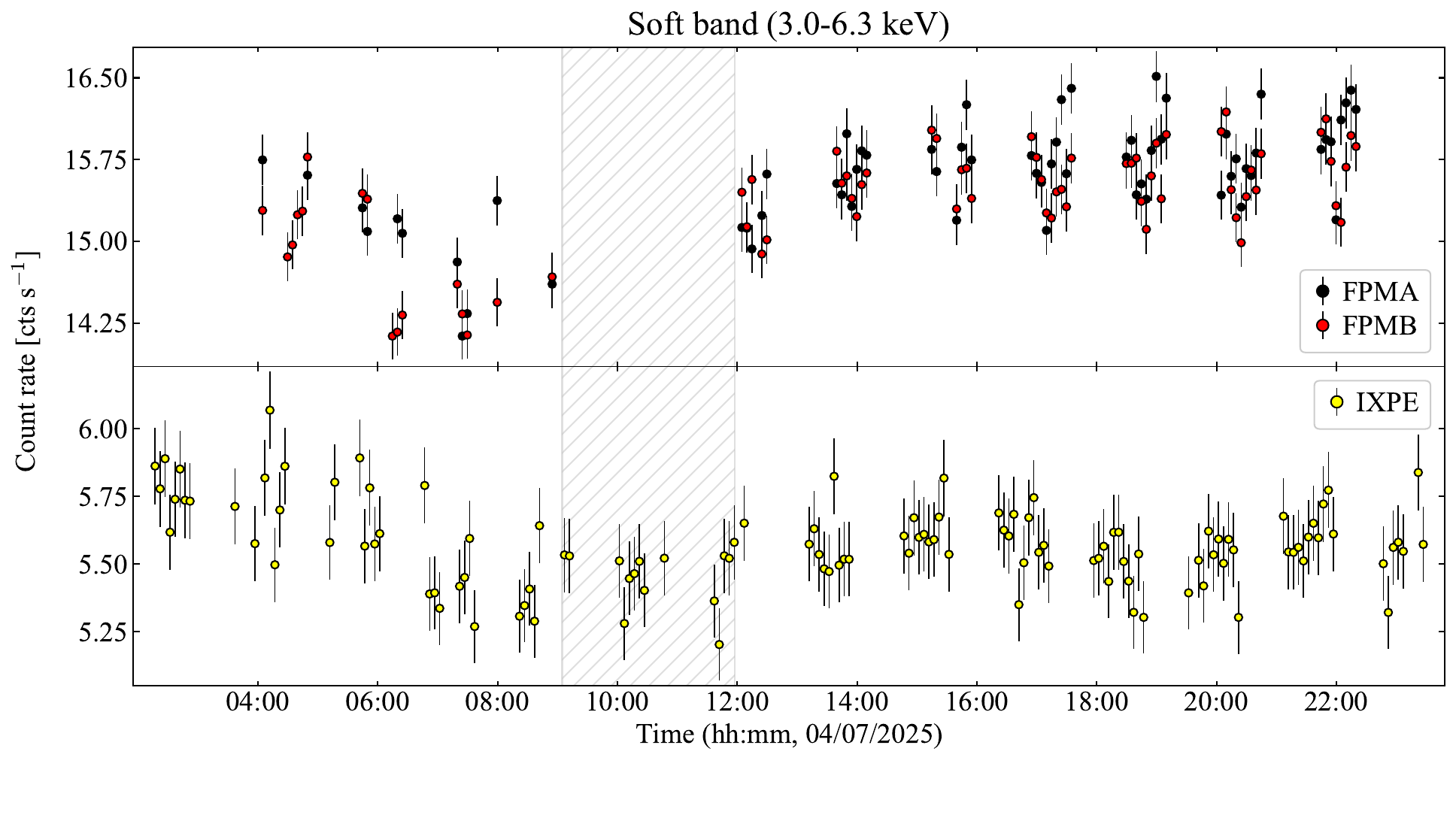}
    \vspace{-0.5cm}
    \caption{Top: \nustar\ FPMA (black) and FPMB (red) light curves from observation ID 31002004006 in the soft ($<6$ keV) band.  Bottom: \ixpe\ light curve in the same energy range (all three DUIs combined).  The shaded area marks the \nustar\ observation gap.  All data was binned in 300s intervals.  The hard X-ray emission appeared to increase significantly partway through the observation (Figure \ref{fig:go3_lc_h}), while the soft X-rays experienced a brief drop in flux just prior to the event. }
    \label{fig:go3_lc_s}
\end{figure}

Figure \ref{fig:go3_lc_s} shows the \nustar\ (top) and \ixpe\ (bottom) light curves spanning the \nustar\ observation period.  Both telescopes show a drop in count rate just prior to the observing gap, though this decrease is more pronounced in the \nustar\ data (probably for statistical reasons; \nustar's higher effective area allows it to collect more photons than \ixpe\ over an equivalent time period).  Due to the relatively large uncertainties in the \ixpe\ count rate, it is difficult to ascertain if/when the soft flux returned to previous levels.  The rise in \nustar's soft count rate following the observing gap may be a spillover effect of the spectral hardening, which \nustar\ is more sensitive to, with increasing effective area up to $\sim10$ keV.

\begin{figure}
    \centering
    \includegraphics[width=0.95\linewidth]{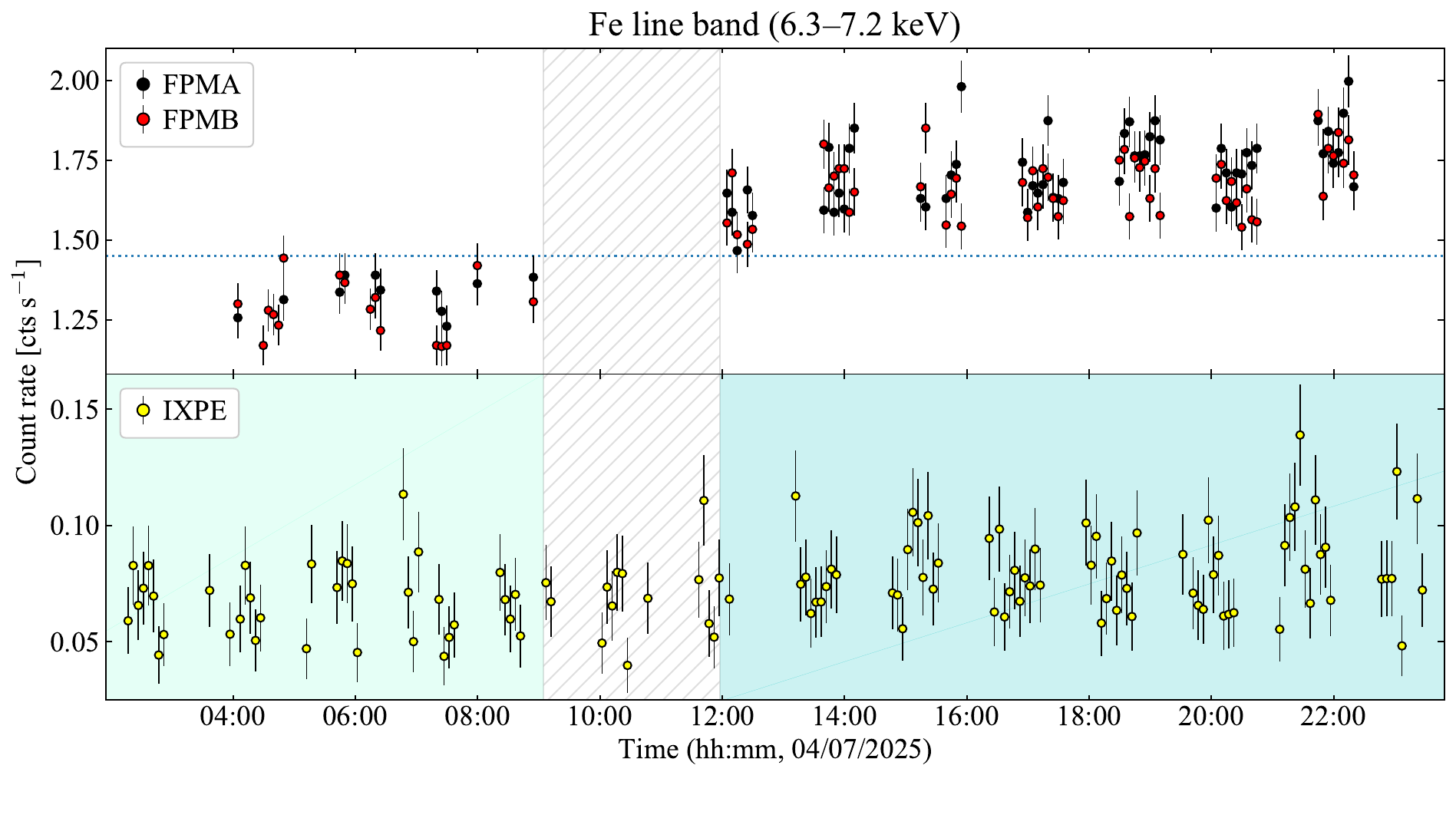}
    \vspace{-0.5cm}
    \caption{Top: \nustar\ FPMA (black) and FPMB (red) light curves from observation ID 31002004006 in the Fe line complex ($6.3-7.2$ keV) band.  Bottom: \ixpe\ light curve in the same energy range (all three DUIs combined).  The shaded silver area marks the \nustar\ observation gap.  All data was binned in 300s intervals.  While the \nustar\ count rate is clearly higher following the gap (blue horizontal dotted line), the evolution of the \ixpe\ light curve is more ambiguous due to poor statistics. The shaded green and blue regions mark the periods that were selected for the K--S test, which indicates that the count rate distributions before (green) and after the gap (blue) are significantly different ($p-$value$=1.8\times10^{-4}$).  }
    \label{fig:lc_nustar+ixpe_fe}
\end{figure}

We also compare the \nustar\ and \ixpe\ light curves in the $6.3-7.2$ keV band, corresponding to the range occupied by the observed Fe emission lines (Figure \ref{fig:lc_nustar+ixpe_fe}).  The \nustar\ data shows significant hardening in the second part of the observation, and this is reflected in the $6.3-7.2$ keV light curve; a blue dotted line in Figure \ref{fig:lc_nustar+ixpe_fe} (top) clearly separates the \nustar\ count rates before and following the observing gap.  The corresponding increase in the \ixpe\ data (bottom) is much less obvious due to the large uncertainties; \ixpe's effective area drops precipitously beyond $\sim6$ keV.  
To quantify the significance of such an increase in the \ixpe\ light curve, we selected two intervals -- one each before (green shaded area) and after (blue shaded area; Figure \ref{fig:lc_nustar+ixpe_fe}) the \nustar\ observing gap, respectively -- and compared the count rate distributions within each interval.  
A K--S test indicates that the two count rate distributions are different beyond the $3\sigma$ level ($p-$value$=1.8\times10^{-4}$ for $\alpha=1\times10^{-3}$; $D_{bef, aft}=0.53$).

\bibliography{ref}{}
\bibliographystyle{aasjournalv7}

\end{document}